\documentclass[mnsc,nonblindrev]{informs3_hide} %

\OneAndAHalfSpacedXI %

\usepackage{amsmath,amsfonts,amssymb}
\usepackage{mathtools}
    \allowdisplaybreaks
\usepackage{bm}
\usepackage[mathscr]{euscript}
\usepackage{enumitem}
\usepackage{xcolor}
\usepackage{graphicx}
\usepackage{multirow}
\usepackage{bigstrut}
\usepackage{booktabs}
\usepackage{microtype}
\usepackage{fix-cm}
\usepackage[ruled,vlined]{algorithm2e}
\usepackage{physics}
\usepackage{xfrac}
\usepackage{nicematrix}
\usepackage{caption}

\DeclareMathOperator{\pr}{\mathbb P}
\DeclareMathOperator{\E}{\mathbb E}
\DeclareMathOperator{\Var}{\mathbb{V}\mathrm{ar}}
\DeclareMathOperator{\VaR}{\mathsf{VaR}}
\DeclareMathOperator{\CVaR}{\mathsf{CVaR}}

\DeclareMathOperator{\RMSE}{\mathrm{RMSE}}

\DeclareMathOperator{\ind}{\mathbb I}
\newcommand{\RR}{\mathbb{R}}
\usepackage{subfigure}

\usepackage{natbib}
 \bibpunct[, ]{(}{)}{,}{a}{}{,}%
 \def\bibfont{\small}%
 \def\BIBand{and}%
\TheoremsNumberedThrough     %
\ECRepeatTheorems

\EquationsNumberedThrough    %

\MANUSCRIPTNO{MS-SMS-22-00204.R2}

\begin{document}

\RUNAUTHOR{Wang, Wang, and Zhang}

\RUNTITLE{Smooth Nested Simulation}

\TITLE{Smooth Nested Simulation: Bridging Cubic and Square Root Convergence Rates in High Dimensions}

\ARTICLEAUTHORS{
\AUTHOR{Wenjia Wang}
\AFF{Data Science and Analytics Thrust, Information Hub, The Hong Kong University of Science and Technology (Guangzhou), Guangzhou, China, \EMAIL{wenjiawang@ust.hk}}
\AUTHOR{Yanyuan Wang, Xiaowei Zhang}
\AFF{Department of Industrial Engineering and Decision Analytics, The Hong Kong University of Science and Technology, Clear Water Bay, Hong Kong Special Administrative Region, China, \EMAIL{yanyuan.wang@connect.ust.hk}, \EMAIL{xiaoweiz@ust.hk}}
}

\ABSTRACT{Nested simulation concerns estimating functionals of a conditional expectation via simulation.
In this paper, we propose a new method based on kernel ridge regression to exploit the smoothness of the conditional expectation as a function of the multidimensional conditioning variable.
Asymptotic analysis shows that the proposed method can effectively alleviate the curse of dimensionality on the convergence rate as the simulation budget increases, provided that the conditional expectation is sufficiently smooth.
The smoothness bridges the gap between the cubic root convergence rate (that is, the optimal rate for the standard nested simulation) and the square root convergence rate (that is, the canonical rate for the standard Monte Carlo simulation).
We demonstrate the performance of the proposed method via numerical examples from portfolio risk management and input uncertainty quantification.
}


\maketitle

\section{Introduction}
Many simulation applications involve estimating a functional of a conditional expectation.
The functional can be in the form of an expectation or a quantile.
Because in general, no analytical expression is available for either the conditional expectation or the functional,
the estimation requires two levels of simulation.
That is, one first simulates in the \emph{outer} level the random variable being conditioned on and then
simulates in the \emph{inner} level some other random object of interest conditioning on each simulated sample---also known as  \emph{scenario}---of the former.
This class of problems is referred to as \emph{nested simulation}.
Specifically, in the present paper we examine the problem of using simulation to estimate quantities of the form
\begin{align}\label{eqgoal}
    \theta = \mathcal{T}(\E [Y|X]),
\end{align}
where $X$ is a $\RR^d$-valued random variable with $d\geq 1$,
$Y$ is a $\RR$-valued random variable,
and  $\mathcal{T}$ is a functional that maps a probability distribution to a real number.

Two representative examples of quantities of the form \eqref{eqgoal} stem from  financial risk management \citep{LanNelsonStaum10,HongJunejaLiu17,DangFengHardy20} and
input uncertainty quantification for stochastic simulation \citep{Barton12,BartonNelsonXie14,XieNelsonBarton14}.

\begin{example}[Portfolio Risk Management]\label{examp:portfolio}
A risk manager is interested in assessing the risk of a portfolio of securities at some future time $T_0$ known as the risk horizon.
The current value of the portfolio, $V_0$, is known to the risk manager.
However, its value at the risk horizon,  $V_{T_0}$, is a random variable that depends on $X$, a collection of risk factors such as interest rates, equity prices, and commodity prices the values of which are realized between time $0$ and time $T_0$.
Moreover, it can usually be expressed as a conditional expectation under a ``risk-neutral measure'': $V_{T_0}(X) = \E[W|X]$, where
$W$ is the discounted cash flow between time $T_0$ and some final horizon $T$ (e.g., the expiration date of the derivatives).
When the portfolio includes complex financial derivatives, as is often the case,
$V_{T_0}(X)$ does not possess an analytical form, and its evaluation relies on Monte Carlo simulation.

Suppose that the portfolio does not generate interim cash flows prior to time $T_0$ and that the risk-free rate is $r$.
The loss of the portfolio at the risk horizon in scenario $X$ is then $Z = V_0 - V_{T_0}(X)$, which can be written as $Z=\E[Y|X]$ if we let $Y=V_0 -  W$.
The risk can be assessed in various ways, such as probability of a large loss $\pr(Z \geq z_0)$, expected  excess loss $\E[\max(Z-z_0, 0)]$, squared tracking error $\E[(Z-z_0)^2]$ for some threshold or target $z_0$, value-at-risk (VaR), or conditional value-at-risk (CVaR) of $Z$ at some risk level $\tau$.
In the first three cases, the functional $\mathcal{T}$ in \eqref{eqgoal} is in the form of $\mathcal{T}(Z) = \E[\eta(Z)]$ for some function $\eta$, whereas in the last two cases, $\mathcal{T}$ represents VaR or CVaR.
Nested simulation is needed to compute these quantities;
one first simulates realizations of $X$ and then---conditional on each realization---evaluates $V_{T_0}(X)$ via simulation.
\hfill$\Box$
\end{example}

\begin{example}[Input Uncertainty Quantification]\label{examp:IUQ}
A decision-maker uses simulation to estimate the performance of a complex service system (e.g., health care facilities or ride-sharing platforms) that is driven by a random input process (e.g., the arrival of patients/customers/drivers).
Suppose that the distribution of the input process is parameterized by some parameter $X$ (e.g., the arrival rates for different times of day of a non-homogeneous Poisson process).
Suppose also that the performance measure of interest can be expressed as $\E[Y|X]$ (e.g., the mean waiting time or the order fulfillment rate), where the expectation is taken with respect to the input distribution given $X$.
However, in general, $X$ is not known and must be estimated from a sample of the input distribution.
This results in the issue of input uncertainty---the uncertainty about $X$---and it often has a substantial impact on the accuracy of the estimated performance of the system.

To quantify the impact of input uncertainty on simulation outputs,
one may adopt the method of Bayesian model averaging \citep{Chick01,Chick06} and compute $\E_{X\sim \mathsf{P}}[\E[Y|X]]$, where $\mathsf{P}$ is the posterior distribution of $X$ given the sample of the input distribution.
In this case,
the functional $\mathcal{T}$ in \eqref{eqgoal} is in the form of  $\mathcal{T}(Z)=\E[Z]$.
One may also construct a $90\%$ credible interval $(l,u)$ for $\E_{X\sim \mathsf{P}}[\E[Y|X]]$,
where $l$ and $u$ are, respectively, the $5\%$ and $95\%$ quantiles of $\E[Y|X]$ under $X\sim\mathsf{P}$.
In this case, $\mathcal{T}$ is in the form of a quantile (or equivalently, VaR). One may use nested simulation to compute these quantities \citep{XieNelsonBarton14,AndradottirGlynn16}.
\hfill$\Box$
\end{example}

In addition to the preceding examples,
there is a connection between nested simulation and conditional Monte Carlo, a general  technique for variance reduction \citep[Chapter~5]{AsmussenGlynn07}.
Any expectation $\E[Y]$ can be written as $\E[\E[Y|X]]$,
and $\Var[\E[Y|X]]\leq \Var[Y]$ due to the law of total variance.
Therefore, $\E[Y|X]$ is an unbiased estimator having a lower variance than $Y$.
It is usually used when  $X$ is strongly correlated with $Y$ and
the conditional expectation can be computed exactly or estimated efficiently.
Conditional Monte Carlo can also be used as a  smoothing technique for gradient estimation \citep{FuHu97,FuHongHu09}.

A central question to address in nested simulation concerns  allocation of the simulation budget in terms of how many outer-level scenarios to simulate and how many inner-level samples to simulate for each outer-level scenario.
In the present paper, we focus on \emph{uniform sampling}, a standard treatment in the literature \citep{GordyJuneja10,BroadieDuMoallemi15,AndradottirGlynn16,ZhuLiuZhou20}.
That is, an equal number of inner-level samples are used for each outer-level scenario.
The structural simplicity  makes it easily parallelizable to leverage modern computing platforms \citep{Lan10}.

Given a simulation budget $\Gamma$,
it can be shown under general conditions that to minimize the root mean squared error (RMSE) the asymptotically optimal outer-level sample size $n$ should grow at a rate of $\Gamma^{2/3}$ (and therefore the inner-level sample size $m$ for each outer-level scenario should grow at a rate of $\Gamma^{1/3}$).
Under this rule of allocating the simulation budget, the convergence rate of the standard nested simulation is $\Gamma^{-1/3}$ in terms of the RMSE \citep{GordyJuneja10,ZhangLiuWang22}.
This cubic root convergence is markedly slower than the square root convergence of a typical Monte Carlo simulation for estimating an expectation without the conditioning.
The deterioration in convergence rate is caused by the outer-level estimation bias, which is introduced by the nonlinear transformation $\mathcal{T}$ taking effect on the error associated with using the inner-level simulation to estimate the conditional expectation.

To reduce the outer-level bias---without increasing the simulation budget---one may seek to utilize the inner-level samples in a more efficient manner based on the following simple insight.
In the standard nested simulation, the inner-level samples that are simulated for an outer-level scenario $\BFx_i$ are used \emph{only} for estimating $\E[Y|X=\BFx_i]$,  the conditional expectation for that scenario.
These inner-level samples, however, may carry information about the conditional expectation associated with another outer-level scenario $\BFx_j$
if we anticipate $f(\BFx)\coloneqq \E[Y|X=\BFx]$ to be \emph{smooth} with respect to $\BFx$.
The standard nested simulation precludes an exchange of information between different outer-level scenarios.
Instead, we may treat the estimation of the conditional expectation via the inner-level simulation as a machine learning task.
When predicting $f(\BFx)$ for any $\BFx$,
this perspective allows us to benefit from the inner-level samples associated with \emph{all} the simulated outer-level scenarios,  even if $\BFx$ itself is not one of them.

\subsection{Main Contributions} \label{sec:contribution}

Our first contribution is to propose a new method for nested simulation.
The new method employs kernel ridge regression (KRR, \citealt{KanHenSejSri18}), a popular machine learning method, to estimate $f$.
Given inner-level samples,
KRR seeks the best function in the reproducing kernel Hilbert space (RKHS) of one's choosing via regularized least squares
in a way similar to ridge regression \citep{Hastie20}; hence the name.
A notable feature of KRR is that it allows one to easily leverage the smoothness (i.e., the degree of differentiability) of $f$,
which is essential for improving the convergence rate.
For implementation of the KRR-driven method, we also develop a new $\mathcal{T}$-dependent cross-validation technique for hyperparameter tuning.
The new technique significantly outperforms the standard cross-validation, and it may be interesting in its own right.

\begin{figure}[ht]
    \FIGURE{\includegraphics[width = \textwidth]{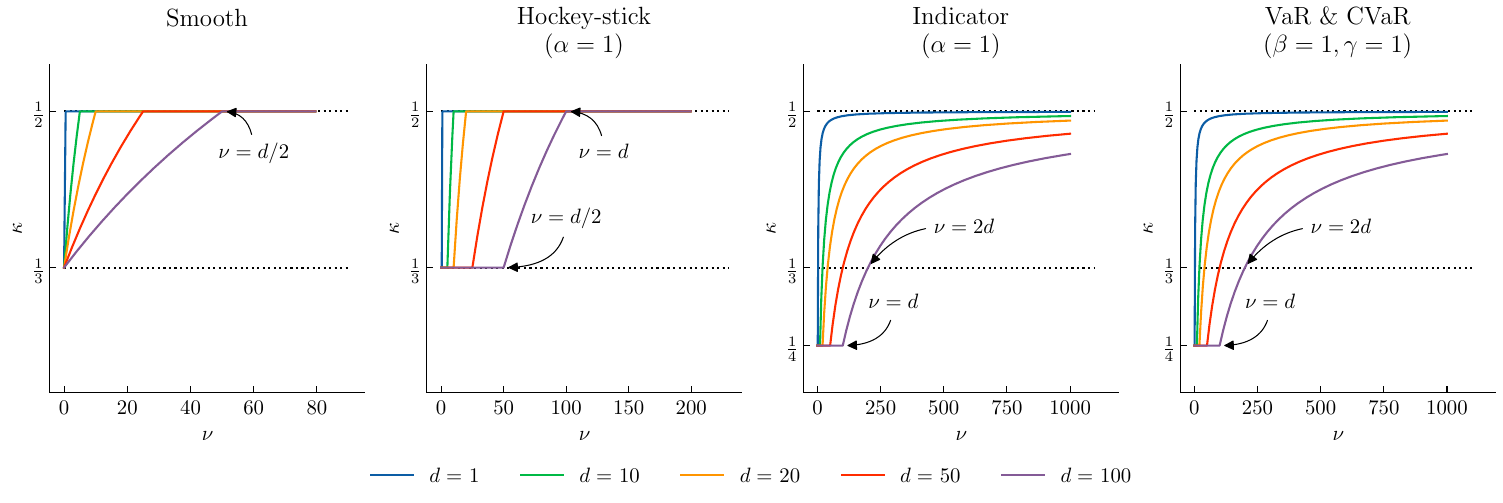}}
{The Parameter $\kappa$ in $ |\hat\theta_{n,m}-\theta| = O_{\pr}(\Gamma^{-\kappa}(\log  \Gamma)^{\tilde\kappa})$. \label{fig:rates} }
{The first three charts correspond to the cases that $\mathcal{T}(\cdot) = \E[\eta(\cdot)]$ with $\eta$ being a smooth (twice-differentiable) function, a hockey-stick function, and an indicator function, respectively.  The last chart corresponds to the case that $\mathcal{T}$ represents VaR or CVaR.
The parameter $\nu$ determines the smoothness of $f$ (see Section~\ref{sec:KRR});
$(\alpha, \beta,\gamma)$ are defined in Assumptions~\ref{assump:margin}--\ref{assump:margin-strong} for regulating the distribution of $f(X)$ and their typical values are~1.
The results hold when the budget allocation rule and the regularization parameter of KRR are properly specified (see Table~\ref{tab:correct-smooth} in Section~\ref{sec:analysis}.)
}
\end{figure}

Our second contribution is the asymptotic analysis of the KRR-driven estimator $\hat\theta_{n,m}$ for nested simulation,
including its convergence rate and the corresponding budget allocation rule $(n,m)$.
These properties show that the use of KRR in nested simulation bridges the gap between the cubic root and square root convergence rates.
Specifically,
we derive probabilistic bounds in the form $|\hat\theta_{n,m}-\theta| = O_{\pr}(\Gamma^{-\kappa}(\log  \Gamma)^{\tilde\kappa})$ for some constants $\kappa$ and $\tilde\kappa$ (see Figure~\ref{fig:rates}). Our choice of error metric (i.e., the absolute error) differs from previous studies \citep{GordyJuneja10,HongJunejaLiu17,ZhangLiuWang22}, which focus on the RMSE.
While our analysis could potentially be modified to derive bounds on the RMSE, such a revision would significantly increase the technical complexity of the analysis; see the Appendix for further comments on this issue.
Additionally, we determine the growth rates of $n$ and $m$ in relation to the increasing budget $\Gamma$.
Because the conditioning variable $X$ is high-dimensional (i.e., $d$ is large) in many applications,
we examine the curse of dimensionality on the performance of the proposed method.
The probabilistic bounds we derive reveal a mitigating effect of the smoothness on the curse of dimensionality.
For any fixed $d$,
the gap between the cubic root and square root convergence rates diminishes gradually  as the smoothness increases, and the convergence rate of the KRR-driven method recovers (or at least approaches) $\Gamma^{-1/2}$.
However, if  the smoothness of $f$ is relatively low, the use of KRR may have a detrimental effect.\footnote{For instance, when $\mathcal{T}$ represents VaR or CVaR (refer to the last chart in Figure~\ref{fig:rates}), the value of $\kappa$ is less than $-1/3$ if the smoothness parameter $\nu < 2d$. This implies a convergence rate that is even slower than the cubic root rate.}
In such situations, using the standard nested simulation may be a more suitable choice.

The theoretical framework that we develop in this paper is general.
Built upon empirical process theory \citep{geer2000empirical},
the framework permits analysis of a variety of forms of the functional $\mathcal{T}$, including not only the expectation of a function but risk measures such as VaR and CVaR.
In particular, the analysis for estimating VaR/CVaR using machine learning-driven nested simulation has not been available in the literature.
In addition,
the framework can potentially be adopted to study the use of other machine learning methods for estimating $f$ in nested simulation.

Our third contribution is that we conduct extensive numerical experiments to assess the performance of the KRR-driven method, using examples from both portfolio risk management and input uncertainty quantification.
The dimensionality of the conditioning variables involved in these examples is as high as 300.
These experiments complement our theoretical analysis and demonstrate that the proposed method is indeed a viable option for nested simulation.

\subsection{Literature Review}

The literature on nested simulation has been growing quickly in recent years \citep{LeeGlynn03,LesnevskiNelsonStaum07,GordyJuneja10,SunApleyStaum11,BroadieDuMoallemi11,AndradottirGlynn16,ZhuLiuZhou20,ZhangLiuWang22}.

\paragraph{Machine Learning in Nested Simulation.}
This paper is part of a line of research focused on enhancing sample efficiency in nested simulation by incorporating machine learning techniques for estimating the conditional expectation $\mathbb{E}[Y|X=\mathbf{x}]$ in the inner-level simulation.
These machine learning techniques can be either parametric or nonparametric in nature.
A common example of parametric methods is the linear basis function model, where basis functions are often selected from a range of options such as polynomials, splines, or radial basis functions.
Additionally, domain-specific basis functions can also be incorporated to refine the model.
\cite{BroadieDuMoallemi15} utilized this parametric model in nested simulation for portfolio risk management applications.
Their approach incorporated both polynomial functions of risk factors and basis functions associated with the payoff functions of the financial derivatives included in the portfolio.
For a similar nested simulation problem but involving insurance products, \cite{LinYang20} employed spline regression to estimate the portfolio risk.

Nonparametric machine learning techniques have found applications in nested simulation as well.
For instance, \cite{LiuStaum10} employed stochastic kriging (also referred to as Gaussian process regression) for estimating the CVaR of a financial portfolio.
Moreover, \cite{BartonNelsonXie14} and \cite{XieNelsonBarton14} also made use of stochastic kriging in nested simulation problems arising from input uncertainty quantification.
It is important to note that stochastic kriging shares a close relationship with KRR. In broad terms, stochastic kriging can be considered the Bayesian counterpart of KRR; further details can be found in Section~\ref{sec:KRR}.

The study most relevant to ours is that of \cite{HongJunejaLiu17}.
They examined the use of kernel smoothing\footnote{The notion of ``kernel'' in kernel smoothing should not be confused with that in kernel ridge regression.
We refer to \citet[Chapter 3]{BerlinetThomas-Agnan04} for a discussion on their differences.} in nested simulation and particularly the curse of dimensionality on the convergence rate, which is shown to be $\Gamma^{-\min\left(\frac{1}{2},\frac{2}{2+d}\right)}$ in terms of the RMSE
This suggests that the use of kernel smoothing is beneficial to nested simulation only for low-dimensional problems, whereas in high dimensions ($d\geq 5$)
it becomes detrimental, yielding a convergence rate even slower than $\Gamma^{-1/3}$ (i.e., the rate of the standard nested simulation).
A root cause for the severe curse of dimensionality is that kernel smoothing is unable to track derivatives of $f$ of an order higher than two, even if they exist.
In contrast, KRR does not suffer from this issue\footnote{
In addition to KRR, another machine learning method that can exploit the smoothness property is local polynomial regression. It generalizes kernel smoothing and approximates $f(\BFx)$ locally with a polynomial in $\BFx$, rather than a constant as kernel smoothing does.
The order of the polynomial should be set in accordance with the degree of differentiability of $f$ (see \citealt[Chapter 5]{GyorfiKohlerKrzyzakWalk02}).
However, unlike KRR, which is nonparametric and parsimonious, it can be challenging to fit a local polynomial regression model in multiple dimensions
because it takes an enormous number of parameters to represent a high-order multivariate polynomial.}.
We show that the curse of dimensionality can be  greatly mitigated by the smoothness of $f$.
Regardless of the value of $d$,
the convergence rate of
the KRR-driven method may recover or get arbitrarily close to $\Gamma^{-1/2}$,
provided that $f$ is sufficiently smooth, which is reasonably the case for typical applications of nested simulation, as demonstrated in our numerical experiments.

\paragraph{Likelihood Ratio Methods in Nested Simulation.}
A key factor contributing to the considerable improvement in sample efficiency in nested simulation using machine learning techniques is the facilitation of information exchange between different outer-level scenarios.
The likelihood ratio method is another approach that also allows for the utilization of all inner-level samples, irrespective of their association with specific outer-level scenarios, to estimate the conditional expectation for any given outer-level scenario.
However, unlike the machine learning approach, the likelihood ratio method achieves the cross-scenario information sharing through the likelihood ratio between different outer-level scenarios for a given inner-level sample.

The likelihood ratio method has been actively investigated in recent years.
For example, \cite{ZhouLiu18}, \cite{LiuZhou19}, and \cite{FengSong19} applied the likelihood ratio method to address input uncertainty quantification.
Among these three studies, the first two focused on an online environment, where data for constructing input models are received sequentially;
in contrast, the last one focused on the conventional offline environment, where all data are available simultaneously.
\cite{FengSong21} and \cite{ZhangFengLiuWang22} further explored this approach,  examining its application to portfolio risk management. In particular, the RMSE convergence rate of the likelihood ratio estimators can reach $O(\Gamma^{-1/2})$.

A potential limitation of the likelihood ratio method is its reliance on the availability of the likelihood ratio in closed form.
This can be problematic when dealing with complex stochastic models, as the likelihood ratio may be unknown or difficult to obtain.
In contrast, the machine learning approach does not face this limitation.
However, the machine learning approach does require specifying a model for the conditional expectation, which could give rise to the issue of model misspecification.
In this regard, employing KRR, a nonparametric technique, offers a potential advantage for our nested simulation method.
By utilizing KRR, our method becomes less susceptible to model misspecification, as it does not rely on a predetermined parametric model for the conditional expectation.

\paragraph{Convergence Rate Analysis of KRR.}
The convergence rate of KRR  has been extensively studied under various assumptions in the machine learning literature \citep{geer2000empirical,CaponnettoVito07,SteinwartHushScovel09,ZhangDuchiWainwright15,TuoWangWu20}.
However, nested simulation presents a unique setting that differs substantially from typical machine learning tasks in two aspects.
First, the objective is different. Whereas KRR is generally used to estimate the unknown function $f$, we aim to estimate $\theta=\mathcal{T}(f(X))$.
The presence of the nonlinear functional $\mathcal{T}$ changes the relative importance of bias and variance in the estimation of $f$.
The estimation error is measured differently when taking $\mathcal{T}$ into account.
Simply plugging the known results of KRR into the nested simulation setting does not yield adequate   performance.
Second, in nested simulation we study budget allocation rules, and
the number of observations of $f(\BFx_i)$ (i.e., inner-level samples) for each outer-level scenario $\BFx_i$ is a key decision variable.
In contrast, in typical KRR settings no repeated observations $f$ are allowed at the same location.
Hence, in order to improve the convergence rate of the KRR-driven method, we need to \emph{jointly} select both
the regularization parameter of KRR---which itself is critical to the convergence rate of KRR---and the inner-level sample size per outer-level scenario.
The existence of these two differences significantly complicates our theoretical analysis,
and we develop new technical results to cope with the complication.

\subsection{Notation and Organization}

Throughout the paper, we use the following notation.
For two positive sequences $a_n$ and $b_n$, we write
$a_n=O(b_n)$ and $a_n\asymp b_n$
if there exist some constants $C,C^\prime >0$ such that
$a_n\leq C b_n$ and
$C^\prime \leq a_n/b_n \leq C $, respectively, for all $n$ large enough.
Moreover,
$a_n = O_{\pr}(b_n)$ means  that for any $\varepsilon>0$, there exists $C>0$ such that
$\pr(a_n >C b_n) <\varepsilon$
for all $n$ large enough.
We also write $a\wedge b$  to denote $\min(a,b)$.
For a vector $v$, which is treated as a column vector by default, we use $v^\intercal$ and $\|v\|\coloneqq \sqrt{v^\intercal v}$ to denote its transpose and its Euclidean norm, respectively.

The remainder of the paper is organized as follows.
In Section~\ref{sec:formulation}, we introduce the background of nested simulation and formulate the research question.
In Section~\ref{sec:KRR}, we propose the KRR-driven method.
In Section~\ref{sec:analysis}, we analyze its asymptotic properties.
In Section~\ref{sec:CV}, we propose the $\mathcal{T}$-dependent cross-validation technique for hyperparameter tuning.
In Section~\ref{sec:numeric}, we conduct numerical experiments to assess the performance of the proposed method.
In Section~\ref{sec:conclusion}, we conclude the paper.
Technical results and proofs are presented in the Appendix and the e-companion to this paper.

\section{Problem Formulation}\label{sec:formulation}

Motivated by typical applications of nested simulation (see Examples~\ref{examp:portfolio} and \ref{examp:IUQ}),
in this paper we consider the following forms of the functional $\mathcal{T}$ in equation \eqref{eqgoal}:
\begin{enumerate}[label=(\roman*)]
\item $\mathcal{T}(Z) = \E[\eta(Z)]$ for some function $\eta:\RR^d\mapsto \RR$. \label{form:exp}
\item $\mathcal{T}(Z) = \VaR_\tau(Z)  \coloneqq \inf\{z\in\RR:\pr(Z\leq z)\geq \tau\}$, that is, the VaR of $Z$ at level $\tau\in(0,1)$. \label{form:var}
\item $\mathcal{T}(Z) = \CVaR_\tau(Z)\coloneqq \E[Z | Z > \VaR_\tau(Z)]$, that is, the CVaR of $Z$ at level $\tau\in(0,1)$. \label{form:cvar}
\end{enumerate}

\subsection{Standard Nested Simulation}

\begin{figure}[ht]
    \FIGURE{\includegraphics[height=0.2\textheight]{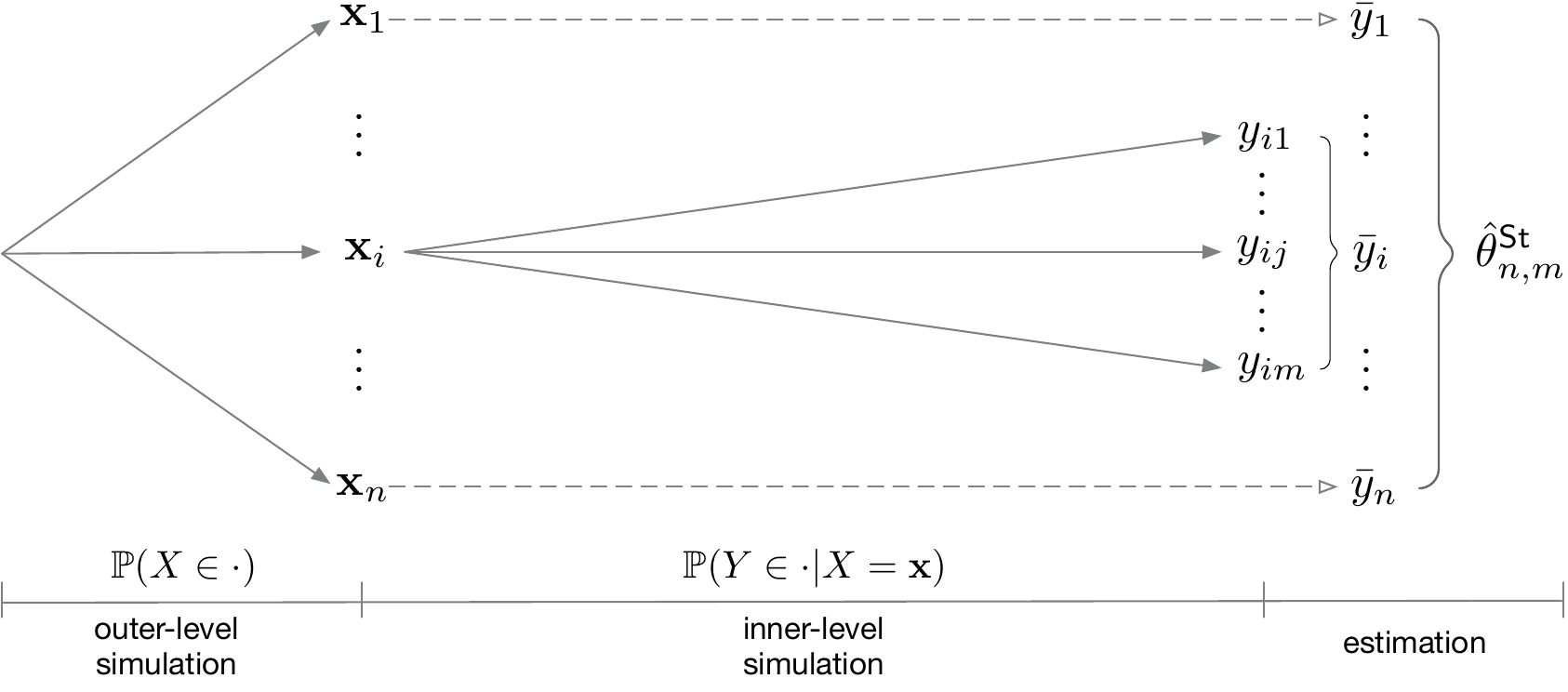}}
{Standard Nested Simulation. \label{fig:NS} }
{}
\end{figure}

The standard nested simulation involves the following two steps to generate data.
(i) Generate $n$ independent and identically distributed (i.i.d.)  outer-level scenarios $\{\BFx_i:i=1,\ldots,n\}$.
(ii) For each $\BFx_i$, generate $m$ i.i.d. inner-level samples from the conditional distribution of $Y$ given $X=\BFx_i$, denoted by $\{y_{ij}:j=1,\ldots,m\}$.
Then, $\theta$ can be  estimated based on the averages $\bar{y}_i$,
where $\bar{y}_i \coloneqq \frac{1}{m}\sum_{j=1}^m y_{ij}$ (see Figure~\ref{fig:NS}).

Specifically, let $\bar{y}_{(1)} \leq \cdots \leq \bar{y}_{(n)}$  denote the order statistics of $\{\bar{y}_1,\ldots, \bar{y}_n\}$.
Then, $\bar{y}_{(\lceil \tau n\rceil)}$ is the sample quantile\footnote{One may use more sophisticated quantile estimators via variance reduction techniques, such as importance sampling \citep{GlassermanHeidelbergerShahabuddin00,JinFuXiong03}. However, the analysis of these extensions in the context of nested simulation is beyond the scope of this paper.} at level $\tau$,
where $\lceil z\rceil$ denotes the least integer greater than or equal to $z$.

The standard nested simulation estimates $\theta$ via
\begin{equation}\label{eq:standard-estimator}
\hat\theta_{n,m}^{\mathsf{St}} \coloneqq
\left\{
\begin{array}{ll}
n^{-1}\sum_{i=1}^n \eta(\bar{y}_i ),     &\quad\mbox{if } \mathcal{T}(\cdot)= \E[\eta(\cdot)],  \\[0.5ex]
\bar{y}_{(\lceil \tau n\rceil)},     &\quad  \mbox{if }\mathcal{T}(\cdot) = \VaR_\tau(\cdot),  \\[0.5ex]
\bar{y}_{(\lceil \tau n\rceil)} + (1-\tau)^{-1}  n^{-1}\sum_{i=1}^n (\bar{y}_i -\bar{y}_{(\lceil \tau n\rceil)})^+,     &\quad  \mbox{if }\mathcal{T}(\cdot) = \CVaR_\tau(\cdot),
\end{array}
\right.
\end{equation}
where $(z)^+=\max(z,0)$.
The estimator for the case of CVaR is valid because it can be shown \citep{RockafellarUryasev02} that
\begin{equation*}
\CVaR_\tau(Z) = \VaR_\tau(Z) + (1-\tau)^{-1}\E[(Z-\VaR_\tau(Z))^+].
\end{equation*}

A central problem for nested simulation is budget allocation.
Given a  simulation budget $\Gamma$,
what are the optimal values for $n$ and $m$ in order to minimize the error in estimating $\theta$?
For typical applications, the simulation time required to generate one inner-level sample is substantially greater---often by orders of magnitude---than that required to generate one outer-level scenario.
Therefore, it is usually assumed in the literature that the latter is negligible relative to the former.
We also adopt this setup and suppose, without loss of generality, that $\Gamma=nm$.

It is shown in \cite{GordyJuneja10} and \cite{ZhangLiuWang22} that
to minimize the asymptotic RMSE of $\hat \theta_{n,m}^{\mathsf{St}}$ in the standard nested simulation,
the simulation budget should be allocated in such way that
$n\asymp \Gamma^{2/3}$ and $m\asymp \Gamma^{1/3}$ as $\Gamma\to\infty$, in which case
\[
\RMSE[\hat \theta_{n,m}^{\mathsf{St}}] =  \left(\E\bigl[(\hat\theta_{n,m}^{\mathsf{St}}-\theta)^2\bigr] \right)^{1/2} \asymp \Gamma^{-1/3}.
\]
Therefore, to achieve an RMSE of size $\epsilon$,
the simulation budget needs to grow like $O(\epsilon^{-3})$.
This stands in clear contrast to the standard Monte Carlo simulation for estimating unconditional expectations,
for which the RMSE diminishes at a rate of $\Gamma^{-1/2}$ and thus the corresponding sample complexity is
$O(\epsilon^{-2})$.

The deterioration from the square root rate of convergence to the cubic root rate is caused by the presence of the additional outer-level simulation.
In the standard nested simulation,
despite the absence of the inner-level estimation bias  ($\E[Y|X=\BFx_i]$ is estimated via $\bar{y}_i$ for each $\BFx_i$),
the nonlinear transform $\mathcal{T}$ that takes effect in the outer level indeed introduces bias in estimating $\mathcal{T}(\E[Y|X])$.
It may also exacerbate the impact of the inner-level estimation variance.
Both these complications demand more simulation samples to be overcome and thus, worsen the convergence rate.

\subsection{Strategies for Enhancement}

Our goal in the present paper is to accelerate nested simulation to achieve the square root convergence rate---the canonical rate of Monte Carlo simulation.
To fulfill the goal, we rely on two strategies.
The first strategy is to view the purpose of the inner-level simulation to be estimating $f(\cdot)=\E[Y|X=\cdot]$ as a whole instead of estimating $f(\BFx_i)$ separately for each $\BFx_i$.
This subtle change in viewpoint reveals that the inner-level estimation is essentially a regression problem, open for various machine learning methods to take on the task.
It further implies that we can and should estimate $f(\BFx_i)$ using not only
the inner-level samples specific to the outer-level scenario $\BFx_i$ but those from \emph{all} scenarios.
The use of machine learning in the inner level also presents us with an opportunity to improve the bias--variance trade-off---specific to the functional $\mathcal{T}$---in the outer level.

The second strategy is to leverage structural information about $f$ to alleviate the curse of dimensionality on the convergence rate, which is common in nonparametric regression and arises when the conditioning variable $X$ is high-dimensional.
Specifically, we assume that the \emph{smoothness}---that is, the degree of differentiability---of $f$ is known.
Knowing the smoothness allows us to properly postulate a function space within which we search for the target $f$.
In particular, the function space induced by the smoothness property is a subspace of the $\mathscr{L}_2$ space (i.e., the set of all square-integrable functions); and the higher the smoothness, the smaller the induced function space.
Therefore, knowing the smoothness may prevent us from searching an unnecessarily vast function space,
which would incur higher sample complexity.
Suppose, for example, $f$ is twice-differentiable.
We may then devise a machine learning method to find the best sample-based approximation to $f$ within---instead of $\mathscr{L}_2$---the set of twice-differentiable functions.
Nevertheless,
if we are unaware of the smoothness information or use a method unable to take advantage of it,
we are essentially seeking ``a needle in a bigger haystack.''

Our choosing to exploit the smoothness of $f$ to accelerate nested simulation is motivated by both practical and theoretical considerations.
Indeed, in typical applications of nested simulation, $f$ usually has high-order derivatives.
Consider, for example, a financial portfolio that consists of a number of options that are written on some assets.
Let $f(\BFx)$ be the sum of the values of these options if the prices of the underlying assets are $\BFx$ at the risk horizon.
Then, $f$ is at least twice-differentiable under typical asset pricing models, such as the Black--Scholes model or the Heston model (see \citealt[Chapter~7]{Glasserman03} for details).
Meanwhile, from a theoretical viewpoint, conditions regarding smoothness may be more general and easier to impose than other structural properties, such as  convexity and additivity.
The above considerations lead to our use of KRR.
Its strong theoretical underpinnings that are built upon the smoothness property will  facilitate our asymptotic analysis when it is used in nested simulation.

\section{A Kernel Ridge Regression Approach}\label{sec:KRR}

KRR seeks a sample-based approximation to $f$ in an RKHS, which is constructed as follows.
We first specify a positive definite kernel $k:\Omega\times \Omega\mapsto \RR$, where $\Omega\subset \RR^d$ is the support of the conditioning variable $X$.
We then define the space $\mathscr{N}^0_k(\Omega)$ of all functions of the form $x\mapsto \sum_{i=1}^n\beta_i k(\cdot,\BFx_i)$ for some $n\geq 1$, $\beta_1,\ldots,\beta_n\in\RR$, and
 $\BFx_1,\ldots,\BFx_n\in\Omega$
and equip $\mathscr{N}^0_k(\Omega)$  with the inner product defined as
\begin{eqnarray*}
\biggl\langle \sum_{i=1}^n\beta_i k(\cdot,\BFx_i), \sum_{j=1}^{\tilde{n}}\tilde{\beta}_j k(\cdot, \tilde{\BFx}_i)\biggr\rangle_{\mathscr{N}^0_k(\Omega)} \coloneqq \sum_{i=1}^n\sum_{j=1}^{\tilde{n}}\beta_i\tilde{\beta}_j k(\BFx_i, \tilde{\BFx}_j).
\end{eqnarray*}
The norm of $\mathscr{N}^0_k(\Omega)$ is induced by the inner product, that is,
$\|g\|_{\mathscr{N}^0_k(\Omega)}^2 \coloneqq \langle g, g\rangle_{\mathscr{N}^0_k(\Omega)}$ for all $g\in \mathscr{N}^0_k(\Omega)$.
Finally, the RKHS induced by $k$, denoted by $\mathscr{N}_k(\Omega)$, is defined as
the closure of $\mathscr{N}^0_k(\Omega)$ with respect to the norm $\|\cdot\|_{\mathscr{N}^0_k(\Omega)}$.
See \cite{BerlinetThomas-Agnan04} for a thorough exposition on RKHSs.

\subsection{RKHSs as Spaces of Smooth Functions}\label{sec:matern}

The choice of kernel $k$ represents one's knowledge about the unknown function $f$.
For example, if $k$ is the linear kernel,
then $\mathscr{N}_k(\Omega)$ is the space of all linear functions (see \citealt[Chapter 7]{BerlinetThomas-Agnan04} for more details).
In the present paper,
we consider the Mat\'ern class of kernels because the RKHSs that they induce consist of functions that possess a prescribed smoothness property,\footnote{This smoothness property leads to a strong connection between RKHSs associated with Mat\'ern kernels and Sobolev spaces. Other classes of kernels, such as the Wendland class and its generalization  \citep{bevilacqua2019estimation}, also exhibit similar properties. Our theoretical results can be extended to accommodate these classes of kernels.} determined by a parameter $\nu>0$.

The Mat\'ern kernel of smoothness $\nu$ is defined as $k(\BFx,\tilde{\BFx}) = \Psi(\BFx-\tilde{\BFx})$, where
\begin{eqnarray}\label{materngai}
\Psi(\BFx)\coloneqq \frac{1}{\mathsf{\Gamma}(\nu)2^{\nu-1}}\biggl(\frac{\sqrt{2\nu} \|\BFx\|}{\ell}\biggr)^{\nu} \mathsf{K}_{\nu}\biggl(\frac{\sqrt{2\nu}\|\BFx\|}{\ell}\biggr), \quad \forall \BFx\in\RR^d,
\end{eqnarray}
where $\ell>0$, $\mathsf{\Gamma}(\cdot)$ is the gamma function, and $\mathsf{K}_{\nu}(\cdot)$ is the modified Bessel function of the second kind of order $\nu$.
In practice, $\nu$ is often taken as a half-integer, that is,
$\nu=p+1/2$ for some nonnegative integer $p$.
In this case, $\Psi(\BFx)$ can be expressed in terms of elementary functions:
\[\Psi(\BFx) = \exp\biggl(\frac{- \sqrt{2p+1}\|\BFx\|}{\ell}\biggr)\frac{p!}{(2p)!} \sum_{i=0}^p \frac{(p+i)!}{i!(p-i)!}\biggl(\frac{2\sqrt{2p+1}\|\BFx\|}{\ell} \biggr)^{p-i},\quad p=0,1,2,\ldots\]
See  \citet[page~85]{Rasmussen2006Gaussian}. For instance,
\[\Psi(\BFx) =
\left\{
\begin{array}{ll}
\displaystyle \exp\biggl(\frac{- \|\BFx\|}{\ell}\biggr), &\quad\mbox{if }\nu=1/2,  \\[1ex]
\displaystyle  \biggl(1+ \frac{\sqrt{3}\|\BFx\|}{\ell}\biggr)\exp\biggl(-\frac{ \sqrt{3}\|\BFx\|}{\ell}\biggr), &\quad\mbox{if }\nu=3/2.
\end{array}
\right.
\]

We denote the corresponding RKHS by $\mathscr{N}_\Psi(\Omega)$.
Its smoothness property is reflected by the fact that $\mathscr{N}_\Psi(\Omega)$
is \emph{norm-equivalent}\footnote{The norm equivalence means that a function
$g \in \mathscr{N}_\Psi(\Omega)$ if and only if $g \in \mathscr{H}^s(\Omega)$, and there exist  some positive constants $C_1, C_2$ such that
$C_1 \| g\|_{\mathscr{H}^s(\Omega)} \leq \| g\|_{\mathscr{N}_\Psi(\Omega)} \leq C_2 \| g\|_{\mathscr{H}^s(\Omega)}$ for all $g\in \mathscr{N}_\Psi(\Omega)$.
} to the \emph{Sobolev space} of order $s=\nu+d/2$, denoted by $\mathscr{H}^s(\Omega)$; see  \citet{KanHenSejSri18} for details.
If $s$ is an integer\footnote{See \citet[Chapter~7]{adams2003sobolev} for the definition of Sobolev spaces of a fractional order.}, then $\mathscr{H}^s(\Omega)$ is
defined as
\[
\mathscr{H}^s(\Omega) \coloneqq
\biggl\{ g\in \mathscr{L}_2(\Omega): \|g\|^2_{\mathscr{H}^s(\Omega)}\coloneqq \sum_{|\BFalpha|\leq s} \|\partial^\BFalpha g \|_{\mathscr{L}_2(\Omega)}^2 < \infty\biggr\},
\]
where
$\BFalpha=(\alpha_1,\ldots,\alpha_d)$ is a multi-index,
$|\BFalpha| = \sum_{j=1}^d \alpha_j$, and $\partial^\BFalpha g$ denotes the \emph{weak}\footnote{The weak differentiability should not be confused with the classic notion of differentiability  (see  \citealt[page 19]{adams2003sobolev}).} partial derivative
$
\partial^\BFalpha g = \frac{\partial^{|\BFalpha|}g}{\partial x_1^{\alpha_1}\cdots \partial x_d^{\alpha_d}}
$.
Hence, if we assume $f\in \mathscr{N}_\Psi(\Omega)$,
we effectively assume that $f$ is square-integrable and is weakly differentiable up to order $\nu+d/2$.

\subsection{KRR-driven  Nested Simulation}\label{sec:KRR-NS}

Suppose that the unknown function $f\in\mathscr{N}_\Psi(\Omega)$  and that
the inner-level samples satisfy
\begin{align}\label{eqrela}
     y_{ij} = f(\BFx_i)+\epsilon_{ij}, \quad i=1,\ldots,n,\; j=1,\ldots,m,
\end{align}
where $\epsilon_{ij}$'s are independent zero-mean random variables that may not be identically distributed.

KRR estimates $f$ via regularized least squares of the following form:
\begin{equation}\label{eq:RLS}
\min_{g\in\mathscr{N}_{\Psi}(\Omega)} \frac{1}{n}\sum_{i=1}^n \left(\bar{y}_i - g(\BFx_i)\right)^2 + \lambda \|g\|^2_{\mathscr{N}_{\Psi}(\Omega)},
\end{equation}
where $\lambda > 0$ is the regularization parameter,
which controls the penalty imposed to the ``model complexity'' of a candidate solution to avoid overfitting.
The optimal solution to \eqref{eq:RLS} is
\begin{equation}\label{KRRest}
\begin{aligned}
\hat{f} \coloneqq {}& \argmin_{g\in \mathscr{N}_{\Psi}(\Omega)}\bigg( \frac{1}{n}\sum_{i=1}^n(\bar{y}_i-g(\BFx_i))^2 + \lambda \|g\|^2_{\mathscr{N}_{\Psi}(\Omega)}\bigg)  \\[0.5ex]
={}& \BFr(\BFx)^\intercal (\BFR+n\lambda \BFI )^{-1}\bar{\BFy},
\end{aligned}
\end{equation}
where $\BFr(\BFx) = (\Psi(\BFx-\BFx_1),\ldots,\Psi(\BFx-\BFx_n))^\intercal$,
$\BFR = (\Psi(\BFx_i-\BFx_j))_{i,j=1}^n\in\RR^{n\times n}$,
and the second identity follows from the representer theorem\footnote{
The representer theorem stipulates that given a finite set of observations of a function in an RKHS, the task of finding the best approximation of the function in the RKHS, which is an infinite-dimensional optimization problem in general,
can be reduced to a finite-dimensional problem that possesses an explicit optimal solution.
} \citep[Section~4.2]{ScholkopfSmola02}.

After computing the KRR estimator $\hat{f}$, we
let $\hat{f}_{(\lceil \tau n\rceil)}$ denote the $\lceil \tau n\rceil$-th order statistic of $\{\hat{f}(\BFx_1),\ldots,\hat{f}(\BFx_n)\}$, where $\lceil z\rceil$ denotes the least integer greater than or equal to $z$.
Formally, we propose the KRR-driven nested simulation estimator for $\theta$ as follows:
\begin{equation}\label{eq:generic-estimator}
\hat\theta_{n,m} \coloneqq
\left\{
\begin{array}{ll}
n^{-1}\sum_{i=1}^n \eta(\hat{f}(\BFx_i)),     &\quad\mbox{if } \mathcal{T}(\cdot)= \E[\eta(\cdot)],  \\[0.5ex]
\hat{f}_{(\lceil \tau n\rceil)},     &\quad \mbox{if }\mathcal{T}(\cdot) = \VaR_\tau(\cdot),  \\[0.5ex]
\hat{f}_{(\lceil \tau n\rceil)} + (1-\tau)^{-1}  n^{-1}\sum_{i=1}^n (\hat{f}(\BFx_i)-\hat{f}_{(\lceil \tau n\rceil)})^+,     &\quad \mbox{if }\mathcal{T}(\cdot) = \CVaR_\tau(\cdot).
\end{array}
\right.
\end{equation}

\begin{figure}[ht]
    \FIGURE{\includegraphics[height=0.2\textheight]{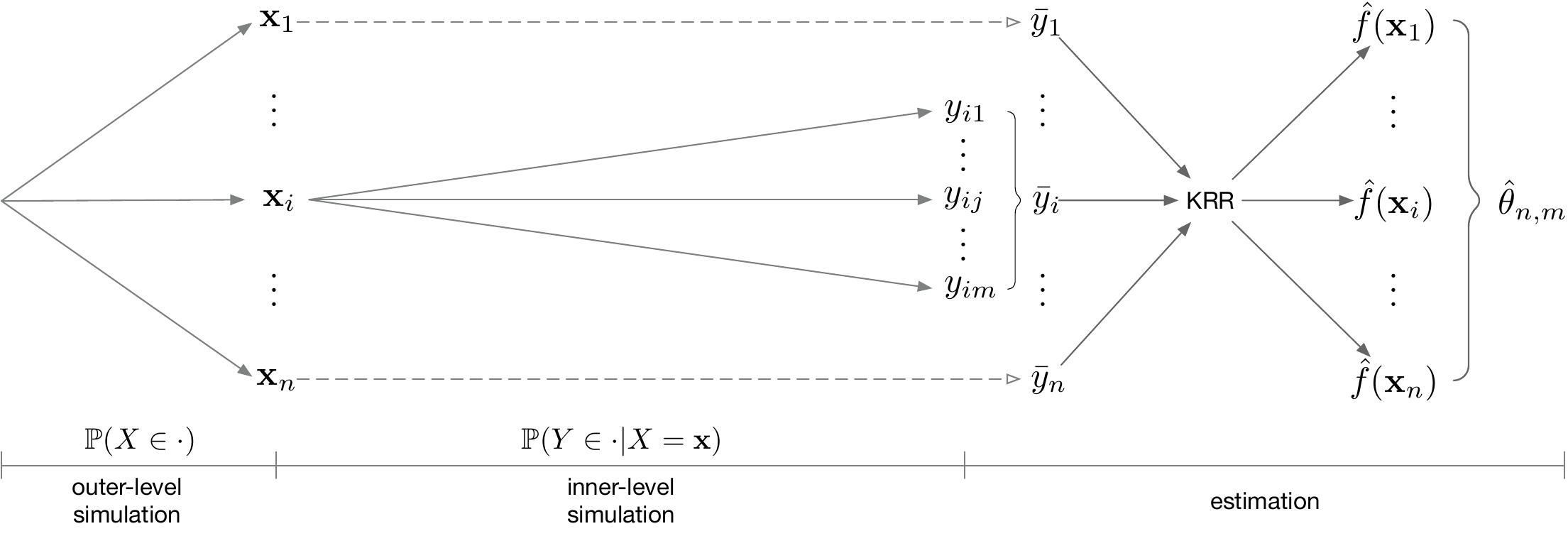}}
{KRR-driven Nested Simulation. \label{fig:NS-KRR} }
{}
\end{figure}

Note that in the standard nested simulation,
$f(\BFx_i)$ is estimated via the average of the inner-level samples for the specific scenario $\BFx_i$; that is,
$
\hat{f}(\BFx_i) = \bar{y}_i$.
In the KRR-driven nested simulation, the estimation of $f(\BFx_i)$  is based on all the data points $\{(\BFx_i, \bar{y}_i):i=1,\ldots,n\}$,
thereby leveraging spatial information globally. See Figure~\ref{fig:NS-KRR} for an illustration.

\begin{remark}
The implementation of  the KRR-driven method
requires---in addition to the budget allocation rule---the specification of three \emph{hyperparameters} $(\nu, \ell, \lambda)$.
Theoretically,
the convergence rate of $\hat{\theta}_{n,m}$ depends critically on $\nu$ and $\lambda$, but it is independent of $\ell$.
In the asymptotic analysis in Section~\ref{sec:analysis},
we assume the smoothness parameter $\nu$ of $f$ is \emph{known}.
Given $\nu$, our analysis reveals the proper order of magnitude of $\lambda$ (but not the exact value) as $\Gamma$ increases.
The independence of the convergence rate on $\ell$ arises from the fact that the RKHSs induced by the Mat\'ern kernels are identical for different values of $\ell$ but the same value of $\nu$.
However, from a practical point of view,
the three hyperparameters all have an impact on the finite-sample performance of $\hat{\theta}_{n,m}$.
They can be specified via cross-validation, with one complication.
The goal of nested simulation is to estimate
$\theta=\mathcal{T}(f(X))$ instead of $f$ itself.
Typical cross-validation focuses on the estimation accuracy of the latter,
which may not be the right metric for assessing the estimator of $f$ in the setting of nested simulation.
Instead,  in Section~\ref{sec:CV} we propose a new technique called $\mathcal{T}$-dependent cross-validation, which
significantly outperforms the standard cross-validation.
\end{remark}

\begin{remark}\label{remark:SK}
KRR is closely related to stochastic kriging \citep{AnkenmanNelsonStaum10},
which has been used in nested simulation with empirical success \citep{LiuStaum10,BartonNelsonXie14,XieNelsonBarton14}.
Stochastic kriging is a Bayesian method; assuming the prior distribution of $f$ is a Gaussian process with mean 0, it  estimates $f$ using the posterior mean
$
\hat{f}_{\mathsf{SK}}(\BFx) \coloneqq  \BFr(\BFx)^\intercal (\BFR + \BFSigma)^{-1}\bar{\BFy},
$
where $\BFSigma$ is the $n\times n$ diagonal matrix where the $j$-th diagonal element is $\Var[\bar{\epsilon}_i]$ with $\bar{\epsilon}_i= m^{-1}\sum_{j=1}^m \epsilon_{ij}$.
Compared to $\hat{f}_{\mathsf{SK}}(\BFx)$,
a particular advantage of the KRR estimator \eqref{KRRest} is the presence of  the regularization parameter $\lambda$, which can be treated as a tuning parameter, providing significant flexibility to improve the estimation accuracy of $f$ and eventually the performance of the nested simulation.
Another advantage of KRR relative to stochastic kriging is that the former requires weaker conditions on the simulation noise.
Whereas the latter requires $\epsilon_{ij}$ to be Gaussian with a \emph{known} variance in order that the posterior of $f$ should remain a Gaussian process,
KRR allows $\epsilon_{ij}$ to be sub-Gaussian (see Definition~\ref{def:subG} in Section~\ref{sec:analysis}) and does not need to assume a known variance.
\end{remark}

\section{Asymptotic Analysis}\label{sec:analysis}
In this section, we analyze the performance of the KRR-driven estimator in a large computational budget asymptotic regime.
We establish upper bounds on the convergence rate of
$|\hat \theta_{n,m}-\theta|$
for the three forms of $\theta$ that are listed in Section~\ref{sec:formulation}.
Before presenting the results, we highlight two main differences between the asymptotic analysis of the KRR-driven nested simulation
and that of KRR in typical machine learning contexts.

First, the objective of nested simulation is to estimate $\theta$, whereas that of KRR is to estimate $f$.
The difference in objective means a different, more careful bias--variance trade-off.
The analysis of KRR in the machine learning literature can be used to handle the bias--variance trade-off in the inner-level estimation of $f(\BFx) = \E[Y|X=\BFx]$.
However, the presence of the nonlinear functional~$\mathcal{T}$---which transforms the distribution of $f(X)$ to $\theta$---essentially redefines bias and variance in the outer-level estimation of $\theta = \mathcal{T}(f(X))$
and thus breaks the inner-level bias--variance trade-off.
Hence, a simple plug-in of existing analysis of KRR from machine learning literature would not suffice in the nested simulation setting.
The multiplicity of the forms of $\mathcal{T}$ also complicates our analysis significantly.
As shown in Theorems~\ref{thm:gsmooth}--\ref{thm:gvarcvar},
the convergence rate of the KRR-driven method varies for different forms of $\mathcal{T}$,
and each is different from the convergence rate of the KRR estimator of $f$.

A second main difference is that
multiple ($m>1$) observations of $f(\BFx_i)$ are allowed in nested simulation, whereas
only one single observation is allowed in typical KRR settings.
Therefore, not only do we need to judiciously choose the regularization parameter $\lambda$---which is the main factor that determines the convergence rate of KRR in typical machine learning settings---we also need to take into account the impact of $m$ on the convergence rate.
Hence, \emph{joint} selection of $\lambda$ and $m$ is needed to improve the convergence rate of the KRR-driven nested simulation.
This further complicates our theoretical analysis.

In the following, we state  in Section~\ref{sec:assump} two basic assumptions that will be imposed throughout the paper.
(Additional assumptions will be introduced later when needed.)
We analyze the form of nested expectation (i.e., $\mathcal{T}(\cdot)=\E[\eta(\cdot)]$) in Section~\ref{sec:analy-nested-expec}, analyze the two risk measures---VaR and CVaR---in  Section~\ref{sec:analy-risk-measure},
and summarize the results in Section~\ref{sec:summary}.
Due to space limitations, we give only  proof sketches of the theoretical results and relegate the complete proofs to the e-companion.

\subsection{Basic Assumptions}\label{sec:assump}
\begin{definition}[Sub-Gaussian Distribution]\label{def:subG}
A random variable $Z$ is said to be \emph{sub-Gaussian} with \emph{variance proxy} $\sigma^2$, denoted by $Z\sim \mathsf{subG}(\sigma^2)$, if
\[
\E \left[e^{t (Z- \E[Z])} \right]  \leq e^{\frac{\sigma^2 t^2}{2}},\quad \forall t\in \RR.
\]
\end{definition}

Typical examples of a sub-Gaussian distribution include Gaussian distribution or distribution with a bounded support.
Note that $\sigma^2$ is not necessarily equal to---but an upper bound on---the variance of $Z$; that is,
$\Var[Z]\leq \sigma^2$ \citep[page 51]{Wainwright19}.

Throughout this paper, we impose the following assumptions.

\begin{assumption}\label{assump:subG}
The noise terms $\{\epsilon_{ij}:1\leq i\leq n, 1\leq j\leq m\}$ are independent, zero-mean $\mathsf{subG}(\sigma^2)$. Moreover, $\epsilon_{i1},\ldots,\epsilon_{im}$ are identically distributed,  for each $i=1,\ldots,n$.
\end{assumption}

\begin{assumption}\label{assump:support}
The support of $X$ is a bounded convex set  $\Omega\subset \RR^d$.
Moreover, $X$ has a probability density function that is bounded above and below away from zero.
\end{assumption}

\begin{assumption}\label{assump:RKHS}
$f\in \mathscr{N}_{\Psi}(\Omega)$,
the RKHS associated with the Mat\'ern kernel of smoothness $\nu$.
\end{assumption}

Assumption~\ref{assump:subG} allows the simulation noise to be heteroscedastic, that is,
the variances at different $\BFx_i$'s are potentially unequal (i.e., $\Var(\epsilon_{i1})$ may not be  constant with respect to $i$).
We can relax the sub-Gaussian assumption for the noise terms to a sub-exponential (i.e., light-tailed) assumption, and then follow an analysis framework akin to the one presented later in this section to derive the convergence rate of $\hat{\theta}_{n,m}$. However, the resulting analysis would be more technically complex.
See \citet[page~168]{geer2000empirical}  for a discussion on this kind of relaxation.
The heavy-tailed case,
which often appears in financial applications \citep{FuhHuHsuWang11},
is significantly more challenging.
Our analysis framework might still be valid, by virtue of a sharper characterization of the corresponding empirical process, such as that in \cite{han2019convergence}.

In Assumption~\ref{assump:support},
the boundedness condition might appear restrictive. However, we impose this condition to simplify the theoretical analysis. In fact, this condition can be relaxed so that $X$ follows a sub-Gaussian or light-tailed distribution. In this case, one may apply the analysis presented in this paper to a ball with a radius that grows logarithmically in $n$, the number of outer-level scenarios. However, this relaxation would render the analysis significantly more complex from a technical standpoint, without adding substantial value to the main ideas of the present paper.\footnote{For example, in our analysis, we employ the Gagliardo–Nirenberg interpolation inequality \citep{BrezisMironescu19} to link error estimates of function approximations under different norms in Sobolev spaces. With the relaxation, the constants in this inequality would depend on $n$. Consequently, we would need to keep track of these constants throughout the analysis, rather than discarding them in the asymptotics. This additional tracking would significantly complicate the calculations but would only alter the current results on the convergence rate up to a logarithmic term.}
Similarly, the convexity of $\Omega$ in Assumption~\ref{assump:support} is also imposed for simplicity. This condition ensures that Sobolev spaces over $\Omega$ are well-defined, and it can be relaxed so that $\Omega$ satisfies Lipschitz-type boundary conditions, as described in Chapter 4 of \citet{adams2003sobolev}.

As discussed in Section~\ref{sec:matern},
Assumption~\ref{assump:RKHS}
is equivalent to the assumption that $f$ lies in the Sobolev space $\mathscr{H}^{\nu+d/2}(\Omega)$,
basically meaning that
$f$ is square-integrable and
is weakly differentiable up to order $\nu+d/2$.

In practice, unfortunately, it is often difficult to verify Assumption~\ref{assump:RKHS}, because both  $f$ and $\nu$ are unknown in general.
    In other words, the issue of smoothness misspecification may arise.
Nevertheless, recent theoretical studies, such as those by \cite{DickerFosterHsu17} and \cite{BlanchardMucke18}, have demonstrated that even with a misspecified kernel, KRR can still perform well as a supervised learning method. It is plausible that their findings could be naturally extended to the KRR-driven method. A theoretical investigation of misspecified smoothness, however, is beyond the scope of this paper. Instead, we conduct a numerical study on this topic in Section~\ref{ec-sec:misspec} of the e-companion.

In the Appendix, we compare our assumptions to those often used in the literature. We also discuss the implications of these differences on the proof techniques employed in this paper.

\subsection{Nested Expectation}\label{sec:analy-nested-expec}

We now consider problems in the form of a nested expectation,
$\theta = \E[\eta(\E[Y|X])]$, for some function $\eta$.
The KRR-driven estimator in \eqref{eq:generic-estimator} is then
$\hat\theta_{n,m} = n^{-1} \sum_{i=1}^n \eta(\hat{f}(\BFx_i))$.
By the triangle inequality,
\begin{align}\label{eq:first-decomp}
    |\hat \theta_{n,m}-\theta|
    \leq \underbrace{\left|\E [\eta(f(X))] -\frac{1}{n}\sum_{i=1}^n \eta(f(\BFx_i))\right|}_{I_1}
    + \underbrace{ \left|\frac{1}{n}\sum_{i=1}^n \bigl[\eta(f(\BFx_i))- \eta(\hat{f}(\BFx_i))\bigr]\right| }_{I_2}.
\end{align}
While we may apply the central limit theorem  to derive $I_1=O_{\pr}(n^{-1/2})$,
asymptotic analysis of $I_2$ is highly nontrivial and categorically depends on the property of $\eta$.
It might be intuitive to anticipate the general tendency that the smoother $f$ is, the faster $\hat{f}$ converges to $f$.
The presence of $\eta$, however, complicates characterization of the specific dependence of the convergence rate on the smoothness of $f$ and the dimensionality of $X$.
In particular,
without knowledge about $\eta$,
it is unclear \emph{a priori} whether $I_2$ renders a square root convergence rate even if $f$ is sufficiently smooth.

In light of Example~\ref{examp:portfolio} in Section~\ref{sec:formulation}, we study the following three cases of $\eta$.
They are standard cases in the literature \citep{BroadieDuMoallemi15,HongJunejaLiu17,ZhangLiuWang22}.
\begin{enumerate}[label=(\roman*)]
\item $\eta$ is twice-differentiable with bounded first- and second-order derivatives; that is,
\[
\sup_{z\in \{f(\BFx): \BFx\in\Omega\}} |\eta'(z)|<\infty \qq{and} \sup_{z\in \{f(\BFx): \BFx\in\Omega\}} |\eta''(z)|<\infty.
\]
\item $\eta$ is a hockey-stick function,  $\eta(z) = (z-z_0)^+$, for some constant $z_0\in \{f(\BFx):\BFx\in\Omega\}$.
\item $\eta$ is an indicator function,  $\eta(z)=\ind{\{z\geq z_0\}}$, for some constant $z_0\in \{f(\BFx):\BFx\in\Omega\}$.
\end{enumerate}

\subsubsection{Smooth Functions}

Assume $\eta$ is twice-differentiable with bounded first- and second-order derivatives.
It follows from  Taylor's expansion and the triangle inequality that
\begin{align} \label{eq:I_2-decomp}
    I_2
    \leq  \underbrace{\left|\frac{1}{n}\sum_{i=1}^n \eta'(f(\BFx_i))(f(\BFx_i)-\hat{f}(\BFx_i))\right|}_{I_{21}}
    + \underbrace{\left|\frac{1}{2n}\sum_{i=1}^n \eta''(\tilde{z}_i)(f(\BFx_i)-\hat{f}(\BFx_i))^2 \right|}_{I_{22}},
\end{align}
where $\tilde{z}_i$ is a value between $f(\BFx_i)$ and $\hat{f}(\BFx_i)$.
We present two technical results below to bound the convergence rates of $I_{21}$ and $I_{22}$, respectively.

\begin{proposition}\label{prop:rate_l1norm}
Suppose
$\varphi:\{f(\BFx):\BFx\in\Omega\}\mapsto\RR$ is bounded, and
Assumptions~\ref{assump:subG}--\ref{assump:RKHS}
hold.
Then,
\[
\left|\frac{1}{n}\sum_{i=1}^n \varphi(f(\BFx_i))(f(\BFx_i)-\hat{f}(\BFx_i))\right| = O_{\pr}(\lambda^{1/2} + (mn)^{-1/2}).
\]
\end{proposition}

\begin{proposition}\label{prop:rate_seminorm}
Suppose Assumptions~\ref{assump:subG}--\ref{assump:RKHS} hold.
Then,
\[\frac{1}{n}\sum_{i=1}^n (f(\BFx_i)-\hat{f}(\BFx_i))^2 = O_{\pr}\left(\lambda + (mn)^{-1}\lambda^{-\frac{d}{2\nu+d}} + (mn)^{-\frac{2\nu+d}{2\nu+2d}}\right).\]
\end{proposition}

\begin{remark}
Combining the Cauchy--Schwarz inequality and Proposition~\ref{prop:rate_seminorm} may  yield an upper bound on $I_{21}$.
However, this bound would not be as tight as that in  Proposition~\ref{prop:rate_l1norm}, which is a result of a refined analysis.
\end{remark}

Because $\eta$ has bounded first- and second-order derivatives,
we can apply
Propositions~\ref{prop:rate_l1norm} and \ref{prop:rate_seminorm} to conclude, with elementary algebraic calculations, that
$|\hat{\theta}_{m,n}-\theta| = O_{\pr}\bigl(n^{-1/2} + \lambda^{1/2} + (mn)^{-1}\lambda^{-\frac{d}{2\nu + d}}\bigr)$.
We then reduce this upper bound as much as possible by careful selection of $n$ and $\lambda$.
This leads to Theorem~\ref{thm:gsmooth}.

\begin{theorem}\label{thm:gsmooth}
Let $\mathcal{T}(\cdot) = \E[\eta(\cdot)]$ and $\eta$ be a twice-differentiable function with bounded first- and second-order derivatives.
Suppose Assumptions~\ref{assump:subG}--\ref{assump:RKHS} hold.
Then,
$|\hat \theta_{n,m}-\theta| = O_{\pr}(\Gamma^{-\kappa})$, where $\kappa$ is specified as follows:
\begin{enumerate}[label=(\roman*)]
    \item If $\nu \geq \frac{d}{2}$,  then $\kappa=\frac{1}{2}$ by setting
    $n \asymp \Gamma$,
    $m\asymp 1$,
    and $\lambda \asymp \Gamma^{-1}$.
    \item If $0 < \nu < \frac{d}{2}$, then $ \kappa = \frac{2\nu+d}{2\nu + 3d}$ by setting
    $n\asymp \Gamma^{\frac{2(2\nu+d)}{2\nu + 3d}}$,
    $m\asymp \Gamma^{\frac{d-2\nu}{2\nu + 3d}}$,
    and $\lambda \asymp \Gamma^{-\frac{2(2\nu+d)}{2\nu+3d}}$.

\end{enumerate}
\end{theorem}

Theorem~\ref{thm:gsmooth} has several implications.
First, it clearly reveals a mitigating effect of $\nu$ on the curse of dimensionality on the convergence rate---the larger $\nu$ is, the faster the rate is.
In particular, in the case of nested expectation with $\eta$ being smooth,
$\hat{\theta}_{m,n}$ achieves the square root convergence rate when $\nu\geq \frac{d}{2}$,
recovering the canonical rate of Monte Carlo simulation.

Second, as $\nu\to 0$, the convergence rate of $\hat{\theta}_{m,n}$ approaches $O_{\pr}(\Gamma^{-1/3})$, and meanwhile, the outer-level sample size becomes $n\asymp \Gamma^{2/3}$.
Both recover the results for the standard nested simulation \citep{GordyJuneja10,ZhangLiuWang22}.
Further, in light of the fact that $\kappa > \frac{1}{3}$ for all $\nu>0$ in Theorem~\ref{thm:gsmooth},
if $\eta$ is smooth, then regardless of the dimensionality,
the use of KRR will have a beneficial effect on the estimation of $\E[\eta(f(X))]$, at least from the perspective of convergence rates.
This is in clear contrast to the effect of using kernel smoothing in the inner-level estimation.
\cite{HongJunejaLiu17} show that
under the same assumptions on $\eta$,
the use of kernel smoothing is beneficial only for low-dimensional problems and becomes detrimental when $d\geq 5$.
That KRR and kernel smoothing have different effects on the convergence rate in high dimensions is mainly because
the former manages to leverage the smoothness of $f$, providing us with a proper function space to perform function estimation.

Third,
as $\nu$ increases,
there exists a ``phase transition'' in the budget allocation rule.
The inner-level sample size should remain constant ($m \asymp 1$)
if $\nu$ is above a threshold ($\frac{d}{2}$ in this case),
whereas it should grow as $\Gamma$ increases
otherwise.
Intuitively, this is because if $f$ is sufficiently smooth we do not anticipate  it to vary substantially over different locations.
Hence,
even if each observation $f(\BFx_i)$ is highly noisy,
the noises would mostly cancel one another and
$f$ would be reasonably estimated,
as long as $f$ is observed at a sufficient number of   locations.
However, if $\nu$ is small
$f$ may exhibit dramatic variations,
and a slight change in an observation of $f$ due to noise may lead to a significant change in the estimate of $f$.
To reduce the impact of noise on the observation of $f$,
we must take multiple replications  at each location. Moreover,
a higher proportion of the budget should be allocated to the inner-level samples the lower the smoothness of $f$.

Lastly, we consider the special case that $\eta(z) = z$,
which occurs in the context of input uncertainty quantification (see Example~\ref{examp:IUQ}).
In this case, the term $I_{22}$ in \eqref{eq:I_2-decomp} vanishes and therefore Proposition~\ref{prop:rate_seminorm} is no longer needed.
With the same proof for Theorem~\ref{thm:gsmooth}, we have $|\hat{\theta}_{m,n}-\theta| = O_{\pr}\bigl(n^{-1/2} + \lambda^{1/2} + (mn)^{-1/2}\bigr)$, which leads to the square root rate for all $\nu>0$.
\begin{corollary}\label{corollary:identity}
Let $\mathcal{T}(\cdot) = \E[\cdot]$.
Suppose Assumptions~\ref{assump:subG}--\ref{assump:RKHS} hold.
Then,
$|\hat \theta_{n,m}-\theta| = O_{\pr}(\Gamma^{-1/2})$ for all $\nu>0$, by setting
    $n \asymp \Gamma$,
    $m\asymp 1$,
    and $\lambda \asymp \Gamma^{-1}$.
\end{corollary}

Numerical studies by \cite{BartonNelsonXie14} and \cite{XieNelsonBarton14} demonstrate that the use of stochastic kriging greatly enhances the accuracy for quantifying the impact of input uncertainty on simulation outputs.
In light of the close connection between KRR and stochastic kriging as elucidated in Remark~\ref{remark:SK}, Corollary~\ref{corollary:identity} sheds light on this empirical success.
Similar results have been established in \citet[Theorem~4.4]{WuZhuZhou18} and \citet[Theorem~2]{WangZhangNg21}.

\subsubsection{Hockey-stick Functions}

Assume $\eta(z) = (z-z_0)^+$.
The non-differentiability of $\eta$ at $z_0$
implies that the magnitude of $I_2$ in the decomposition \eqref{eq:first-decomp}, and therefore the accuracy of $\hat\theta_{n,m}$, is potentially sensitive relative to the accuracy of $\hat{f}$.
Namely,
a slight change in $\hat{f}$ may result in a significant change in $\hat\theta_{n,m}$.
Because
the non-differentiability takes effect only when $f(\BFx)$ falls in the vicinity of $z_0$,
we impose the following assumption to characterize the likelihood of this event.
\begin{assumption}\label{assump:margin}
There exist positive constants $C$, $t_0$, and $\alpha\leq 1$ such that
\begin{align*}
\pr(|f(X)-z_0|\leq t)\leq Ct^\alpha,\quad \forall t\in(0,t_0].
\end{align*}
\end{assumption}

Assumption~\ref{assump:margin} is similar to the Tsybakov margin condition \citep{Tsybakov04}, which is widely used in the machine learning literature to study classification algorithms.
A large value of $\alpha$ means that $|f(\BFx) - z_0|$ is bounded away from zero  with a high probability.
Conversely, if $\alpha$ is close to zero, Assumption~\ref{assump:margin} is essentially void---because it is satisfied by \emph{any} probability distribution of $X$ if we set $\alpha=0$ and $C=1$---and $|f(\BFx)-z_0|$ can be arbitrarily close to zero,
making the non-differentiability of $\eta$ negatively affect nearly all $\BFx\in\Omega$.
The typical case\footnote{In general, $\alpha$ may take values greater than one if $f$ (not $\eta$) exhibits non-differentiability in the region $\{\BFx:f(\BFx)=z_0\}$ (e.g., $|x-z_0|^{1/2}$ if $d=1$).
However, as the present paper focuses on scenarios where $f$ is smooth,
we anticipate $\alpha\leq 1$ in our theoretical framework.
See \cite{PerchetRigollet13} for a discussion on a similar tension between the smoothness of functions in H\"older spaces and the value of $\alpha$ in the Tsybakov margin condition.
} in practice is $\alpha=1$.
This, for example, can be easily shown via Taylor's expansion
if $X$ has a density that is bounded above and below away from zero (Assumption~\ref{assump:support})
and $f$ has bounded first- and second-order derivatives with $\|\nabla f(\BFx_0)\|> 0$ for all $\BFx_0$ such that $f(\BFx_0)=z_0$.

Due to
its non-differentiability,
Taylor's expansion does not apply to $\eta$, and therefore we cannot use \eqref{eq:I_2-decomp} to analyze the error term $I_2$.
Instead, we adopt the treatment in \cite{HongJunejaLiu17} and construct a twice-differentiable function $\eta_\delta$, which is parameterized by $\delta$ and approximates $\eta$ as $\delta\to 0$.
We decompose $I_2$ with $\eta_\delta$ being an intermediate step:
\begin{equation}\label{eq:I_2-decomp-new}
   I_2
    \leq
    \underbrace{\left| \frac{1}{n}\sum_{i=1}^n \bigl[ \eta(f(\BFx_i))-  \eta_\delta(f(\BFx_i)) \bigr]\right|}_{J_1}     + \underbrace{\left|\frac{1}{n}\sum_{i=1}^n \bigl[\eta_\delta(f(\BFx_i)) -   \eta_\delta(\hat{f}(\BFx_i)) \bigr] \right|}_{J_2}   + \underbrace{ \left|\frac{1}{n}\sum_{i=1}^n \bigl[ \eta_\delta(\hat{f}(\BFx_i))-  \eta(\hat{f}(\BFx_i))\bigr]\right|}_{J_3}.
\end{equation}

Assuming, without loss of generality, that $z_0=0$,
the key properties of $\eta_\delta$ include
(i) $|\eta(z)-\eta_\delta(z)|  = O(\delta \ind{\{z\in [-\delta, \delta]\}})$, (ii) $|\eta_\delta'(z)|$ is bounded uniformly for all $\delta$, and (iii)  $|\eta_\delta''(z)| = O(\delta^{-1} \ind{\{z\in [-\delta, \delta]\}})$.
First, applying property (i) to $J_1$, we have that for some constant $C_1>0$,
\begin{align}
    J_1 \leq \left| \frac{C_1}{n}\sum_{i=1}^n \delta \ind{\{f(\BFx_i)\in [-\delta,\delta]\}}\right|
    \leq{}&  C_1\delta \left| \frac{1}{n}\sum_{i=1}^n \ind{\{f(\BFx_i)\in [-\delta,\delta]\}}- \E \ind{\{f(\BFx_i)\in [-\delta,\delta]\}}\right|\nonumber  \\
    & + C_1 \delta \E\bigl[ \ind{\{f(\BFx_i)\in [-\delta,\delta] \}} \bigr]
    = O_{\pr}(\delta n^{-1/2} + \delta^{\alpha+1}), \label{eq:J_1-bound}
\end{align}
where the last step follows from the central limit theorem and  Assumption~\ref{assump:margin}.

Next, $J_3$ can be analyzed in the same fashion, except that the relevant event here is $\{\hat f(\BFx_i)\in[-\delta,\delta]\}$.
The complication is addressed by considering the larger event
$\{f(\BFx_i)\in [-\delta-\rho_n,\delta+\rho_n]\}$,
where $\rho_n \coloneqq \max_{1\leq i\leq n}|f(\BFx_i) - \hat{f}(\BFx_i)|$.
This leads us to     $J_3 = O_{\pr}(\delta n^{-1/2}+\delta(\delta+\rho_n)^\alpha)$.
The quantity $\rho_n$ is critical and is also involved in the analysis of $J_2$.

Lastly, similar to \eqref{eq:I_2-decomp}, we may decompose $J_2$ via Taylor's expansion:
\begin{equation} \label{eq:J_2-decomp}
    J_2 =\underbrace{\left|\frac{1}{n}\sum_{i=1}^n \eta_\delta'(f(\BFx_i))(f(\BFx_i)-\hat{f}(\BFx_i))\right|}_{J_{21}}
    + \underbrace{\left| \frac{1}{2n}\sum_{i=1}^n \eta_\delta''(\breve{z}_i)(f(\BFx_i)-\hat{f}(\BFx_i))^2 \right|}_{J_{22}},
\end{equation}
where $\breve{z}_i$ is a value between $f(\BFx_i)$ and $\hat{f}(\BFx_i)$.
Because $|\eta_\delta'(z)|$ is bounded uniformly for all $\delta$,
$J_{21}$ can be bounded using Proposition~\ref{prop:rate_l1norm}.
However, unlike our treatment of $I_{22}$ in \eqref{eq:I_2-decomp},
applying Proposition~\ref{prop:rate_seminorm} to $J_{22}$ would yield a loose bound
because $|\eta_\delta''(z)|=O(\delta^{-1}\ind{\{z\in [-\delta, \delta]\}})$ is not uniformly bounded as $\delta\to 0$.
Instead, we bound $J_{22}$ in terms of $\rho_n$:
\begin{align*}
    J_{22}
    \leq{}& \left| \frac{1}{n}\sum_{i=1}^n C_2 \delta^{-1} \ind{\{\breve{z}_i \in [-\delta, \delta]\}}\rho_n^2 \right|
    \leq \left| \frac{1}{n} \sum_{i=1}^n  C_2 \delta^{-1}\rho_n^2\ind{\{f(\BFx_i)\in[-\delta - \rho_n, \delta + \rho_n]\}}\right|  \\
    ={}& O_{\pr}\left(\delta^{-1} \rho_n^2 \bigl(n^{-1/2} + (\delta+\rho_n)^\alpha \bigr)\right),
\end{align*}
for some constant $C_2>0$,
where the second step holds because $\breve{z}_i$ is a value between $f(\BFx_i)$ and $\hat{f}(\BFx_i)$,
and the last step can be shown with the same argument used for $J_1$.
Putting these bounds for $J_1$, $J_2$, and $J_3$ together yields a bound for $|\hat \theta_{n,m}-\theta| $ that involves both $\delta$ and $\rho_n$. If we further set $\delta = \rho_n$, then the bound is reduced to
$|\hat \theta_{n,m} - \theta | =  O_{\pr}\left(n^{-1/2} + \rho_n^{\alpha+1}\right)$.
Then, applying the asymptotic property of $\rho_n$ in Proposition~\ref{prop:rho_n}, we can prove
Theorem~\ref{thm:gnonsmooth} via straightforward calculations.

\begin{proposition}\label{prop:rho_n}
Suppose Assumptions~\ref{assump:subG}--\ref{assump:RKHS} hold.
If $\lambda\asymp \Gamma^{-1}$, then
\begin{align*}
    \max_{1\leq i\leq n}|f(\BFx_i) - \hat{f}(\BFx_i)| = O_{\pr}\left(\bigl( n^{-\frac{\nu}{2\nu + 2d}} \wedge  m^{-1/2}\bigr)(\log n)^{1/2}  \right).
\end{align*}
\end{proposition}

\begin{theorem}\label{thm:gnonsmooth}
Let $\mathcal{T}(\cdot) = \E[\eta(\cdot)]$ and $\eta$ be a hockey-stick function.
Suppose Assumptions~\ref{assump:subG}--\ref{assump:margin} hold.
Then, $|\hat \theta_{n,m}-\theta| = O_{\pr}(\Gamma^{-\kappa}(\log \Gamma)^{\tilde\kappa})$, where $\kappa$ and $\tilde\kappa$ are specified as follows:
\begin{enumerate}[label=(\roman*)]
\item If $\nu \geq \frac{d}{\alpha+1}$, then
$\kappa =\frac{1}{2}\wedge\frac{\nu(\alpha+1)}{2(\nu+d)}$
and $\tilde\kappa=\frac{\alpha+1}{2}\ind{\{\nu \alpha <d \}}$ by setting
$n\asymp\Gamma$,
$m\asymp 1$,
and $\lambda \asymp \Gamma^{-1}$.
\item If $\nu < \frac{d}{\alpha+1}$, then $\kappa  =\frac{\alpha+1}{2(\alpha+2)}$ and $\tilde\kappa=\frac{\alpha+1}{2}$ by setting
$n\asymp \Gamma^{\frac{\alpha+1}{\alpha+2}}$,
$m\asymp \Gamma^{\frac{1}{\alpha+2}}$,
and $\lambda \asymp \Gamma^{-1}$.
\end{enumerate}

\end{theorem}

Similar to Theorem~\ref{thm:gsmooth},
Theorem~\ref{thm:gnonsmooth} manifests the mitigating effect of the smoothness on the curse of dimensionality,
as well as the phase transition in the budget allocation rule.
In addition, it shows that
if $\eta$ is a hockey-stick function,
a larger value of $\alpha$, through inducing a smaller probability of $f(\BFX)$ falling  near the point $z_0$ where $\eta$ is non-differentiable,
leads to a higher convergence rate of $\hat{\theta}_{m,n}$ for estimating $\E[\eta(f(X))]$.

In the typical scenario $\alpha=1$, the rate achieves $O_{\pr}(\Gamma^{-1/2})$ if $\nu\geq d$ (and thus $\frac{\nu(\alpha+1)}{2(\nu+d)}\geq \frac{1}{2}$).
Meanwhile, if $\nu < \frac{d}{2}$, the rate becomes $O_{\pr}(\Gamma^{-1/3} (\log \Gamma))$, which is nearly (discarding the logarithmic factor) identical  to that of the standard nested simulation. See Figure~\ref{fig:rates} for an illustration.

In the worst scenario $\alpha=0$, the square root rate cannot be fully recovered (because $\frac{\nu(\alpha+1)}{2(\nu+d)}\leq \frac{1}{2}$ for all $\nu$ in this case), but it can be approached arbitrarily close to, as $\nu\to\infty$.
Meanwhile,  if $\nu < 2d$, the rate is slower than $O_{\pr}(\Gamma^{-1/3})$, and therefore the standard nested simulation is preferable to the KRR-driven method.

\subsubsection{Indicator Functions}

Assume $\eta(z) = \ind{\{z \geq z_0\}}$.
The discontinuity at $z_0$ poses an even bigger challenge to the convergence rate of $\hat\theta_{n,m}$ compared to the case for hockey-stick functions.
In particular, it renders the smooth approximation approach employed for hockey-stick functions ineffective.
This is because for any differentiable function $\tilde{\eta}_\delta$ that converges to $\eta$ uniformly as $\delta\to 0$,
$|\tilde{\eta}_\delta'(z_0)|$ would blow up as $\delta \to 0$.
Therefore, if we base our analysis on the decomposition \eqref{eq:J_2-decomp}, we would end up with an undesirable bound.

Instead, we directly work  on $I_2$ in \eqref{eq:first-decomp} without further decomposing it. Again, assume $z_0 = 0$ without loss of generality.
Note that if $\ind{\{f(\BFx_i) \geq 0\}} \neq \ind{\{\hat{f}(\BFx_i) \geq 0\}}$, then we must have $f(\BFx_i)\in [-\rho_n,\rho_n]$, where $\rho_n = \max_{1\leq i\leq n}|f(\BFx_i) - \hat{f}(\BFx_i)|$.
It follows that
\begin{align*}
    I_2 ={}& \left|\frac{1}{n}\sum_{i=1}^n \left(\eta(f(\BFx_i))-\eta(\hat{f}(\BFx_i))\right)\ind{\{f(\BFx_i)\in [-\rho_n,\rho_n]\}}\right| \\
    \leq{}& \frac{1}{n}\sum_{i=1}^n \ind{\{f(\BFx_i)\in [-\rho_n,\rho_n]\}}
    = O_{\pr}(n^{-1/2}+\rho_n^\alpha),
\end{align*}
where the last step follows the same argument as \eqref{eq:J_1-bound}.
We then can apply Proposition~\ref{prop:rho_n} to derive Theorem~\ref{thm:gnonsmoothin} below.

\begin{theorem}\label{thm:gnonsmoothin}
Let $\mathcal{T}(\cdot) = \E[\eta(\cdot)]$ and $\eta$ be an indicator function.
Suppose Assumptions~\ref{assump:subG}--\ref{assump:margin} hold.
Then,
$|\hat \theta_{n,m}-\theta| = O_{\pr}(\Gamma^{-\kappa}(\log \Gamma)^{\tilde\kappa})$, where $\kappa$ and $\tilde\kappa$ are specified as follows:
\begin{enumerate}[label=(\roman*)]
\item
If $\nu \geq \frac{d}{\alpha}$, then $\kappa = \frac{\nu\alpha}{2(\nu+d)}$ and
$\tilde\kappa=\frac{\alpha}{2}$ by setting $n\asymp\Gamma$,
$m\asymp 1$,
and $\lambda \asymp \Gamma^{-1}$.
\item If $\nu < \frac{d}{\alpha}$, then $\kappa=\frac{\alpha}{2(\alpha+1)}$ and $\tilde\kappa=\frac{\alpha}{2}$ by setting $n\asymp \Gamma^{\frac{\alpha}{\alpha+1}}$,
$m\asymp \Gamma^{\frac{1}{\alpha+1}}$,
and $\lambda \asymp \Gamma^{-1}$.
\end{enumerate}

\end{theorem}

Note that the final calculations that lead to Theorems~\ref{thm:gnonsmooth} and \ref{thm:gnonsmoothin} are the same, except that the bound
$O_{\pr}(n^{-1/2}+\rho_n^{\alpha+1})$ for the former is replaced with $O_{\pr}(n^{-1/2}+\rho_n^\alpha)$ for the latter.
Therefore,
if $\eta$ is an indicator function and $\alpha=1$ (the typical value),
the convergence rate of $\hat\theta_{n,m}$ is the same as
the case in which $\eta$ is a hockey-stick function and $\alpha=0$,
meaning the square root rate cannot be fully recovered, but it can be approached arbitrarily close to, as $\nu\to\infty$
(see Figure~\ref{fig:rates}).

In addition, if $\alpha=0$, Theorem~\ref{thm:gnonsmoothin} asserts that $|\hat \theta_{n,m}-\theta| =  O_{\pr}(1)$ for all $\nu>0$.
Consider the following simple example.
Suppose that $z_0=0$ and $f(\BFx) \equiv 0$ for all $\BFx\in\Omega $.
Then,
$\theta = \E[\ind{\{f(X) \geq 0\}}] = 1$,
and Assumption~\ref{assump:margin} is not satisfied for any $\alpha>0$, as $\pr(|f(X) |\leq t) = 1$ for all $t>0$;
moreover, $\hat{f}(\BFx) =  \BFr(\BFx)^\intercal(\BFR+n\lambda \BFI )^{-1}\bar \BFepsilon$.
Suppose also that $\epsilon_{i,j}$'s are i.i.d.  normal random variables.
Then, given $\mathcal{D}=\{(\BFx_i,\bar{y}_i):i=1,\ldots,n\}$, $\hat{f}(\BFx) $ has a normal distribution with a mean of zero.
It follows that
$\pr(\hat{f}(\BFx) \geq 0 | \mathcal{D} ) = \frac{1}{2}$ for all $\BFx\in\Omega$,
and  therefore
\[
\E\bigl[\hat{\theta}_{n,m} \,\big|\, \mathcal{D} \bigr] = \E\bigl[\ind{\{\hat{f}(X) \geq z_0\}}\,\big|\,\mathcal{D}  \bigr] =
\E\bigl[\pr\bigl(\hat{f}(X) \geq z_0 \,\big|\, \mathcal{D}, X\bigr)\,\big|\,\mathcal{D} \bigr]
= \frac{1}{2}.
\]
This implies that $\E[\hat{\theta}_{n,m} ] = \frac{1}{2}$ for all $n$ and $m$,
so $\hat{\theta}_{n,m}$ does not converge to $\theta$.
Recall that $\alpha=0$ effectively nullifies Assumption~\ref{assump:margin}.
The preceding discussion suggests that Assumption~\ref{assump:margin} (with $\alpha>0$) is necessary to ensure the consistency of $\hat{\theta}_{m,n}$ if $\theta = \E[\eta(f(\BFX) )]$ and $\eta$ is an indicator function.

\subsection{Risk Measures}\label{sec:analy-risk-measure}

We now examine the KRR-driven method for the case that $\mathcal{T}$ represents VaR or CVaR, two popular risk measures.
Let $ \mathtt{G}(z) \coloneqq \pr(f(X)\leq z)$  denote the cumulative distribution function (CDF) of $f(X)$ and
$\mathtt{G}^{-1}(q) \coloneqq \{z: \mathtt{G}(z) \geq q\}$ denote
its quantile function.
Fixing an arbitrary risk level $\tau\in(0,1)$, we define $\zeta_{\VaR} \coloneqq \VaR_\tau(f(X)) = \mathtt{G}^{-1}(\tau)$
and let
$\hat{\zeta} \coloneqq  \hat{f}_{(\lceil \tau n\rceil)}$ be its KRR-driven estimator defined in  \eqref{eq:generic-estimator}.
We will analyze the convergence rate  of $\hat{\zeta} - \zeta_{\VaR}$
by analyzing that of  $\mathtt{G}(\hat{\zeta}) - \mathtt{G}(\zeta_{\VaR})$.
To that end, we impose the following assumption to regularize the behaviors of both $\mathtt{G}$ and $\mathtt{G}^{-1}$.

\begin{assumption}\label{assump:margin-strong}
There exist positive constants $C_1$, $C_2$, $t_0$, and $\beta\leq \gamma$ such that
\begin{align*}
C_2 t^{\gamma} \leq \pr(|f(X)-z|\leq t)\leq C_1 t^{\beta},\quad
\forall t\in(0,t_0],\;\forall z\in \{f(\BFx):\BFx\in\Omega \}.
\end{align*}
\end{assumption}

\begin{remark}\label{remark:holder}
Note that $\pr(|f(X)-z|\leq t) = \mathtt{G}(z + t) - \mathtt{G}(z-t)$
and that $\mathtt{G}$ is a non-decreasing function by definition.
Therefore,
Assumption~\ref{assump:margin-strong} is equivalent to
\begin{align}
    & |\mathtt{G}(z) - \mathtt{G}(\tilde{z})| \leq \tilde{C}_1 |z - \tilde{z}|^\beta, \label{eq:upper-bound-beta}\\
    & |\mathtt{G}(z) - \mathtt{G}(\tilde{z})| \geq \tilde{C}_2 |z - \tilde{z}|^\gamma, \label{eq:lower-bound-gamma}
\end{align}
for all $z,\tilde{z}\in\{f(\BFx):\BFx\in\Omega\}$,
where $\tilde{C}_1$ and $\tilde{C}_2$ are some positive constants.
The condition \eqref{eq:upper-bound-beta} means that $\mathtt{G}$ is $\beta$-H{\"o}lder continuous,
which includes  Lipschitz continuous ($\beta=1$) as a special case.
The condition \eqref{eq:lower-bound-gamma} can be interpreted as follows.
Let $\mathtt{G}(z) = q$ and $\mathtt{G}(\tilde{z}) = \tilde{q}$.
If we further assume $\mathtt{G}^{-1}$ is continuous,
then the condition \eqref{eq:lower-bound-gamma} can be rewritten as
$|q - \tilde{q}| \geq \tilde{C}_2 |\mathtt{G}^{-1}(q) - \mathtt{G}^{-1}(\tilde{q}) |^\gamma $,
meaning $\mathtt{G}^{-1}$ is $\gamma^{-1}$-H{\"o}lder continuous.
It is known that if $\beta>1$, then a $\beta$-H{\"o}lder continuous function on an interval is a constant.
However, neither $\mathtt{G}$ nor $\mathtt{G}^{-1}$ is a constant by definition.
Hence, Assumption~\ref{assump:margin-strong} implicitly implies
$\beta \leq 1 \leq \gamma $.
\end{remark}

To analyze the convergence rate of $|\zeta_{\VaR} - \hat{\zeta}|$, we first note that by  Assumption~\ref{assump:margin-strong},
\begin{align*}
    |\mathtt{G}(\hat{\zeta}) - \mathtt{G}(\zeta_{\VaR})|
    ={}&
    \pr\Bigl( \min(\hat{\zeta},\zeta_{\VaR}) \leq f(X) \leq \max(\hat{\zeta},\zeta_{\VaR})\Bigr) \\
    ={}& \pr\Bigl(\Bigl|f(X) - \frac{(\hat{\zeta}+\zeta_{\VaR})}{2}\Bigr|\leq \frac{|\hat{\zeta}-\zeta_{\VaR}|}{2}\Bigr)
    \geq   \tilde C_2 \frac{|\hat{\zeta}-\zeta_{\VaR}|^{\gamma}}{2^\gamma}.
\end{align*}
Therefore, $|\hat{\zeta}-\zeta_{\VaR}| = O\bigl(|\mathtt{G}(\hat{\zeta}) - \mathtt{G}(\zeta_{\VaR})|^{1/\gamma}\bigr)$,
and we may focus on the convergence rate of $|\mathtt{G}(\hat{\zeta}) - \mathtt{G}(\zeta_{\VaR})|$ in the sequel.
Let $\mathtt{G}_n(z) \coloneqq n^{-1}\sum_{i=1}^n \ind{\{f(\BFx_i)\leq z\}}$ denote
the empirical CDF of $f(X)$ and
$\hat{\mathtt{G}}_n(z)  \coloneqq  n^{-1}\sum_{i=1}^n \ind{\{\hat{f}(\BFx_i)\leq z\}}$ denote the empirical CDF of $\hat{f}(X)$.
Then,
\[
    |\mathtt{G}(\hat{\zeta}) - \mathtt{G}(\zeta_{\VaR})|
    \leq \underbrace{|\mathtt{G}(\hat{\zeta}) - \mathtt{G}_n(\hat{\zeta})| }_{V_1}
    + \underbrace{| \mathtt{G}_n(\hat{\zeta}) - \hat{\mathtt{G}}_n(\hat{\zeta})| }_{V_2}
    + \underbrace{| \hat{\mathtt{G}}_n(\hat{\zeta}) - \mathtt{G}(\zeta_{\VaR})|}_{V_3}.
\]

The term $V_1$ can be bounded using the Dvoretzky--Kiefer--Wolfowitz inequality \citep{Massart90},
which bounds the difference between a CDF and its empirical counterpart and
states that $\sup_z |\mathtt{G}(z) - \mathtt{G}_n(z) | = O_{\pr}(n^{-1/2})$.
The term $V_3$ is also easy to handle;
by definition, $\hat{\mathtt{G}}_n(\hat{\zeta}) = \frac{\lceil \tau n\rceil}{n}$ and $\mathtt{G}(\zeta_{\VaR}) = \tau$.
Hence, $V_3 \leq n^{-1}$.
The analysis of the term $V_2$ is technically more involved.
We will establish a (uniform) bound on  $\sup_{z}|\mathtt{G}_n(z)-\hat{\mathtt{G}}_n(z)|$.
This is done via
the \emph{chaining} method \citep[Chapter 5]{Wainwright19}.
It basically reduces the analysis of $\sup_{z}|\mathtt{G}_n(z)-\hat{\mathtt{G}}_n(z)|$ to that of the maximum of random variables over a finite set.

Having completed the rate analysis for the case of VaR,
it is not difficult to analyze the case of CVaR because
the KRR-driven estimator for $\CVaR_\tau(f(X))$, given by \eqref{eq:generic-estimator}, is calculated based on $\hat{\zeta}$.

\begin{theorem}\label{thm:gvarcvar}
Let $\mathcal{T}(\cdot) = \VaR_\tau(\cdot)$ or  $\mathcal{T}(\cdot) = \CVaR_\tau(\cdot)$ for some $\tau\in(0,1)$.
Suppose Assumptions~\ref{assump:subG}--\ref{assump:RKHS} and \ref{assump:margin-strong} hold.
Then,
$|\hat \theta_{n,m}-\theta| = O_{\pr}(\Gamma^{-\kappa}(\log \Gamma)^{\tilde\kappa})$, where $\kappa$ and $\tilde\kappa$ are specified as follows:
    \begin{enumerate}[label=(\roman*)]
        \item If $\nu \geq \frac{d}{\beta}$,
        then $\kappa=\frac{\nu\beta}{2\gamma(\nu+d)}$ and $\tilde\kappa=\frac{\beta}{2\gamma}$
        by setting $n\asymp\Gamma$,
        $m\asymp 1$,
        and $\lambda \asymp \Gamma^{-1}$.
        \item If $\nu < \frac{d}{\beta}$, then
        $\kappa = \frac{\beta}{2\gamma(\beta+1)}$ and $\tilde\kappa=\frac{\beta}{2\gamma}$
        by setting $n\asymp \Gamma^{\frac{\beta}{\beta+1}}$,
        $m\asymp \Gamma^{\frac{1}{\beta+1}}$,
        and $\lambda \asymp \Gamma^{-1}$.
    \end{enumerate}
\end{theorem}

Theorem~\ref{thm:gvarcvar} indicates that
a larger value of $\beta$ or $\gamma^{-1}$ leads to a faster convergence rate of $\hat{\theta}_{n,m}$ if    $\mathcal{T}(\cdot) = \VaR_\tau(\cdot)$ or  $\CVaR_\tau(\cdot)$.
As discussed in Remark~\ref{remark:holder},
$\beta$ and $\gamma^{-1}$ are interpreted as the parameters with which $\mathtt{G}$ and $\mathtt{G}^{-1}$ satisfy the H{\"o}lder condition, determining their degrees of smoothness.
Hence,  for the cases of VaR and CVaR,
Theorem~\ref{thm:gvarcvar} reveals the dependence of the performance of the KRR-driven method on the smoothness of both the CDF of $f(X)$ and its quantile function,
in addition to the known dependence on the smoothness of $f$ and the dimensionality.
Furthermore,
because $\beta\leq 1\leq \gamma$, which is implied implicitly by Assumption~\ref{assump:margin-strong},
the best scenario is $\beta=\gamma=1$
(i.e., both $\mathtt{G}$ and $\mathtt{G}^{-1}$ are Lipschitz continuous).
In this scenario, the convergence rate in Theorem~\ref{thm:gvarcvar} is the same as that in Theorem~\ref{thm:gnonsmoothin} with $\alpha=1$;
that is, the square root rate cannot be fully recovered, but it can be approached arbitrarily close to, as $\nu\to\infty$ (see Figure~\ref{fig:rates}).

\begin{remark}
It can be shown via Taylor's expansion that if $\|\nabla f(\BFx)\|$ is bounded below away from zero for all $\BFx\in\Omega$, then Assumption~\ref{assump:margin-strong} is satisfied with $\beta=\gamma=1$.
However, if $f$ has a stationary point $\BFx_0\in\Omega$ (i.e., $\|\nabla f(\BFx_0)\|=0$),
then we may have $\beta<1$ or $\gamma>1$,
yielding a slower convergence rate in Theorem~\ref{thm:gvarcvar}.
It turns out that our result can be enhanced in this scenario by virtue of ``localization''.
For a given risk level $\tau$ and its associated VaR, $\zeta_{\VaR} = \VaR_\tau(f(X))$,
if we suppose Assumption~\ref{assump:margin-strong} holds \emph{and} suppose there exist positive constants $C_1^\prime$, $C_2^\prime$, $t_0$, $\delta$, and $ \beta^\prime \leq \gamma^\prime$ such that
\[C_2^\prime t^{\gamma^\prime} \leq \pr(|f(X)-z|\leq t)\leq C_1^\prime t^{\beta^\prime},\quad
\forall t\in(0,t_0],\;\forall z\in (\zeta_{\VaR}-\delta,\zeta_{\VaR}+\delta),
\]
then a similar but refined analysis can establish the same result as in Theorem~\ref{thm:gvarcvar} with $(\beta,\gamma)$ being replaced with $(\beta^\prime, \gamma^\prime)$.
Therefore, even if Assumption~\ref{assump:margin-strong} is satisfied with $\beta<1$ or $\gamma>1$,
we may still have an improved rate result,
provided that the inequalities in the assumption are satisfied with $\beta^\prime \geq \beta$ and $\gamma^\prime\leq \gamma$ in a neighborhood of $\zeta_{\VaR}$.
In particular, if $\|\nabla  f(\BFx)\| $ is bounded below away from zero for all $\BFx$ such that $| f(\BFx) -  \zeta_{\VaR}| < \delta$,
we have $\beta^\prime = \gamma^\prime = 1$.
\end{remark}

\subsection{Summary}\label{sec:summary}

We summarize in Table~\ref{tab:correct-smooth} the upper bounds on the convergence rates for the various forms of $\mathcal{T}$ (see Section~\ref{sec:test-function} for a numerical examination regarding the tightness of these bounds).

\begin{table}[ht]
\TABLE
{The Convergence Rates $|\hat \theta_{n,m}-\theta| = O_{\pr}(\Gamma^{-\kappa}(\log\Gamma)^{\tilde\kappa}) $ in Theorems~\ref{thm:gsmooth}--\ref{thm:gvarcvar}.\label{tab:correct-smooth}}
{
\begin{tabular}{@{}clclllclclclc@{}}
\toprule
$\mathcal{T}$                                          &  & $\eta$               &  & $\nu$                                    &  & $\kappa$ & & $\tilde{\kappa}$  && $m$ && $\lambda $
\\
\midrule
\multicolumn{1}{c}{\multirow{8.2}{*}{$\E[\eta(\cdot)]$}} &  & \multirow{2.5}{*}{Smooth}      &  & $\nu\geq \frac{d}{2}$                    &  & $\frac{1}{2}$   && 0        && $1$ && $\Gamma^{-1}$                     \\
\addlinespace[0.5ex]
\multicolumn{1}{c}{}                                   &  &                              &  & $\nu<\frac{d}{2}$                                     &  & $\frac{2\nu+d}{2\nu + 3d}$  &  & 0         && $\Gamma^{\frac{d-2\nu}{2\nu + 3d}}$ && $\Gamma^{-\frac{2(2\nu+d)}{2\nu + 3d}}$           \\
\cmidrule(r){3-13}
\multicolumn{1}{c}{}                                   &  & \multirow{2.5}{*}{Hockey-stick} &  &  $\nu \geq \frac{d}{\alpha+1}$                &  & $\frac{1}{2}\wedge\frac{\nu(\alpha+1)}{2(\nu+d)}$  &  &  $\frac{\alpha+1}{2}\ind{\{\nu\alpha < d\}}$ && $1$  && \multirow{2.5}{*}{$\Gamma^{-1}$} \\
\addlinespace[0.5ex]
\multicolumn{1}{c}{}                                   &  &                              &  & $\nu < \frac{d}{\alpha+1}$                   &  & $\frac{\alpha+1}{2(\alpha+2)}$       & & $\frac{\alpha+1}{2}$   && $\Gamma^{\frac{1}{\alpha+2}}$ &&                           \\
\cmidrule(r){3-13}
\multicolumn{1}{c}{}                                   &  & \multirow{2.5}{*}{Indicator}   &  &  $\nu \geq \frac{d}{\alpha}$            &  & $\frac{\nu\alpha}{2(\nu+d)}$  &  & $\frac{\alpha}{2}$    && $1$  && \multirow{2.5}{*}{$\Gamma^{-1}$}      \\\addlinespace[0.5ex]
\multicolumn{1}{c}{}                                   &  &                              &  & $\nu< \frac{d}{\alpha}$                         &  & $\frac{\alpha}{2(\alpha+1)}$   & & $\frac{\alpha}{2}$    &&  $\Gamma^{\frac{1}{\alpha+1}}$ &&                                      \\
\cmidrule{1-13}
\multirow{2.5}{*}{VaR \& CVaR}                                   &  &                              &  & $\nu \geq \frac{d}{\beta}$                &  & $\frac{\nu\beta}{2\gamma(\nu+d)}$  & & $\frac{\beta}{2\gamma}$ && $1$  && \multirow{2.5}{*}{$\Gamma^{-1}$} \\
\addlinespace[0.5ex]
    &  &                              &  & $\nu < \frac{d}{\beta}$                  &  & $\frac{\beta}{2\gamma(\beta+1)}$  &  & $\frac{\beta}{2\gamma}$    &&   $\Gamma^{\frac{1}{\beta+1}}$ &&                      \\
\bottomrule
\end{tabular}
}
{ $0<\alpha\leq 1$ and $0<\beta\leq 1\leq \gamma$.}
\end{table}

First, there exist two thresholds with respect to the value of the smoothness parameter $\nu$. One threshold determines whether the convergence rate of $\hat{\theta}_{n,m}$ exceeds $O_{\pr}(\Gamma^{-1/3})$; that is, this threshold determines whether the use of KRR in the inner-level estimation is beneficial relative to the standard nested simulation.
The other threshold determines whether the rate achieves $O_{\pr}(\Gamma^{-1/2})$, thereby
recovering the canonical rate for Monte Carlo simulation.
The values of these two thresholds depend on the form of $\mathcal{T}$ and other relevant parameters.
They can be easily calculated based on the results in Table~\ref{tab:correct-smooth} and are presented in Table~\ref{tab:thresholds} (see also Figure~\ref{fig:rates}).
For any given dimensionality $d$,
the KRR-driven nested simulation enjoys a faster convergence rate than the standard nested simulation for most of the cases covered by our analysis,
and in many cases it can even achieve or at least approach the square root rate,
provided that $\nu$ is sufficiently large.
This feature is different from previous studies in the literature  that suggest
the use of machine learning in nested simulation is beneficial only for low-dimensional ($d<5$) problems.
The difference stems from two facts about KRR. (i) The information about the smoothness of $f$ allows us to postulate a proper function space to construct an estimate.
(ii) It can leverage the spatial information in all the inner-level samples on a global scale.

\begin{table}[ht]
\TABLE
{Smoothness Thresholds for the Cubic and Square Root Convergence Rates.\label{tab:thresholds}}
{
\begin{tabular}{@{}clclcllll@{}}
\toprule
$\mathcal{T}$                                          &  & $\eta$      &  & $(\alpha,\beta,\gamma)$                                        &  & $ \kappa \geq \frac{1}{3}$   & & $\kappa = \frac{1}{2}$
\\
\midrule
\multicolumn{1}{c}{\multirow{7}{*}{$\E[\eta(\cdot)]$}} &  & Smooth      & &      && $\nu>0$                 &  & $\nu\geq \frac{d}{2}$                              \\
\cmidrule(r){3-9}
\multicolumn{1}{c}{}                                   &  & \multirow{2.2}{*}{Hockey-stick}      &  &  $\alpha=1$ &&  $\nu>0$ && $\nu\geq d$   \\
\addlinespace[0.5ex]
&& && $0<\alpha < 1$    &  & $\nu\geq \frac{2d}{3\alpha + 1}$ & & $\nu \geq  \frac{d}{\alpha}$
\\
\cmidrule(r){3-9}
\multicolumn{1}{c}{}                                   &  & \multirow{3.5}{*}{Indicator}    &  &  $\alpha=1$ && $\nu \geq 2d$      &  & $\nu \to \infty$           \\
\addlinespace[0.5ex]
  && && $\frac{2}{3} <  \alpha < 1 $ && $\nu \geq \frac{2d}{3\alpha-2}$  && n.a. \\
\addlinespace[0.5ex]
  && &&  $0 < \alpha \leq \frac{2}{3}$ && n.a. && n.a. \\
\cmidrule{1-9}
\multirow{3.5}{*}{VaR \& CVaR}                   & &                &  &    $\frac{\beta}{\gamma}=1$          &  & $\nu \geq 2d$                     &  &  $\nu \to \infty$           \\
\addlinespace[0.5ex]
  && && $\frac{2}{3} <  \frac{\beta}{\gamma} < 1 $ && $\nu \geq \frac{2d}{3(\beta/\gamma)-2}$  && n.a. \\
\addlinespace[0.5ex]
  && &&  $0 < \frac{\beta}{\gamma} \leq \frac{2}{3}$ && n.a. && n.a. \\
\bottomrule
\end{tabular}
}
{ $0<\alpha\leq 1$ and $0<\beta\leq 1\leq \gamma$. The results hold if $n$, $m$, and $\lambda$ are properly specified.  }
\end{table}

Second, our results give refined guidelines with regard to the inner-level sample size.
The existing literature suggests that
when using parametric regression \citep{BroadieDuMoallemi15} or kernel smoothing \citep{HongJunejaLiu17} in nested simulation,
$m$ should be fixed relative to the  simulation budget $\Gamma$.
We find, however, that this decision should depend on the smoothness of $f$.
In general, $m$ should be fixed if $\nu$ is sufficiently large;
otherwise, it should grow properly without bound as $\Gamma$ increases.
The idea is that the accuracy in estimating less smooth functions is more sensitive to the sample noise, thereby requiring more replications from the inner-level simulation.

Third, the choice of the regularization parameter in all the cases that we consider is different from that when KRR is used in typical machine learning tasks.
It is known that if $\BFx_1,\ldots,\BFx_n$ are i.i.d., $m\asymp 1$, and a Mat\'ern kernel is used,
then one should set $\lambda \asymp \Gamma^{-\frac{2\nu+d}{2(\nu+d)}}$, which is clearly different from our specifications in Table~\ref{tab:correct-smooth},
in order to minimize $\E\|\hat{f} - f\|^2_{\mathscr{L}_2(\Omega)}$ \citep[Chapter 10]{geer2000empirical}.
Indeed, using the standard choice of $\lambda$ would lead to a significantly slower convergence rate of $\hat{\theta}_{n,m}$ in the setting of nested simulation.
For example, it would lead to a rate of $\Gamma^{-\frac{2\nu+d}{4(\nu+d)}}$ if $\mathcal{T}(\cdot) = \E[\eta(\cdot)]$ with $\eta$ being twice-differentiable with bounded first- and second-order derivatives,
whereas Theorem~\ref{thm:gsmooth} gives a rate of $\Gamma^{-\kappa}$ with $\kappa = \max(\frac{1}{2}, \frac{2\nu+d}{2\nu+3d})$.
For other forms of $\mathcal{T}$, the standard choice of $\lambda$ cannot even ensure the convergence of $\hat{\theta}_{n,m}$.
Note that our choice of $\lambda$ diminishes at a faster rate than the standard choice.
Also note that with everything else the same, using a smaller value of $\lambda$ in KRR
results in a smaller bias but a larger variance in estimating $f$.
Therefore, in the setting of nested simulation,
it is more important to reduce the bias than to reduce the variance in the inner-level estimation to improve the estimation quality of $\theta$.

Lastly, in this paper, we analyze the convergence rate of the KRR-driven method in terms of the absolute error, expressed as $|\hat\theta_{n,m} - \theta|$. This is in contrast to prior studies on nested simulation, such as those by \cite{GordyJuneja10}, \cite{HongJunejaLiu17}, and \cite{ZhangLiuWang22}, which typically focus on the RMSE, expressed as $\bigl(\E\bigl[(\hat\theta_{n,m}-\theta)^2\bigr]\bigr)^{1/2}$.
Our analysis framework potentially allows for the probabilistic bounds on the absolute error in Theorems \ref{thm:gsmooth}--\ref{thm:gvarcvar} to be converted into bounds on the RMSE. However, this conversion would necessitate a significantly more intricate analysis from a technical perspective, potentially obscuring the main ideas presented in this paper; see the Appendix for details. We reserve this extension for future research.

\begin{remark}
    There are two sets of hyperparameters affecting the performance of the KRR-driven method: one related to the problem instance (i.e., the unknown function $f$ and the distribution of the conditioning variable $X$), and the other related to the implementation of the method.
    The former set includes $\alpha$, $\beta$, $\gamma$, and the smoothness parameter $\nu$. The latter set includes $\nu$, the length scale parameter $\ell$ in the Mat\'ern kernel, and  the regularization parameter $\lambda$.
    While the convergence rate of the KRR-driven method is jointly determined by both sets of hyperparameters, only the second set is relevant in terms of practical implementation. In other words, there is no need to estimate $\alpha$, $\beta$, and $\gamma$. Instead, we recommend practitioners use the cross-validation approach presented in Section~5 to determine suitable values for the second set of hyperparameters $(\nu,\ell,\lambda)$.
\end{remark}

\section{$\mathcal{T}$-dependent Cross-validation}\label{sec:CV}

Given a simulation budget,
the performance of the KRR-driven estimator depends critically on---in addition to the sample allocation rule---the selection of hyperparameters.
First,
the regularization parameter $\lambda$ plays a significant role in light of the asymptotic analysis in Section~\ref{sec:analysis}.
Moreover, for a given nested simulation problem we may not know precisely the smoothness of the unknown function $f$, and therefore it is common practice to treat $\nu$ as a hyperparameter \citep{SalemiStaumNelson19}.
The performance of our method may also depend on the value of $\ell$ that specifies the Mat\'ern kernel \eqref{materngai}.
For notational simplicity, let $\Xi=(\lambda,\nu,\ell)$ denote the collection of these hyperparameters.

A standard approach for selecting hyperparameters of a machine learning model is cross-validation.
The basic idea is to divide the dataset $\mathcal{D}=\{(\BFx_i, \bar{y}_i):i=1,\ldots,n\}$ into two disjoint subsets.
One subset is used, for a given value of $\Xi$, to train the machine learning model $\hat{f}$ via some loss function $L$ that measures the discrepancy between the predicted value $\hat{f}(\BFx_i)$ and the actual observation $\bar{y}_i$.
(For example, $L(\hat{y}, y) = (\hat{y}- y)^2$ for many machine learning models,  including KRR.)
The other subset is the validation set, and it is used to assess the said value of $\Xi$
via the \emph{same} loss function.

However, in the context of nested simulation, training KRR to estimate $f(\BFx) = \E[Y|X=\BFx]$ is merely an intermediate step.
The eventual goal is to estimate $\theta = \mathcal{T}(f(X))$.
The nonlinear functional $\mathcal{T}$ transforms the distribution of $X$ to a scalar
and in the process changes the relative importance of different regions of the support of $X$.
Consequently, in order to align with the goal of estimating $\theta$, the quality of $\hat{f}$ ought to be evaluated (on the validation set) via a metric that adapts to the functional $\mathcal{T}$.

Specifically, when selecting $\Xi$, our goal is to minimize the generalization error. This refers to the expected error in estimating $\theta$ that arises from using the KRR estimator $\hat{f}$ on previously unseen data points.
Using the squared loss function to measure the discrepancy between $\hat{\theta}$ and $\theta$---which is not the same as the discrepancy between $\hat{f}$ and $f$---the generalization error is
\begin{equation} \label{eq:gen-error}
\bigl[\mathcal{T}(\hat{f}(X)) - \mathcal{T}(f(X)) \bigr]^2,
\end{equation}
where $\hat{f}$ is taken as given.
To approximate $\mathcal{T}(\hat{f}(X))$, it is necessary to obtain a sample of $X$ that is independent of the training set, in order to avoid any dependency on $\hat{f}$. The validation set effectively serves this purpose. We utilize the validation set to construct an estimate of $\mathcal{T}(\hat{f}(X))$, which we denote as $\hat\theta(\Xi, \mathcal{T}, I^{\mathsf{Tr}}, I^{\mathsf{Va}})$.
Here, $I^{\mathsf{Tr}}$ and $I^{\mathsf{Va}}$ are the sets of indices for the training set and the validation set, respectively.

Additionally, we construct an estimate of $\theta = \mathcal{T}(f(X))$ by applying the standard nested simulation method, which is ``model-free'', to the validation set. It is crucial to use the validation set instead of the training set in this case, as using the training set would introduce dependency with $\hat\theta(\Xi, \mathcal{T}, I^{\mathsf{Tr}}, I^{\mathsf{Va}})$, potentially leading to overfitting issues. We denote the estimate obtained using the standard nested simulation method on the validation set as $\hat\theta^{\mathsf{St}}(\mathcal{T}, I^{\mathsf{Va}})$.
Detailed expressions of $\hat\theta(\Xi, \mathcal{T}, I^{\mathsf{Tr}}, I^{\mathsf{Va}})$ and $\hat\theta^{\mathsf{St}}(\mathcal{T}, I^{\mathsf{Va}})$, in terms of $\Xi$, $\mathcal{T}$, $I^{\mathsf{Tr}}$, and $I^{\mathsf{Va}}$, can be found in Section~\ref{sec:metric-expression} of the e-companion.

The generalization error, as defined in \eqref{eq:gen-error}, can be approximated using the following expression:
\begin{equation}\label{eq:new-metric}
\bigl(\hat\theta(\Xi,  \mathcal{T}, I^{\mathsf{Tr}}, I^{\mathsf{Va}}) - \hat\theta^{\mathsf{St}}(\mathcal{T}, I^{\mathsf{Va}})\bigr)^2.
\end{equation}
This approximation measures the discrepancy between $\hat{f}(\BFx_i)$ and $\bar{y}_i$, while taking into account the unique characteristics of the functional $\mathcal{T}$.
In contrast, the standard cross-validation simply measures the discrepancy via the mean squared error, $|I^{\mathsf{Va}}|^{-1}\sum_{i\in I^{\mathsf{Va}}}(\hat{f}(\BFx_i)-\bar{y}_i)^2$, which is independent of $\mathcal{T}$.

To reduce the variance for computing the metric \eqref{eq:new-metric},
one may use the $K$-fold cross-validation setting.
The dataset $\mathcal{D}$ is divided into $K$ disjoint roughly equal-sized parts.
One of them is taken as the validation set, while the remaining $K-1$ parts are merged as the training set.
The procedure is repeated $K$ times, each time with a different part as the validation set.
The performance of $\Xi$ is evaluated via the  average value of \eqref{eq:new-metric} over the $K$ validation sets:
\[
\mathrm{CV}(\Xi, \mathcal{T}) \coloneqq \frac{1}{K} \sum_{l=1}^K \bigl( \hat\theta(\Xi, \mathcal{T},   \mathcal{D}\setminus I_l, I_l) - \hat\theta^{\mathsf{St}}( \mathcal{T}, I_l) \bigr)^2,
\]
where $I_l$ is the set of indices for the $l$-th part.
Lastly, for a given functional $\mathcal{T}$, we determine the optimal value of $\Xi$ by minimizing $\mathrm{CV}(\Xi, \mathcal{T})$.
Namely, $\Xi$ is selected in a $\mathcal{T}$-dependent manner.

In general, the larger $K$, the higher the computational cost of $K$-fold cross-validation because one needs to train the machine learning model of concern $K$ times, each on a different dataset.
However, when the machine learning method involved is KRR, the extreme case that $K=n$---that is, leave-one-out cross-validation (LOOCV)---can make use of a well-known trick that greatly simplifies the expression of $\mathrm{CV}(\Xi, \mathcal{T})$, thereby leading to substantial savings in computational cost relative to the case of a smaller $K$  (see Section~\ref{sec:LOO} of the e-companion for details).
We use the LOOCV setting in the numerical experiments in Section~\ref{sec:numeric}.

\begin{remark}\label{remark:equiv-for-quantile}

To see the connection between the $\mathcal{T}$-dependent LOOCV and the standard (i.e., $\mathcal{T}$-independent) LOOCV, let us consider three special cases: (i) $\mathcal{T}(\cdot) = \E[\cdot]$, (ii) $\mathcal{T}$ represents VaR; and (iii) $\mathcal{T}$ represents CVaR.
Then, it is easy to derive that
\[
\hat\theta(\Xi, \mathcal{T},   \mathcal{D}\setminus I_l, I_l) =  \hat{f}^{-l}(\BFx_l;\Xi) \qq{and}
\hat\theta^{\mathsf{St}}( \mathcal{T}, I_l) = \bar{y}_l,
\]
where $I_l = \{\BFx_l\}$ and $\hat{f}^{-l}(\cdot;\Xi)$ denotes the KRR estimator trained with $\BFx_l$ removed from $\mathcal{D}$ and with the hyperparameters being $\Xi$; see Section~\ref{sec:LOO} of the e-companion.
Therefore, in these three particular cases, the $\mathcal{T}$-dependent LOOCV is reduced to the standard LOOCV that uses the squared loss between $\hat{f}$ and $f$ as the measure of fit. Additionally, the expression for $\mathrm{CV}(\Xi, \mathcal{T})$ does not depend on the risk level $\tau$ that defines VaR or CVaR. Thus, if one is interested in evaluating both VaR and CVaR across multiple risk levels, it is only necessary to perform hyperparameter tuning once, rather than separately for each risk measure and risk level.

\end{remark}

\begin{remark}
A potential drawback of the $\mathcal{T}$-dependent LOOCV is its increased computational cost when considering multiple functionals simultaneously, as compared to the standard LOOCV.
This increased cost is not due to the $\mathcal{T}$-dependent LOOCV being inherently slower than its standard counterpart; rather, it stems from the necessity to execute the $\mathcal{T}$-dependent LOOCV multiple times, performing it once for each functional.
In contrast, the standard LOOCV allows for a single estimation of the hyperparameter, which can then be applied to all functionals.
Nevertheless, our numerical experiments demonstrate that the $\mathcal{T}$-dependent cross-validation consistently outperforms the standard cross-validation by a significant margin. This advantage outweighs the computational overhead, particularly when there are not many functionals to consider simultaneously. \citet[page~669]{HongJunejaLiu17} also documented similar observations in their numerical experiments; however, they did not offer a principled method to leverage the dependence of $\mathcal{T}$. See more discussion in Section~\ref{ec-sec:dependence-vs-indep} of the e-companion.
\end{remark}

\begin{remark}
Similar to hyperparameter tuning tasks via cross-validation in machine learning, minimizing $\mathrm{CV}(\Xi, \mathcal{T})$ presents a black-box optimization problem. The objective function is non-convex, computationally demanding to evaluate, and lacks both an analytical expression and an efficient mechanism for computing gradients. In recent years, Bayesian optimization \citep{Frazier18} has emerged as a successful approach to hyperparameter tuning problems; see, e.g., \cite{ShahriariSwerskyWangAdamsdeFreitas16} and \cite{TurnerErikssonMcCourtKiiliLaaksonenXuGuyon21}. We apply this approach to determine proper values of the hyperparameters $\Xi$.

\end{remark}

\section{Numerical Experiments}\label{sec:numeric}

We now numerically evaluate the KRR-driven estimator.
All of the experiments are implemented in Python on a computer with 2.90GHz Intel Xeon CPU and 64GB of RAM.
In Section~\ref{sec:test-function}, we study the mitigating effect of smoothness on the curse of dimensionality using test functions with known smoothness.
Practical examples arising from portfolio risk management and input uncertainty quantification are discussed in Section \ref{sec:prm} and Section \ref{sec:uq}, respectively.
In Section~\ref{sec:KRR-suitable}, we discuss when the KRR-driven method is suitable for nested simulation.

\subsection{Smoothness versus Dimensionality}\label{sec:test-function}

The asymptotic analysis in Section~\ref{sec:analysis} provides upper bounds on the convergence rate of the KRR-driven estimator for various forms of the functional $\mathcal{T}$.
These upper bounds reveal the mitigating effect of the smoothness on the curse of dimensionality.
We numerically examine such an effect, as well as how tight these upper bounds are,
using test functions with known smoothness.
Let $f(\BFx) = \E[Y|X=\BFx]$ be the conditional expectation in the inner level.
We assume
\begin{equation} \label{eq:test-fun}
f(\BFx) = \sum_{i=1}^N c_i \Psi(\BFx- \tilde{\BFx}_i),\quad \BFx\in\Omega,
\end{equation}
where
$N= 10$, {$\Omega = [-1,1]^d$}, and
$\Psi$ is defined by \eqref{materngai}
in which {$\ell= 10$} and $\nu$ is specified later.
Both $c_i$ and $\tilde{\BFx}_i$ are randomly generated---with the former from {$\mathsf{Uniform}[0,100]$}  (i.e., the uniform distribution on {$[0,100]$}) and the latter from {$\mathsf{Normal}(0, 1)$}  (i.e., {the standard normal distribution})---and then fixed.
According to the definition of RKHSs, $f$ is a function in $\mathscr{N}_\Psi(\Omega)$ and has a smoothness parameter $\nu$.

We generate each outer-level scenario $\BFx=(x_1,\ldots,x_d)$ as follows.
For each $l=1,\ldots,d$,
$x_l$ independently follows a truncated normal distribution, {$\mathsf{Normal}(0, 1)$} with truncated range {$[-1,1]$}.
Given an outer-level scenario $\BFx_i$,
the inner-level samples are simulated via $y_{ij} = f(\BFx_i) + \varepsilon_{ij}$, where $\varepsilon_{ij}$'s are i.i.d. {$\mathsf{Normal}(0, 1)$} random variables.

\begin{figure}[ht]
    \FIGURE{
    $
    \begin{array}{c}
    \includegraphics[width=\textwidth]{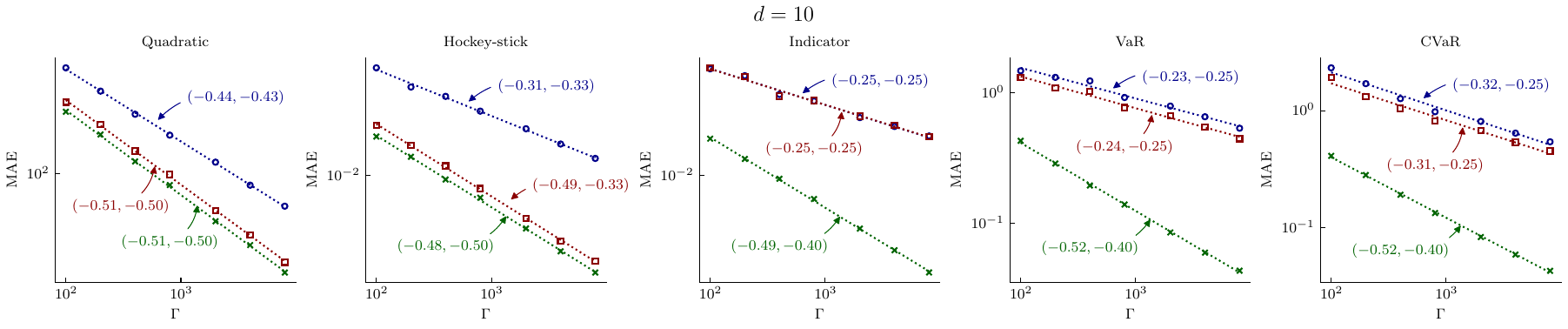} \\
    \includegraphics[width=\textwidth]{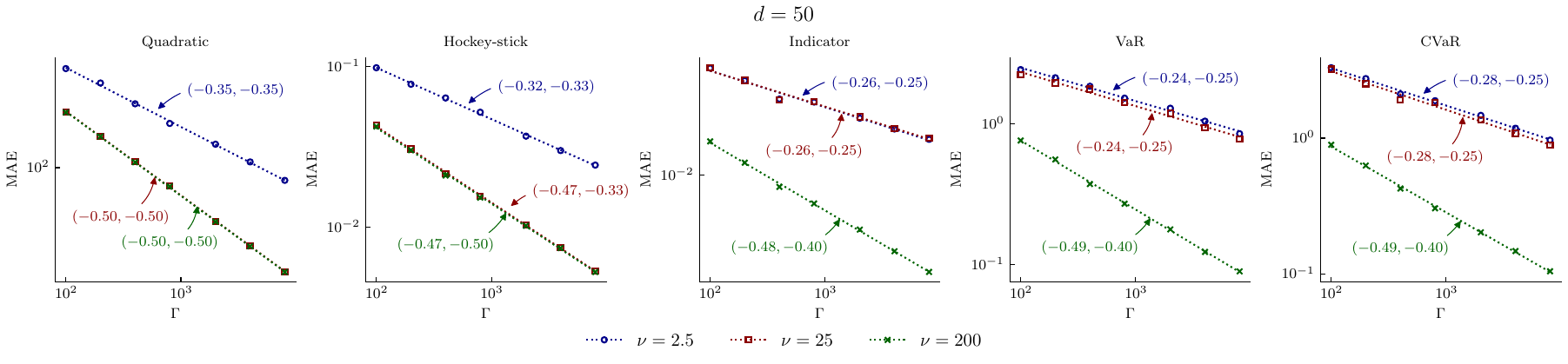}
    \end{array}
    $
    }
    {MAEs for the KRR-driven Nested Simulation.   \label{fig:test-fun}
    }
    {The data are plotted on a logarithmic scale. Each dashed line is drawn via linear regression between $\log(\mathrm{MAE})$ and $\log(\Gamma)$. The first number within each pair of parentheses is the estimated slope of the corresponding line, whereas the second number is the value of $-\kappa$.}
\end{figure}

We examine five different forms of $\mathcal{T}$: $\mathcal{T}(\cdot) = \E[\eta(\cdot)]$ with $\eta(z)=z^2$, $\eta(z)=(z-z_0)^+$, and $\eta(z)=\ind{\{z\geq z_0\}}$,  $\mathcal{T}(\cdot)=\VaR_\tau(\cdot)$, and $\mathcal{T}(\cdot)=\CVaR_\tau(\cdot)$.
Here, the threshold $z_0$ is set to be the {95\%} quantile of the distribution of $f(\BFx)$, and
the risk level $\tau$ is set to be $95\%$.
The true value of $\theta=\mathcal{T}(f(X))$ is estimated based on {$10^6$} i.i.d. copies of $X$.
The budget allocation $(n,m)$ and the regularization parameter $\lambda$ are specified according to our asymptotic results in Table \ref{tab:correct-smooth}.
For each $d\in\{10, 50\}$, $\nu\in \{5/2, d/2, 4d\}$, and a budget of $\Gamma$ ranging from $10^2$ to $10^4$,
we estimate the mean absolute error (MAE), defined as $|\hat \theta_{n,m} - \theta|$, of the KRR-driven method using 1,000 macro-replications.
In Figure~\ref{fig:test-fun}, we plot  the MAE against the budget on a logarithmic scale and compare the slope of the resulting line with the parameter $\kappa$ in the upper bound $|\hat \theta_{n,m}-\theta| = O_{\pr}(\Gamma^{-\kappa}(\log\Gamma)^{\tilde\kappa})$, which is specified in Table~\ref{tab:correct-smooth} (it is easy to show that Assumptions~\ref{assump:subG}--\ref{assump:RKHS} hold,
Assumption~\ref{assump:margin} holds with $\alpha=1$,
and  Assumption~\ref{assump:margin-strong} holds with $\beta=\gamma=1$).

We have several observations from Figure~\ref{fig:test-fun}.
First, the mitigating effect of the smoothness on the curse of dimensionality is clearly demonstrated.
Everything being equal, a higher smoothness yields a faster decay rate of the MAE.
For instance, when $\mathcal{T}=\VaR$, $d=10$, and we examine the fourth subplot on the first row, the estimated slopes of the three dashed lines are {$-0.23$, $-0.24$, and $-0.52$} for $\nu=5/2$, $d/2$, and $4d$, respectively. These values indicate that the decay rate of the MAE is approximately $\Gamma^{-1/4}$ for $\nu=5/2$ or $d/2$, while it is approximately $\Gamma^{-1/2}$ when $\nu=4d$.

Second,
upon examining the pair of numbers associated with each dashed line, we observe that the first number is generally no greater than the second. In many instances, the two values are nearly identical. The first number represents the slope of the linear regression line between $\log(\text{MAE})$ and $\log(\Gamma)$, while the second number is equal to $-\kappa$. This observation aligns with our theoretical results in Theorems~\ref{thm:gsmooth}--\ref{thm:gvarcvar}, which assert that the MAE is asymptotically upper bounded by $\Gamma^{-\kappa}(\log\Gamma)^{\tilde\kappa}$.

Third, in certain instances, the estimated slope of a dashed line is noticeably smaller than $-\kappa$. For example, when $d=50$, $\nu=4d$, and $\mathcal{T}=\CVaR$, these two values are $-0.49$ and $-0.40$, respectively.
Nevertheless, we emphasize that our upper bounds hold for \emph{all} functions $f$ in the RKHS $\mathscr{N}_\Psi(\Omega)$,
while the  results in Figure~\ref{fig:test-fun} are computed for a \emph{particular} function $f$  defined by \eqref{eq:test-fun}.
Therefore, although it suggests that there \emph{may} be room for improvement in our upper bounds,
the mismatch is not conclusive in itself.
To fully address the question regarding the tightness of our upper bounds may require  minimax lower bounds \citep[Chapter~3]{GyorfiKohlerKrzyzakWalk02} on the convergence rate.
We leave that investigation for future research.

\subsection{Portfolio Risk Management}\label{sec:prm}

Consider a portfolio consisting of options that derive their values from the performance of $q$ underlying assets.
Suppose that  these options  share the same expiration date $T$.
The manager of the portfolio is interested in assessing the risk at some future time $T_0<T$ in terms of five metrics:
expected quadratic loss $\E[Z^2]$,
expected  excess loss $\E[(Z-z_0)^+]$,
probability of a large loss $\pr(Z \geq z_0)$,  and
VaR and CVaR of $Z$ at some risk level $\tau$,
where $Z$ is the portfolio loss at time $T_0$ and $z_0$ is some threshold.
The first three of these metrics correspond to
$\mathcal{T}(\cdot) = \E[\eta(\cdot)]$  with $\eta(z) = z^2$, $\eta(z) = (z-z_0)^+$, and $\eta(z) = \ind{\{z\geq z_0\}}$, respectively.

Let $\BFS(t) = (S_1(t), \ldots, S_q(t))$ denote the vector of the asset prices at time $t$.
In the framework of nested simulation,
we simulate $\BFS(t)$ up to time $T_0$ and the sample paths constitute outer-level scenarios.
Then, in the inner-level simulation, we simulate $\BFS(t)$ from $T_0$ to $T$ to estimate the prices of the options in the portfolio.
A subtlety here is that the probability measures for the outer- and inner-level simulations are different; the former uses the real-world measure, whereas the latter uses a risk-neutral measure.
Suppose $\BFS(t)$ follows a $q$-dimensional geometric Brownian motion (GBM):
\[
\frac{\mathrm{d} S_i(t)}{S_i(t)} = \mu_i  \mathrm{d} t + \sum_{j=1}^q \sigma_{ij}  \mathrm{d} B_j(t), \quad i = 1, \ldots, q,
\]
where $B_1(t),\ldots, B_q(t)$ are independent standard one-dimensional Brownian motions, the matrix $(\sigma_{ij})_{i,j=1}^q$ is lower-triangular (i.e., $\sigma_{ij}=0$ for all $i>j$),
and $\mu_i$ takes different values depending on which probability measure is used.
We assume, for simplicity, that each underlying asset has the same return $\mu$ (i.e., $\mu_i = \mu$ for all $i$) under the  real-world measure.
However, under the risk-neutral measure, it should be identical to the risk-free interest rate $r$ (i.e., $\mu_i=r$ for all $i$).

The portfolio consists of six options written on each underlying asset (so the portfolio consists of $6q$ options in total).
These options include three geometric Asian call options with discrete monitoring and three up-and-out barrier call options with continuous monitoring that can be knocked out anytime between $0$ and $T$.
Note that both types of options have path-dependent payoffs.
For an underlying asset with price $S(t)$,
the payoff of
a geometric Asian call option with monitoring at the times $0=t_0 < t_1 < \cdots < t_M = T$ is
$\bigl(\bigl(\prod_{k=1}^M S(t_k)\bigr)^{1/M}-K\bigr)^+$, where $K$ is the strike price,
and the payoff of an up-and-out barrier option with barrier $H$ is
$(S(T)-K)^+\ind{\{\max_{0\leq t\leq T}S(t) \leq H \}}$.
We assume that
the strike prices of the three Asian options written on each asset are $K_1$, $K_2$, and $K_3$, and the same holds for the three barrier options; moreover, these barrier options have the same barrier $H$.
We also assume that the risk horizon $T_0=t_{M_0}$ for some $M_0=1,\ldots,M$.
According to the theory of  derivatives pricing \citep[Chapter~1]{Glasserman03},
the value of each option at time $T_0$ equals its expected discounted payoff.
Therefore, the value of the portfolio at time $T_0$ is
\begin{align*}
V_{T_0}(X) = \E\biggl[\underbrace{e^{-r(T-T_0)} \sum_{i=1}^q \sum_{l = 1}^3 \biggl( \Bigl(\prod_{k=1}^M S_i(t_k)\Bigr)^{1/M} - K_l\Bigr)^+
 + (S_i(T) - K_l)^+ \ind{\Bigl\{\max_{0\leq t\leq T}S_i(t) \leq H \Bigr\}} }_{W} \biggr)
\,\bigg|\, X\biggr],
\end{align*}
where $X \in \RR^{3q}$ is a vector of risk factors defined as
\[
X = \biggl(S_1(T_0), \ldots, S_q(T_0), \Bigl(\prod_{k=1}^{M_0} S_1(t_k) \Bigr)^{1/M_0}, \ldots, \Bigl(\prod_{k=1}^{M_0} S_q(t_k) \Bigr)^{1/M_0},
\max_{0\leq t \leq T_0} S_1(t), \ldots, \max_{0\leq t \leq T_0} S_q(t)\biggr).
\]
The portfolio loss at time $T_0$ is then $Z \coloneqq V_0 - V_{T_0}(X)$, where $V_0$ is the value of the portfolio at time $0$  and is calculated in the same way as $V_{T_0}$, except with $T_0=0$.
Further, it can be expressed as $Z=\E[Y|X]$, where $Y= V_0 - W$.

To assess the performance of a nested simulation method, we need to compute the true value of $\theta$,
which is done as follows.
Under the assumption that $\BFS(t)$ follows a GBM, $V_{T_0}(X)$ can be calculated in closed form (see, e.g., \citealt[Chapter~4]{Haug07}).
We generate $10^8$ i.i.d. copies of $X$ and calculate the corresponding values of $V_0-V_{T_0}(X)$,
which are i.i.d. copies of $Z$ and can be used to accurately estimate $\theta=\mathcal{T}(Z)$. (The relative standard deviations of the estimates for different forms of  $\mathcal{T}$ and different values of $q$ are consistently around $10^{-4}$; see Section~\ref{ec-sec:brute-force} of the e-companion for details.)
We compare different methods based on the relative root mean squared error (RRMSE) defined as
the ratio of the RMSE to this accurate estimate of $\theta$.
For each problem instance,
the RRMSE is estimated via 1,000 macro-replications.

The other parameters involved are specified as follows.
The expiration date of each option in the portfolio is $T=1$,
and the risk horizon is  $T_0=3/50$.
In the GBM model,
the initial price of each asset is $S_i(0)=100$ for all $i=1,\ldots,q$,
the return of each asset under the real-world measure is $\mu=8\%$, the risk-free rate is $r = 5\%$,
and the volatility term $\sigma_{ij}$ is generated randomly (see Section~\ref{ec-sec:vol} of the e-companion for details)
and then fixed if $i\leq j$ and is $\sigma_{ij}=0$ otherwise.
We vary $q\in\{10, 20, 50, 100\}$ (note that the dimensionality of the conditioning variable $X$ is $d=3q$).\footnote{As pointed out by \cite{HongJunejaLiu17}, taking advantage of the additive structure of $V_{T_0}(X)$ and the fact that each option is written on a single underlying asset, one may decompose this $3q$-dimensional problem into $6q$  two-dimensional sub-problems---because the portfolio is comprised of $6q$ options, each having two risk factors---and
apply the KRR-driven method to each sub-problem separately,
resulting in implementing the method $6q$ times.
However, each run of the KRR-driven method incurs a non-negligible computational overhead, mainly due to matrix inversion and selection of hyperparameters. Furthermore, the size of the matrices to invert is $n\times n$ and is independent of the dimensionality of the problem. Therefore, using the decomposition scheme would result in a computational overhead that is essentially $6q$ times larger than not using the scheme. }

For the options in the portfolio,
the three different strike prices are $K_1=90$, $K_2=100$, and $K_3=110$.
Moreover, the monitoring times of each Asian option are $\{t_k=kT/M:k=1,\ldots,M\}$ with $M=50$,
and the barrier of each barrier option is $H=150$.

When implementing nested simulation methods, it is necessary to simulate the running maximum $\max_{0\leq t\leq T} S_i(t)$ to generate the payoffs of the continuously monitored barrier options. However, exact simulation of this running maximum is unavailable.
Instead, we use the commonly adopted approach of Brownian bridge approximation. See Section~\ref{ec-sec:run-max} of the e-companion for details.

Given a simulation budget of $\Gamma=10^4$ or $10^5$,
we compare three methods for nested simulation:
\begin{enumerate}[label=(\roman*)]
    \item The standard method. We try 10 different values of the inner-level sample size $m$ ranging from 5 to 2,000 and report the best performance (i.e., lowest RRMSE) among them.\footnote{The optimal value of $m$ is generally unknown and problem-dependent. \cite{ZhangLiuWang22} develop a bootstrap-based approach to estimate the optimal value of $m$ using a small proportion of the total budget. They report (and we confirm in our experiments) that the performance of the bootstrap-based approach is often close to, but slightly worse than, the best performance among those associated with a wide range of values of $m$.}

    \item The KRR-driven method. Similarly, we try different values of $m$ and report the best performance.
    Given $\Gamma$ and $m$, we apply the $\mathcal{T}$-dependent LOOCV to select the hyperparameters.

    \item The regression-driven method.
    This method is similar to the KRR-driven method, except that the conditional expectation $\E[Y|X]$ is modeled as a linear combination of basis functions in $X$.
    In addition to different values of $m$, we also try different sets of basis functions, including polynomials and orthogonal polynomials (the Legendre, Laguerre, Hermite, and Chebyshev polynomials).
    For each class of polynomials, we cap the highest allowable order at $5$, excluding any interaction terms.
    This configuration leads to $5\times 5 = 25$ distinct sets of basis functions.
    Furthermore, for each of these sets, we consider two variations depending on whether we include the European option
price under the GBM model, which has a closed-form solution.
Hence, in total, we have $50=2\times 50$ sets of basis functions to examine.
    We report the best performance among all the combinations of different values of $m$ and sets of basis functions.

\end{enumerate}

\begin{table}[htbp]
    \TABLE{RRMSE ($\%$) for the Portfolio Risk Management Problem. \label{tab:prm-results-matern}}
    {
    \begin{tabular}{@{\extracolsep{10pt}} l @{\extracolsep{10pt}} l @{\extracolsep{10pt}} r r @{\extracolsep{10pt}} r r @{\extracolsep{10pt}} r r}
    \toprule
    & \multirow{2.5}{*}{$\mathcal{T}$} & \multicolumn{3}{c}{$\Gamma=10^4$} & \multicolumn{3}{c}{$\Gamma=10^5$} \\
    \cmidrule{3-8}
    & & KRR & Regression & Standard & KRR & Regression & Standard \\
    \cmidrule{1-2} \cmidrule{3-5} \cmidrule{6-8}
    \multirow{5}{*}{$q = 10$} & Quadratic & 3.67 & 8.11 & 19.56 & 1.21 & 2.31 & 8.64 \\
    & Hockey-stick & 3.23 & 6.30 & 15.07 & 1.44 & 1.94 & 7.23 \\
    & Indicator & 2.44 & 3.59 & 5.84 & 0.77 & 1.13 & 2.75 \\
    & VaR & 2.64 & 5.16 & 10.13 & 0.91 & 1.58 & 5.50 \\
    & CVaR & 2.83 & 5.57 & 11.32 & 0.83 & 1.65 & 5.81 \\
    \cmidrule{1-2} \cmidrule{3-5} \cmidrule{6-8}
    \multirow{5}{*}{$q = 20$} & Quadratic & 3.46 & 10.96 & 18.75 & 1.13 & 2.55 & 8.29 \\
    & Hockey-stick & 3.92 & 9.43 & 20.18 & 1.23 & 2.54 & 8.83 \\
    & Indicator & 2.93 & 4.30 & 8.36 & 0.93 & 1.37 & 4.15 \\
    & VaR & 3.07 & 6.71 & 12.02 & 0.99 & 1.75 & 6.73 \\
    & CVaR & 3.28 & 6.94 & 13.31 & 1.33 & 1.80 & 6.62 \\
    \cmidrule{1-2} \cmidrule{3-5} \cmidrule{6-8}
    \multirow{5}{*}{$q = 50$} & Quadratic & 3.30 & 18.29 & 18.88   & 1.08 & 2.74 & 8.00 \\
    & Hockey-stick & 4.23 & 13.31 & 20.61 & 1.35 & 2.97 & 9.93 \\
    & Indicator & 3.10 & 4.67 & 9.86 & 1.00 & 1.52 & 4.93 \\
    & VaR & 3.33 & 9.84 & 11.46 & 1.09 & 1.93 & 6.02 \\
    & CVaR & 3.45 & 10.42 & 12.97 & 1.13 & 1.95 & 6.39 \\
    \cmidrule{1-2} \cmidrule{3-5} \cmidrule{6-8}
    \multirow{5}{*}{$q = 100$} & Quadratic & 3.37 & 24.50 & 17.67 & 1.10 & 3.88 & 8.08 \\
    & Hockey-stick & 4.82 & 27.75 & 25.72 & 1.57 & 4.80 & 11.68 \\
    & Indicator & 3.71 & 9.19 & 13.57 & 1.19 & 2.26 & 6.32 \\
    & VaR & 3.52 & 15.29 & 12.44 & 1.13 & 2.65 & 6.24 \\
    & CVaR & 3.82 & 15.77 & 14.03 & 1.17 & 2.69 & 6.78 \\
    \bottomrule
    \end{tabular}
    }
    {The dimensionality of the conditioning variable is $d=3q$. For  hockey-stick and indicator functions, the threshold  is  $z_0=0.02 V_0$. For VaR and CVaR, the risk level is $\tau=99\%$. }
\end{table}

Table~\ref{tab:prm-results-matern} presents the RRMSE results, where $m$ is chosen to be $5$ for the regression-driven and KRR-driven methods. On the contrary, the standard method is more dependent on the budget allocation, and requires for more inner-level samples to achieve the best performance. The values of $m$ used include $25$, $50$, $100$ and $200$.
The KRR-driven method consistently outperforms the other two methods by a significant margin.
For example, when $q=10$, $\mathcal{T}(Z) = \E[Z^2]$, and $\Gamma=10^4$, the RRMSE for the KRR-driven method is 3.67\%, whereas it is {8.11\%} for the regression-driven method;
in contrast, the standard method has a significantly higher RRMSE ({8.64\%}) even with a 10-times larger budget ($\Gamma=10^5$).

Moreover, the dimensionality does not significantly affect the performance of
the KRR-driven method.
As $q$ increases from $10$ to $100$ (i.e., $d$ increases from $30$ to $300$), the RRMSEs that the KRR-driven method achieves do not grow rapidly, remaining at a relatively low level compared to the other two methods.
For example, when $q=100$, $\mathcal{T}$ is VaR, and $\Gamma=10^5$, the RRMSEs for the three methods are {1.13\%} (KRR-driven), {2.65\%} (regression-driven), and {6.24\%} (standard), respectively.
This may be attributed to the fact that the estimated value of $\nu$ is large, indicating that  $\E[Y|X]$ is highly smooth with respect to $X$.

\begin{table}[ht]
        \TABLE{Running Time (sec.) for the Portfolio Risk Management Problem. \label{tab:prm-wct}}
        {
        \begin{tabular}{@{\extracolsep{10pt}} l @{\extracolsep{10pt}} r r @{\extracolsep{10pt}} r r @{\extracolsep{10pt}} r r}
        \toprule
        & \multicolumn{6}{c}{Simulation Running Time} \\
        \cmidrule{2-7}
        & \multicolumn{3}{c}{$\Gamma = 10^4$} & \multicolumn{3}{c}{$\Gamma = 10^5$} \\
        \cmidrule{2-7}
        & KRR & Regression & Standard & KRR & Regression & Standard  \\
        \cmidrule{1-1} \cmidrule{2-4}  \cmidrule{5-7}
        $q = 10$ & \multicolumn{3}{c}{75.10} & \multicolumn{3}{c}{763.16} \\
        $q = 20$  & \multicolumn{3}{c}{76.88} & \multicolumn{3}{c}{794.76} \\
        $q = 50$  & \multicolumn{3}{c}{81.91} & \multicolumn{3}{c}{832.44} \\
        $q = 100$ & \multicolumn{3}{c}{90.31} & \multicolumn{3}{c}{914.77} \\
        \midrule
        & \multicolumn{6}{c}{Estimation Running Time} \\
        \cmidrule{2-7}
        & \multicolumn{3}{c}{$\Gamma = 10^4$} & \multicolumn{3}{c}{$\Gamma = 10^5$} \\
        \cmidrule{2-7}
        & KRR & Regression & Standard & KRR & Regression & Standard  \\
        \cmidrule{1-1} \cmidrule{2-4}  \cmidrule{5-7}
        $q = 10$ & 2.03 & 8.05 & $\approx 0$  & 294.09 & 97.97 & $\approx 0$ \\
        $q = 20$ & 2.63 & 8.78 & $\approx 0$  & 341.53 & 98.33 & $\approx 0$ \\
        $q = 50$ & 4.05 & 9.48 & $\approx 0$ & 488.05 & 111.08 & $\approx 0$ \\
        $q = 100$ & 6.44 & 12.84 & $\approx 0$ & 876.58 & 126.96 & $\approx 0$ \\
        \bottomrule
        \end{tabular}
        }
        {The total running time of a nested simulation method is the sum of its simulation time and estimation time. Given $\Gamma$ and $q$, the simulation time is independent of the form of $\mathcal{T}$ and the nested simulation method. In contrast, the estimation time for the KRR-driven (or regression-driven) method is mainly determined by $n$. We report the result corresponding to the value of $n$ that yields the lowest RRMSE, which is consistent with Table~\ref{tab:prm-results-matern}. Each reported result is an average over macro-replications of the experiment and different forms of $\mathcal{T}$. For the KRR-driven method, the estimation running time involves 10 searches over the hyperparameter space. Similarly, for the regression-driven method, it involves a  search over all possible combinations of basis functions, as specified in the experimental setup.
        }
\end{table}

While the KRR-driven method offers superior estimation accuracy, it may come at the cost of higher computational complexity compared to the other two methods. Table~\ref{tab:prm-wct} presents the running times of the three methods, which accounts for two computational steps: simulation and estimation.
The simulation step generates data, including outer-level scenarios and inner-level samples. The computational cost of this step is mainly determined by the budget $\Gamma$ and is the same across different nested simulation methods.
However, the estimation step, which processes the generated data to construct an estimate of $\theta$, varies significantly for different methods. For the standard method, the estimation step takes negligible time (so the method's total running time is virtually identical to its simulation time). For the regression-driven method, it consists of running linear regression and selecting the set of basis functions. For the KRR-driven method, it involves running KRR and selecting hyperparameters.
As a result, the latter two methods both require a significantly higher computational cost than the standard method, and the gap widens as the problem's dimensionality or the simulation budget increases.

When $\Gamma=10^4$, the running time for estimation in the regression-driven method is longer than that of the KRR-driven method. This is because the regression-driven method requires trying many sets of basis functions, some of which may have a large number of basis functions, leading to a significantly longer time for computation.
However, when $\Gamma=10^5$, the KRR-driven method is significantly slower. This is because its computation involves numerical inversion of $n\times n$ matrices, which is an intensive computational task for large $n$.

Despite the substantial computational cost of the KRR-driven method in a large-budget setting, its use is warranted because of the low RRMSE it achieves, as shown in Table~\ref{tab:prm-results-matern}, as well as the lengthy running time for simulation common to all three methods.
For example, consider the high-dimensional case ($q=100$) with a large budget ($\Gamma=10^5$). In this case, the ratio of the total running time of the KRR-driven method to the standard method is 2.0 ($\approx (914.77 + 876.58) / 914.77$), the highest among all cases considered. According to Table~\ref{tab:prm-results-matern}, with the same $\Gamma$ and $q$, the ratio of RRMSE of the standard method to the KRR-driven method is at least {$5.3$ ($\approx 6.32/1.19$)}, when $\mathcal{T}$ is associated with an indicator function).
This suggests that due to its cubic convergence rate, the standard method would need roughly {$5.3^3 \approx 149$} times more simulation budget (or equivalently, running time) to achieve the same level of RRMSE as the KRR-driven method. A similar argument can be made regarding the comparison between the KRR-driven and regression-driven methods.
In other words, the KRR-driven method can achieve substantial savings in running time when targeting a high level of estimation accuracy of $\theta$, especially for high-dimensional problems and when the simulation samples are expensive.

However, it is well known that KRR rapidly becomes computationally prohibitive when $n$ exceeds $10^5$, as matrix inversion requires a time complexity of order $\mathcal{O}(n^3)$. Consequently, the KRR-driven method, in its present form, is unsuitable for situations requiring the simulation of a large number of outer-level scenarios.
In such cases, one may employ low-rank matrix approximation techniques
including the Nystr\"om method \citep{LuRudiBorgonovoRosasco20} and the random features method \citep{LiuHuangChenSuykens22}. This approach can considerably reduce the computational complexity of KRR without significantly diminishing estimation accuracy. Nonetheless, integrating these methods into nested simulation would demand a careful design of the matrix approximation level.
Striking a balance between computational efficiency and the estimation accuracy of $\theta$ would likely depend on the specific form of $\mathcal{T}$.
This analysis lies beyond the scope of the present paper, and we leave it for future research.

\subsection{Input Uncertainty Quantification}\label{sec:uq}

Consider a newsvendor problem with multiple products.
For product $i=1,\ldots,d$,
let $p_i$ denote its unit selling price, $c_i$ its unit procurement cost, and
$D_i$ its demand.
Given a price vector $( p_1, \ldots, p_d )$,
the demand of these products is driven by a multinomial logit choice model \citep{AydinPorteus08}:
$D_i =  v_i \varepsilon_i$, where
$v_i = e^{\alpha_i - p_i}/( 1 + \sum_{i=1}^d e^{\alpha_i - p_i})$,
and
$\varepsilon_i$ is an independent $\mathsf{Uniform}[a,b]$ random variable.
The parameter $\alpha_i$ can be interpreted as a customer's expected utility for product $i$,
and it needs to be estimated from real data, which gives rise to the issue of input uncertainty.
We assume that the distribution of $\alpha_i$ is  $\mathsf{Normal}(\mu_i, \sigma_i^2)$. Denote $X = (\alpha_1,\ldots,\alpha_d)$.

Let $q_i$ denote the order quantity of product $i$.
Given a realized value of $X$,
the newsvendor's objective is to maximize its expected profit
$ \E[ \sum_{i=1}^d  p_i (D_i \wedge q_i)  - c_i q_i ]$
by optimizing the order quantities.
It is easy to show that the optimal order quantities are
$q_i^*= F_{\varepsilon_i}^{-1} \left( \frac{p_i - c_i}{p_i} \right) v_i = \left( (b-a) \left( \frac{p_i - c_i}{p_i} \right) + a \right) v_i,$
where $F_{\varepsilon_i}^{-1}(\cdot)$ is the inverse CDF (i.e., the quantile function) of $\varepsilon_i$.
Thus, the optimal profit is
\[Z \coloneqq \E\biggl[\underbrace{\sum_{i=1}^d p_i (D_i \wedge q_i^*)  - c_i q_i^*}_{Y}\,\bigg|\, X\biggr].\]

To quantify the impact of input uncertainty (i.e., the uncertainty about $\alpha_i$'s) on the calculation of the newsvendor's optimal profit,
we use nested simulation to construct $100(1 - \tau)\%$ credible intervals (CrIs) of the form
$[\VaR_{\tau/2}(Z), \VaR_{1-\tau/2}(Z)]$ for the distribution of $Z$.
We assess the performance of a nested simulation method using both the probability content (PC) and the width of the CrI that the method constructs.
Both of them are estimated based on 1,000 macro-replications.
In each macro-replication,
we construct a CrI, say $(l,u)$, for a particular method and estimate its PC as follows.
We generate {$10^6$} i.i.d. copies of $X$ and calculate the corresponding values of $Z$ (which can be done in closed form).
We then use these copies of $Z$ to accurately approximate $\pr(Z\in(l,u))$.
We vary $d\in\{10, 20, 50, 100\}$ and $\tau\in\{0.1, 0.05, 0.01\}$.
The other parameters are specified as follows:
$a = 100$, $b = 500$,
$p_i = 0.2i + 3$, $c_i = 2$, $\mu_i = 0.3i + 5$, and $\sigma_i = 1$ for $i=1,\ldots,d$.

\begin{figure}[ht]
    \FIGURE{
    $
    \begin{array}{c}
    \includegraphics[width=0.85\textwidth]{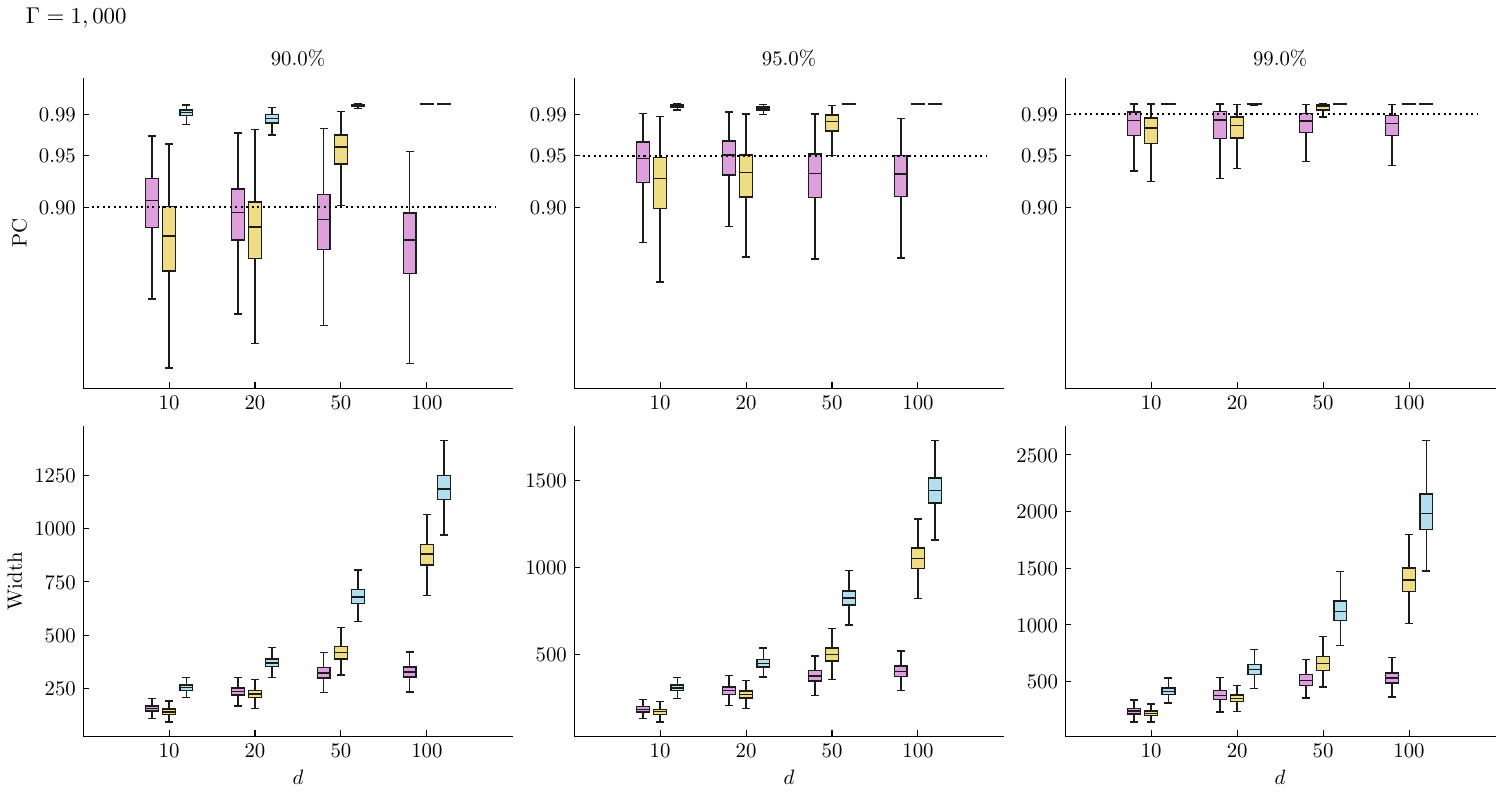} \\
    \includegraphics[width=0.85\textwidth]{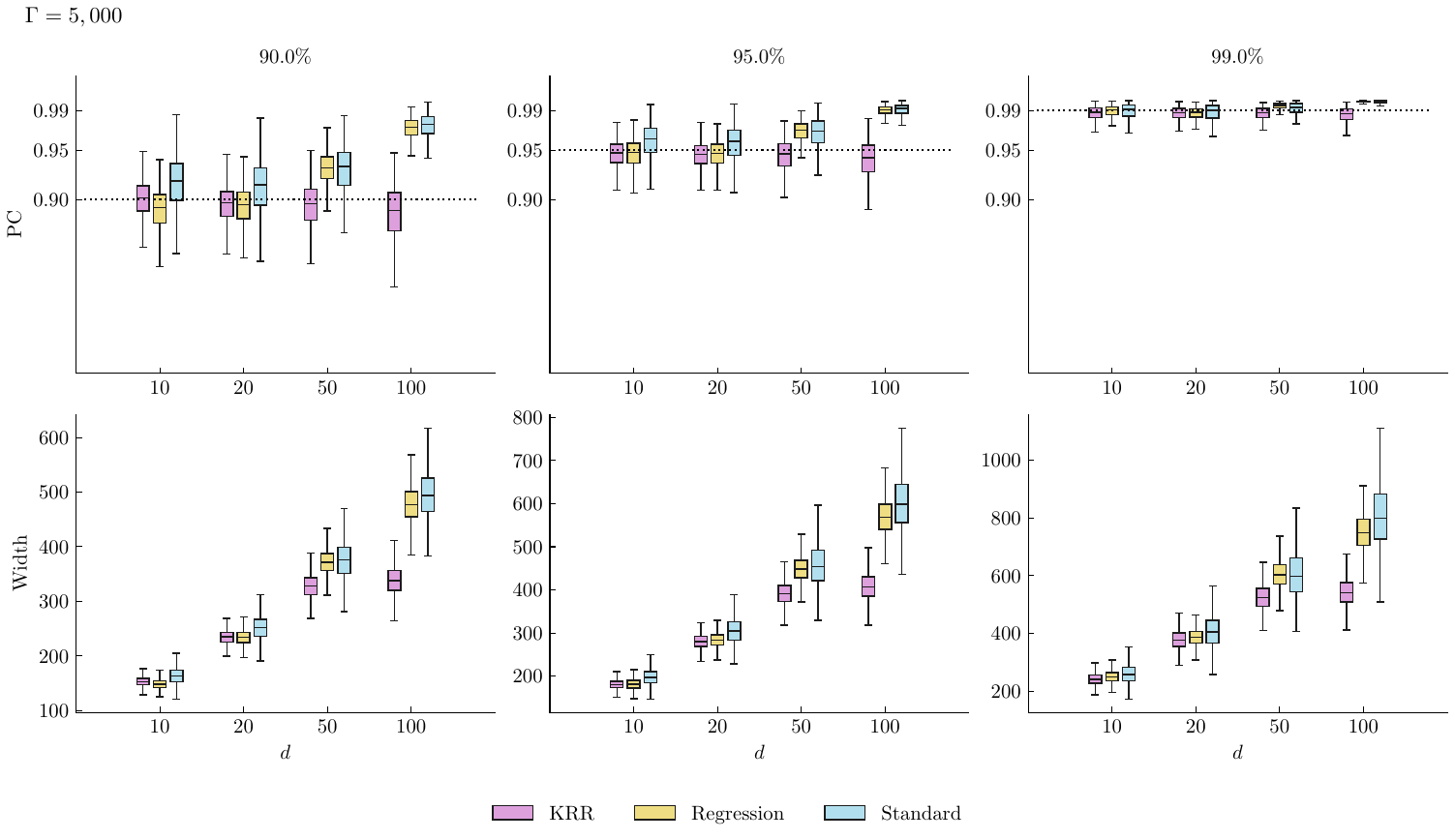}
    \end{array}
    $
    }
    {PC and Width of CrIs for the Input Uncertainty Quantification Problem.    \label{fig:IUQ}
    }
    {The boxplots are made based on 1,000 macro-replications. The percentages are the CrIs' credible levels.}
\end{figure}

Given a simulation budget $\Gamma=1,000$ or $\Gamma=5,000$,
similar to Section~\ref{sec:prm},
for each of the three methods (standard, KRR-driven, and regression-driven) we report the best performance among different values of the inner-level sample size $m$ (and different sets of basis functions in the case of the regression-driven method). Specifically, all three methods yield the best performance with $m = 5$ given the budget of $\Gamma = 1,000$. Under a lager budget, namely $\Gamma = 5,000$, the regression-driven and KRR-driven methods perform optimally with $m = 10$, while the best value of $m$ for the standard method is $50$.
 The numerical results are presented in Figure~\ref{fig:IUQ}.

First, if the budget is small ($\Gamma=1,000$),
the standard method basically fails;
the CrIs it constructed are all excessively wide and have significant over-coverage.\footnote{The notion of ``over-coverage'' here means that the PC of a CrI is larger than the nominal value $100(1-\tau)\%$.
This is different from the over-coverage of a confidence interval from a frequentist perspective.}
The regression-driven method produces accurate CrIs, having a PC close to the corresponding nominal level  for $d=10,20$;
however, it suffers from a severe over-coverage issue in higher dimensions.
Compared to the regression-driven method,
the performance of the KRR-driven method is similar for $d=10$ or $20$ and somewhat better for $d=50$.
For $d=100$,
albeit showing mild under-coverage,
the CrIs constructed by the KRR-driven method are significantly better---in terms of width---than those constructed by the other two methods.

Second, if the budget is large ($\Gamma=5,000$),
the standard and the regression-driven methods have  comparable performances.
Both can construct accurate CrIs for $d=10,20$.
However, their performances begin to deteriorate for $d=50$
and become unacceptable for $d=100$.
In contrast,
the KRR-driven method consistently produce accurate CrIs.
For example, for $d=100$,
the 95\% CrIs constructed by the KRR-driven, regression-driven, and standard methods have an estimated PC of
94.05\%, 98.97\%, and 99.01\%, respectively.
This shows that
the KRR-driven method is a viable option for high-dimensional settings,
which aligns with the experiment results in Section~\ref{sec:prm}.

\begin{table}[ht]
    \TABLE{Running Time (sec.) for the Input Uncertainty Problem. \label{tab:uq-wct}}
    {
    \begin{tabular}{@{\extracolsep{15pt}} l @{\extracolsep{15pt}} r r @{\extracolsep{10pt}} r r @{\extracolsep{10pt}} r r}
    \toprule
    & \multicolumn{6}{c}{Simulation Running Time} \\
        \cmidrule{2-7}
    & \multicolumn{3}{c}{$\Gamma = 1000$} & \multicolumn{3}{c}{$\Gamma = 5000$} \\
        \cmidrule{2-7}
        & KRR & Regression & Standard & KRR & Regression & Standard  \\
        \cmidrule{1-1} \cmidrule{2-4}  \cmidrule{5-7}
    $d = 10$   & \multicolumn{3}{c}{0.07} & \multicolumn{3}{c}{0.35}\\
    $d = 20$   & \multicolumn{3}{c}{0.09} & \multicolumn{3}{c}{0.51}\\
    $d = 50$ & \multicolumn{3}{c}{0.16} & \multicolumn{3}{c}{0.99}\\
    $d = 100$  & \multicolumn{3}{c}{0.26} & \multicolumn{3}{c}{1.77}\\
    \midrule
    & \multicolumn{6}{c}{Estimation Running Time} \\
        \cmidrule{2-7}
    & \multicolumn{3}{c}{$\Gamma = 1000$} & \multicolumn{3}{c}{$\Gamma = 5000$} \\
        \cmidrule{2-7}
        & KRR & Regression & Standard & KRR & Regression & Standard  \\
        \cmidrule{1-1} \cmidrule{2-4}  \cmidrule{5-7}
    $d = 10$ & 0.12 & 0.08 & $\approx 0$ & 0.65 & 0.12 & $\approx 0$  \\
    $d = 20$ & 0.14 & 0.11 & $\approx 0$ & 0.82 & 0.16 & $\approx 0$ \\
    $d = 50$ & 0.15 & 0.29 & $\approx 0$ & 0.88 & 0.35 & $\approx 0$ \\
    $d = 100$ & 0.16 & 0.94 & $\approx 0$ & 0.94 & 1.02 & $\approx 0$ \\
    \bottomrule
    \end{tabular}
    }
    {The setup is the same as that of Table~\ref{tab:prm-wct}, except each reported result is an average over macro-replications of the experiment and different values of $\mathcal{\tau}$.}
\end{table}

Table~\ref{tab:uq-wct} presents the running times of the three methods for the input uncertainty quantification problem.
Notably, the running times for estimation in the regression-driven and KRR-driven methods are at most one second or so (which occurs when $d=100$ and $\Gamma=5000$). This is substantially shorter than the running times reported in Table~\ref{tab:prm-wct} for the portfolio risk management problem. The reason for this difference is that the simulation budgets considered here are much smaller, so it takes these two methods much less time to process the simulation samples.

We connect the quality of the CrIs depicted in Figure~\ref{fig:IUQ} with the running times presented in Table~\ref{tab:uq-wct}. For $d=10$ or $20$ and a budget of $\Gamma=1000$, both the KRR-driven and regression-driven methods produce accurate CrIs, and their running times are similar. As a result, neither method clearly outperforms the other in these cases. Nevertheless, the regression-driven method offers a benefit in terms of implementation simplicity, as long as a suitable set of basis functions is selected.
On the other hand, in higher-dimensional settings ($d=50$ or $100$) with a budget of $\Gamma=5000$, the KRR-driven method exhibits a notably higher degree of accuracy in generating CrIs when compared to the regression-driven method. Although this method demands a longer running time, it stays within an reasonable range. This finding emphasizes the competitiveness of the KRR-driven method for high-dimensional applications.

\subsection{Summarizing Remarks: When is the KRR-driven Method Suitable?} \label{sec:KRR-suitable}

Sections~\ref{sec:prm} and \ref{sec:uq} introduce two problems with distinct simulation costs and budgets, which are crucial factors in deciding the suitability of using the KRR-driven method for nested simulation. The computational overhead (i.e., the estimation step) of this method, beyond the simulation itself, pertains to processing the simulation samples to construct an estimate of $\theta$. This process involves running KRR multiple times for hyperparameter tuning with a running time that does not depend on the simulation cost but is mainly determined by the simulation budget $\Gamma$   due to its effect on the outer-level sample size $n$. Our numerical experiments show that selecting $m = 5$ or $10$ (so $n = \Gamma/5$ or $n=\Gamma/10$) typically results in satisfactory performance when using the KRR-driven method.\footnote{The observation that the choice of $m$ remains relatively constant, regardless of the dimensionality and the simulation budget, implies that $\E[Y|X]$ exhibits a high degree of smoothness with respect to $X$.
Alternatively, it could be well approximated by a function that is highly smooth.
As a reference, our theoretical results outlined in Table~\ref{tab:correct-smooth} propose that $m$ should remain constant as $\Gamma$ increases, provided the smoothness parameter $\nu$ is sufficiently large.
}

When $\Gamma$ ranges from $10^3$ to $10^4$, as in the input uncertainty quantification problem, the KRR-driven method's computational overhead is relatively minor, taking merely seconds in a standard computing environment. Given the substantial improvement in estimation accuracy this method provides, it is generally suitable for nested simulation within this simulation budget.

When $\Gamma$ ranges from $10^4$ to $10^5$,  the computational overhead becomes considerable, taking up to hundreds of seconds. In cases where the simulation cost is significant, such as the portfolio risk management problem and many other practical applications, the use of the KRR-driven method is justifiable despite the computational overhead due to its enhanced rate of convergence, particularly in high-dimensional settings.
However, if the simulation cost is sufficiently low, the hundreds of seconds of computational overhead could be devoted to generating a significantly larger number of simulation samples. This additional sampling could potentially compensate for the slow convergence of the standard method for nested simulation. In such situations, the KRR-driven method may not be the optimal choice.

Lastly, when $\Gamma$ exceeds $10^5$, the computational overhead may become prohibitive due to the demanding task of inverting large matrices, which has a time complexity cubic in the matrix size. As a result, the KRR-driven method in its current form is unsuitable for this large-budget setting.
Nevertheless, it is possible to explore kernel approximation methods that have been extensively studied in the machine learning literature \citep[Chapter~8]{Rasmussen2006Gaussian} as a means to significantly reduce KRR's computational cost.
Incorporating these enhancements would expand the scope of applications for the KRR-driven method even further.

\section{Conclusions}\label{sec:conclusion}
In this paper, we develop a new method based on KRR for nested simulation.
We show that for various forms of nested simulation, the new method
may recover (or at least approach) the square root convergence rate---that is, the canonical rate for the standard Monte Carlo simulation---provided that the conditional expectation as a function of the conditioning variable is sufficiently smooth  with respect to its dimensionality.
In other words, the new method can substantially alleviate the curse of dimensionality by exploiting the smoothness.
Our theoretical framework for convergence rate analysis is general.
Not only does it incorporate different forms of nested simulation,
but it may also be applied to examine the use of kernels other than the Mat\'ern class in KRR or even the use of machine learning methods other than KRR in nested simulation.

Our work can be extended in several ways.
First, the numerical experiments in Section~\ref{sec:test-function} suggest that
there may exist room for improvement in our upper bounds on the convergence rates.
It is interesting, albeit challenging, to identify lower bounds on the convergence rates.
They would address the question as to whether our upper bounds are optimal in a minimax sense.

Second, our theory prescribes rules for budget allocation $(n,m)$ in an asymptotic sense.
Although our numerical experiments suggest that these asymptotic rules indeed serve as a good guideline,
a more refined rule may further facilitate the use of the KRR-driven method.
Given the asymptotic orders of magnitude of $n$ and $m$,
one might adopt the idea of \cite{ZhangLiuWang22} to use a small proportion of the total simulation budget to compute a bootstrap-based estimate of the leading constants in the asymptotic orders.
This would conceivably yield a reasonable choice of $(n,m)$.

Lastly, the computation of KRR involves matrix inversion,
and it may become computationally challenging as the matrix size---which equals the outer-level sample size---grows.
Numerous approximation methods \citep{LuRudiBorgonovoRosasco20,LiuHuangChenSuykens22}
have been developed to address this issue.
These methods are designed to strike a balance between the approximation's computational efficiency and KRR's prediction accuracy.
To use them in the KRR-driven nested simulation, however,
this kind of balance needs to be adjusted because KRR's prediction accuracy is measured differently in nested simulation relative to  typical machine learning tasks.
The adjustment would make the KRR-driven method applicable for large-scale problems.

\begin{APPENDIX}{Error Measure: Absolute Error versus RMSE}
\section{Convergence Rate Results} \label{app:results}

The probabilistic bounds on the absolute error in Theorems \ref{thm:gsmooth}--\ref{thm:gvarcvar} are given in the form $|\hat \theta_{n,m}-\theta| = O_{\pr}(\Gamma^{- \kappa}(\log \Gamma)^{\tilde\kappa})$, where $\kappa$ and $\tilde\kappa$ are constants.
A closer inspection of the proofs for these theorems reveals that, through a refined analysis, the probabilistic bounds can be strengthened to the following statement:
\begin{equation}\label{eq:stronger-statement}
\pr\left(|\hat \theta_{n,m}-\theta| \leq C_3 t \Gamma^{- \kappa}(\log \Gamma)^{\tilde\kappa}\right) \geq 1 - C_1\exp(-C_2t^{c}),
\end{equation}
for all $t>1/C_3$, where $c$, $C_1$, $C_2$, and $C_3$ are some constants.
The refined analysis is facilitated by the application of empirical process theory and Assumption~\ref{assump:subG} (i.e., the sub-Gaussian assumption on the noise in the inner-level simulation).

By applying a change of variables, the statement~\eqref{eq:stronger-statement} can be equivalently expressed as
\[\pr\left(|\hat \theta_{n,m}-\theta|\leq u\right) \geq 1-C_4\exp(-C_5u^{c}\Gamma^{\kappa c}(\log \Gamma)^{-\tilde\kappa c}),
\]
for all $u>\Gamma^{-\kappa }(\log \Gamma)^{\tilde\kappa}$, where $C_4$ and $C_5$ are some constants.
Then, we can derive the following bound on the RMSE:
\begin{align*}
    \left(\E\bigl[(\hat\theta_{n,m}-\theta)^2\bigr]\right)^{1/2} ={} & \left(\int_{0}^\infty \mathbb{P}\left(|\hat \theta_{n,m}-\theta|\geq t^{1/2}\right){\rm d}t\right)^{1/2}    \\
    ={} & \left(\int_0^{\Gamma^{-2\kappa }(\log \Gamma)^{2\tilde\kappa }} \mathbb{P}\left(|\hat \theta_{n,m}-\theta|\geq t^{1/2}\right){\rm d}t + \int_{\Gamma^{-2\kappa }(\log \Gamma)^{2\tilde\kappa }}^\infty \mathbb{P}\left(|\hat \theta_{n,m}-\theta|\geq t^{1/2}\right){\rm d}t\right)^{1/2}\\
    \leq{} & \left(\Gamma^{-2\kappa }(\log \Gamma)^{2\tilde\kappa } + \int_{\Gamma^{-2\kappa }(\log \Gamma)^{2\tilde\kappa }}^\infty\mathbb{P}\left(|\hat \theta_{n,m}-\theta|\geq t^{1/2}\right){\rm d}t\right)^{1/2}\\
    \leq{} & \left(\Gamma^{-2\kappa }(\log \Gamma)^{2\tilde\kappa } + \int_{\Gamma^{-2\kappa }(\log \Gamma)^{2\tilde\kappa }}^\infty C_4\exp(-C_5t^{c/2}\Gamma^{\kappa c}(\log \Gamma)^{-\tilde\kappa c}){\rm d}t\right)^{1/2}\\
    \leq{} & C_6\Gamma^{-\kappa }(\log \Gamma)^{\tilde\kappa },
\end{align*}
for some constant $C_6$.
Consequently, within the scope of the present paper, the probabilistic bounds $|\hat \theta_{n,m}-\theta| = O_{\pr}(\Gamma^{- \kappa}(\log \Gamma)^{\tilde\kappa})$ can be converted to the RMSE bounds $ \left(\E\bigl[(\hat\theta_{n,m}-\theta)^2\bigr]\right)^{1/2} = O(\Gamma^{- \kappa}(\log \Gamma)^{\tilde\kappa})$.

However, the refined analysis leading to the statement \eqref{eq:stronger-statement} is significantly more complex from a technical perspective. This complexity arises from the need to strengthen numerous intermediate results, which are currently written as $O_p$-bounds, into statements similar to \eqref{eq:stronger-statement}.
Such strengthening involves expressing tail probabilities in the form of $C_1\exp(-C_2t^{c})$, which in turn calls for intricate calculations substantially lengthier than those in the present paper.
An additional technical challenge lies in the fact that the relevant tail probabilities, particularly the value of the exponent $c$, may differ for these intermediate results. As a result,  it is essential to track these exponents before integrating them into the final result, as shown in \eqref{eq:stronger-statement}.
Furthermore, this tracking must be performed separately for different types of functionals $\mathcal{T}$ analyzed in Theorems \ref{thm:gsmooth}--\ref{thm:gvarcvar}.
Therefore, we leave the RMSE analysis for future research.

\section{Proof Techniques}

The study most closely related to the present paper is that of \cite{HongJunejaLiu17}. Like our approach, they also applied a machine learning method to estimate the unknown function $f$. However, they employed kernel smoothing (KS) as their chosen technique. Let us denote their KS estimator of $f$ as $\hat{f}_n^{\mathsf{KS}}$.
Moreover, let $Z_n(x) \coloneqq \sqrt{n h^d}(\hat{f}_n^{\mathsf{KS}}(x)- f(x))$ denote the normalized estimator, where $h$ is the bandwidth parameter involved in KS.
It is worth noting that their analysis primarily focused on the case where $m=1$, resulting in
$n=\Gamma$.

The main assumptions in their study include
\begin{enumerate}[label=(\roman*)]
    \item a 4th-moment condition: $\sup_n\E[|Z_n(X) |^4]<\infty$,
    \item a central limit theorem (CLT): $Z_n(x)$ converges to a normal distribution for each $x\in\Omega$, and
    \item regularity conditions  with complex forms concerning the joint density of $X$ and certain random variables associated with the asymptotic properties of $\hat{f}_n^{\mathsf{KS}}$.
\end{enumerate}
Under these three assumptions, along with several additional regularity conditions, they derived the convergence rate of their KS-driven method in terms of the RMSE. Notably, their analysis focused on the three forms of $\mathcal{T}$ that we examine in Section~\ref{sec:analy-nested-expec} and did not include cases where $\mathcal{T}$ represents  VaR or CVaR.

Upon comparing the theoretical analysis in our work with that of \cite{HongJunejaLiu17}, it becomes evident that our assumptions are more primitive, presented in a simpler form, and as a result, easier to interpret. For instance, we impose conditions on the noise $\epsilon_{ij}$ (Assumption \ref{assump:subG}), the distribution of $X$ (Assumption \ref{assump:support}), and the underlying function $f$ (Assumption \ref{assump:RKHS}).
In contrast, they imposed conditions on the normalized KS estimator and regularity conditions with complex forms, as previously mentioned.
We acknowledge that their 4th-moment condition is indeed weaker than our Assumption \ref{assump:subG}. Specifically, if $\epsilon_{ij}$ is sub-Gaussian, then the normalized KS estimator must satisfy the 4th-moment condition.
However, their CLT assumption is considerably stronger as explained below.

The technical proofs of \cite{HongJunejaLiu17} are less sophisticated than ours. In particular, they did not employ advanced technical tools from empirical process theory, as we have done in our work. The crux of the difference lies in the CLT assumption.
It is well known in the nonparametric statistics literature that the KS estimator satisfies a CLT under various sets of regularity conditions.
However, to the best of our knowledge, a CLT for the KRR estimator remains a challenging open problem and has not yet been established.
This is precisely why our proofs are more technically demanding than theirs, as we cannot conveniently rely on CLTs that would characterize the exact asymptotic orders of the bias and variance of the KRR estimator for $f$. Instead, we must instead build our arguments upon more primitive conditions. For the same reason, the convergence rate results of \cite{HongJunejaLiu17} are presented in a stronger form than ours. They derived convergence rates in terms of the RMSE, whereas we have only established probabilistic bounds on the convergence rates (in terms of the absolute error, although these bounds could potentially be converted into bounds on the RMSE, as discussed in Appendix~\ref{app:results}).

\end{APPENDIX}

\bibliographystyle{informs2014} %
\bibliography{ref} %

\ECSwitch

\EquationsNumberedBySection

\ECHead{Supplemental Material}

\section{$\mathcal{T}$-dependent LOOCV}\label{app:LOO}
\subsection{Discrepancy Metric}\label{sec:metric-expression}
We present the expressions of
$\hat\theta^{\mathsf{St}}(\mathcal{T}, I^{\mathsf{Va}})$ and $\hat\theta(\Xi,  \mathcal{T}, I^{\mathsf{Tr}}, I^{\mathsf{Va}})$ in the metric \eqref{eq:new-metric} as follows.
Let $n^{\mathsf{Va}} = |I^{\mathsf{Va}}|$ denote the size of the set $I^{\mathsf{Va}}$. Then,
\begin{equation}\label{eq:st_theta}
\hat\theta^{\mathsf{St}}(\mathcal{T}, I^{\mathsf{Va}}) \coloneqq
\left\{
\begin{array}{ll}
\displaystyle \frac{1}{n^{\mathsf{Va}}}\sum_{i\in I^{\mathsf{Va}}} \eta(\bar{y}_i ),     &\quad\mbox{if } \mathcal{T}(\cdot)= \E[\eta(\cdot)],  \\
\displaystyle  \bar{y}_{(\lceil \tau n^{\mathsf{Va}}\rceil)},     &\quad  \mbox{if }\mathcal{T}(\cdot) = \VaR_\tau(\cdot),  \\
\displaystyle  \bar{y}_{(\lceil \tau n^{\mathsf{Va}}\rceil)} + \frac{1}{(1-\tau) n^{\mathsf{Va}}}\sum_{i\in I^{\mathsf{Va}}} (\bar{y}_i -\bar{y}_{(\lceil \tau n^{\mathsf{Va}}\rceil)})^+,     &\quad  \mbox{if }\mathcal{T}(\cdot) = \CVaR_\tau(\cdot),
\end{array}
\right.
\end{equation}
where $\bar{y}_{(1)} \leq \cdots \leq \bar{y}_{(n^{\mathsf{Va}})}$  denote the order statistics of $\{\bar{y}_i: i\in I^{\mathsf{Va}}\}$.

In addition, let $\hat{f}^{\Xi,I^{\mathsf{Tr}}}$ denote the KRR estimator trained using the hyperparameters $\Xi$ and the data associated with $I^{\mathsf{Tr}}$. Then,
\begin{equation}\label{eq:KRR_theta}
\hat\theta(\Xi,  \mathcal{T}, I^{\mathsf{Tr}}, I^{\mathsf{Va}}) \coloneqq
\left\{
\begin{array}{ll}
\displaystyle \frac{1}{n^{\mathsf{Va}}}\sum_{i\in I^{\mathsf{Va}}} \eta\Bigl(\hat{f}^{\Xi,I^{\mathsf{Tr}}}(\BFx_i)\Bigr), &\quad \mbox{if } \mathcal{T}(\cdot)= \E[\eta(\cdot)],  \\[1ex]
\displaystyle  \hat{f}^{\Xi,I^{\mathsf{Tr}}}_{(\lceil \tau n^{\mathsf{Va}}\rceil)},     &\quad  \mbox{if }\mathcal{T}(\cdot) = \VaR_\tau(\cdot),  \\[.5ex]
\displaystyle  \hat{f}^{\Xi,I^{\mathsf{Tr}}}_{(\lceil \tau n^{\mathsf{Va}}\rceil)} + \frac{1}{(1-\tau) n^{\mathsf{Va}}}\sum_{i\in I^{\mathsf{Va}}} \Bigl(\hat{f}^{\Xi,I^{\mathsf{Tr}}}(\BFx_i)-\hat{f}^{\Xi,I^{\mathsf{Tr}}}_{(\lceil \tau n^{\mathsf{Va}}\rceil)}\Bigr)^+,     &\quad  \mbox{if }\mathcal{T}(\cdot) = \CVaR_\tau(\cdot),
\end{array}
\right.
\end{equation}
where $\hat{f}^{\Xi,I^{\mathsf{Tr}}}_{(1)} \leq \cdots \leq \hat{f}^{\Xi,I^{\mathsf{Tr}}}_{(n^{\mathsf{Va}})} $ denote the order statistics of $\{\hat{f}^{\Xi,I^{\mathsf{Tr}}}(\BFx_i):i\in I^{\mathsf{Va}}\}$.

\subsection{LOOCV}\label{sec:LOO}

In the LOOCV setting, the dataset $\mathcal{D}=\{(\BFx_i,\bar{y}_i:i=1,\ldots,n\}\}$ is divided into $n$ parts, each having exactly one point.
In the $l$-th iteration, KRR is trained using the data  $\mathcal{D}\setminus \{\BFx_l\}$.
The discrepancy metric of the $\mathcal{T}$-dependent LOOCV is
\begin{align*}
\mathrm{CV}^{\mathsf{LOO}}(\Xi, \mathcal{T}) \coloneqq{}& \frac{1}{n} \sum_{l=1}^n \bigl( \hat\theta(\Xi, \mathcal{T},   \mathcal{D}\setminus \{\BFx_l\}, \{\BFx_l\}) - \hat\theta^{\mathsf{St}}( \mathcal{T}, \{\BFx_l\}) \bigr)^2,
\end{align*}
where, according to \eqref{eq:st_theta} and \eqref{eq:KRR_theta},
\[
\hat\theta^{\mathsf{St}}( \mathcal{T},\{\BFx_l\}) =
\left\{
\begin{array}{ll}
     \eta(\bar{y}_l), &\quad \mbox{if } \mathcal{T}(\cdot)= \E[\eta(\cdot)],  \\[1ex]
     \bar{y}_l, &\quad  \mbox{if }\mathcal{T}(\cdot) = \VaR_\tau(\cdot)\mbox{ or }\CVaR_\tau(\cdot),
\end{array}
\right.
\]
and
\[
\hat\theta(\Xi,  \mathcal{T}, \mathcal{D}\setminus \{\BFx_l\}, \{\BFx_l\}) =
\left\{
\begin{array}{ll}
     \eta(\hat{f}^{-l}(\BFx_l;\Xi)), &\quad \mbox{if } \mathcal{T}(\cdot)= \E[\eta(\cdot)],  \\[1ex]
     \hat{f}^{-l}(\BFx_l;\Xi), &\quad  \mbox{if }\mathcal{T}(\cdot) = \VaR_\tau(\cdot)\mbox{ or }\CVaR_\tau(\cdot).
\end{array}
\right.
\]
Here, $\hat{f}^{-l}(\cdot;\Xi)$ denotes the KRR estimator trained with $\mathcal{D}\setminus \{\BFx_l\}$ and with the hyperparameters being $\Xi$.

For each $l$, let $\bar{\BFy}^{-l} \in \RR^d$ denote the vector of which the $l$-th entry is $\hat{f}^{-l}(\BFx_l; \Xi)$ and the $i$-th entry is $\bar{y}_i$ for all $i\neq l$.
It can be shown (see, e.g., \citealt{RifkinLippert07}) that
\begin{equation}\label{eq:LOOV}
    \hat{f}^{-l}(\BFx_l; \Xi) = \frac{\left( \BFR \left( \BFR + n \lambda \BFI \right)^{-1} \bar{\BFy} \right)_l - \left( \BFR \left( \BFR + n \lambda \BFI \right)^{-1} \right)_{ll} \bar{y}_l}{1 - \left( \BFR \left( \BFR + n \lambda \BFI \right)^{-1} \right)_{ll}}.
\end{equation}

The key point here is that according to the expression \eqref{eq:LOOV}, to compute $\mathrm{CV}^{\mathsf{LOO}}(\Xi, \mathcal{T})$ for given $\Xi$ and $\mathcal{T}$,
we need to compute the inverse matrix $\BFR \left( \BFR + n \lambda \BFI \right)^{-1} $ only \emph{once}, the computational time complexity of which is $O(n^3)$.
In contrast,
if we use $K$-fold cross-validation (with $K$ usually set to be 5--20),
then
the computational trick \eqref{eq:LOOV} is not applicable, and therefore
we would need to train KRR separately on each training set of size $(1-1/K)n$ from scratch.
The computational cost involved in matrix inversion in the training process would  amount to $O(K n^3)$, which is substantially higher than LOOCV.

\section{Additional Numerical Results} \label{ec-sec:numerical}

 \subsection{Impact of Hyperparameter Tuning}\label{ec-sec:hyper-tuning}

In this paper, we assume that the smoothness parameter $\nu$ is known and, thus, correctly specified for theoretical analysis. In practice, however, $\nu$ is typically unknown. We propose in Section~\ref{sec:CV} a cross-validation approach for estimating $\nu$, along with other hyperparameters such as the length scale  parameter $\ell$ in the Mat\'ern kernel and the regularization parameter $\lambda$ when implementing KRR.
We now proceed to numerically investigate the impact of hyperparameter tuning.

\begin{figure}[ht]
    \FIGURE{
    $
    \begin{array}{c}
    \includegraphics[width=\textwidth]{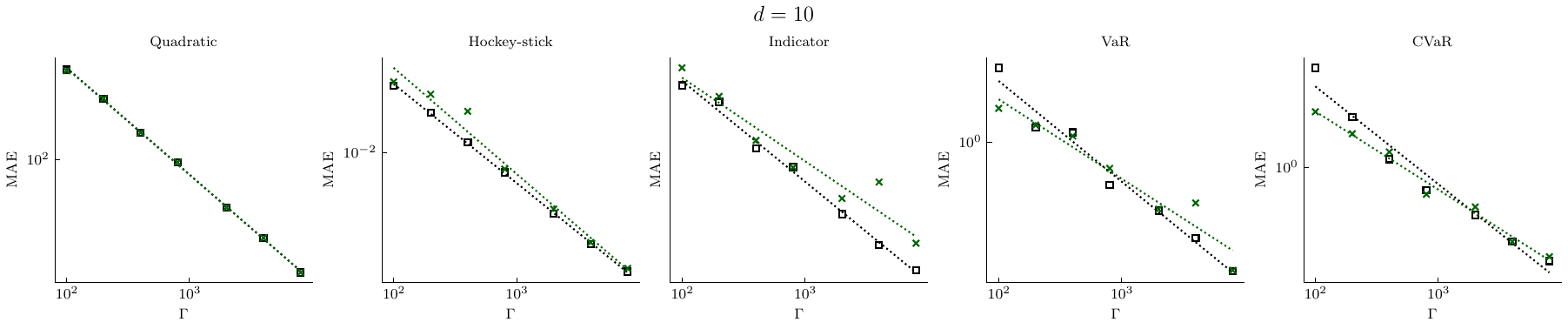} \\
        \includegraphics[width=\textwidth]{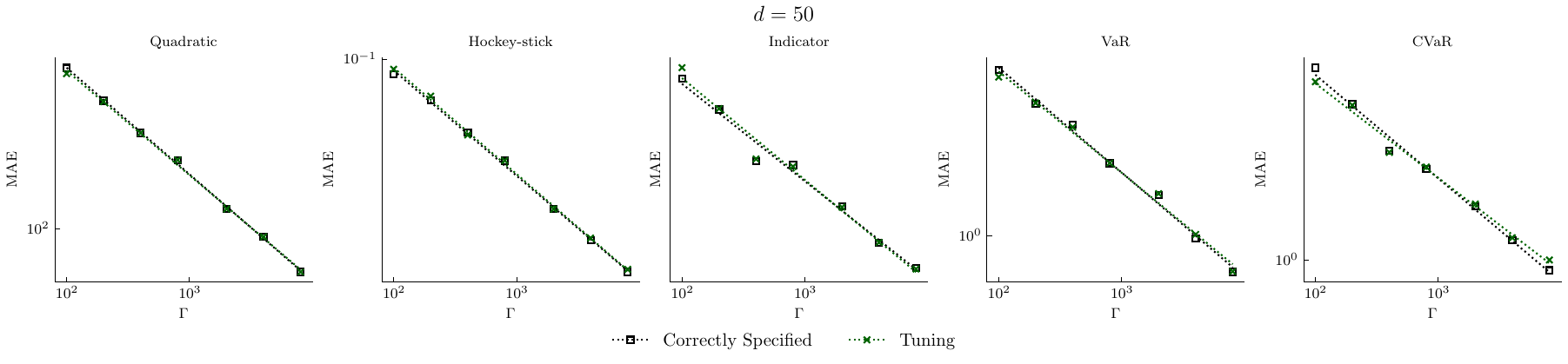}
    \end{array}
    $
    }
    {Impact of Hyperparameter Tuning. \label{fig:sec6.1-tuning-finalize}
    }
    {$\nu= 5.5$ and $\ell=10$. }
\end{figure}

We follow the same experimental setup as in Section~\ref{sec:test-function}. Specifically, the test functions are randomly generated based on the Matérn kernel with $\nu=5.5$ and $\ell=10$. Figure~\ref{fig:sec6.1-tuning-finalize} displays the MAEs of the KRR-driven method when hyperparameters $(\nu,\ell,\lambda)$ are correctly specified, in contrast to when they are treated as unknown and estimated using $\mathcal{T}$-dependent LOOCV.
By ``correctly specified'', we mean that $\nu$ and $\ell$ are set as the true values, and $\lambda$ is specified according to the theoretical analysis suggestions in Table~\ref{tab:correct-smooth}. Note that $\lambda$ is not involved in defining problem instances (i.e., the underlying function $f$ and the distribution of the conditioning variable $X$). Instead, $\lambda$ is a tuning parameter used in KRR by definition,  so the notion of ``true value''  is not applicable to it.

Figure~\ref{fig:sec6.1-tuning-finalize} illustrates that, as anticipated, the necessity to estimate unknown hyperparameters leads to a deterioration in the convergence rate of the KRR-driven method. However, the decline in performance does not seem to be significant in most cases, except when $d=10$ and $\mathcal{T}(\cdot) = \E[\eta(\cdot)]$ with $\eta$ being an indicator function, VaR or CVaR. This indicates that our cross-validation approach effectively addresses the issue of unknown hyperparameters.

Furthermore, in a higher dimension setting ($d=50$), employing estimated hyperparameters results in nearly the same performance as using correctly specified hyperparameters. This indicates that the KRR-driven method, with the assistance of $\mathcal{T}$-dependent LOOCV, can be a viable option for nested simulation in high dimensions, providing reassurance for practitioners.
Intuitively, this finding may be attributed to the fact that the smoothness and dimensionality jointly affect the KRR's convergence rate in the form of a ratio between them, with the former appearing in the numerator and the latter in the denominator; see, e.g., \cite{TuoWangWu20_ec}. Consequently, given the same smoothness, a higher dimensionality implies a weaker impact of smoothness misspecification on the convergence rate.

\subsection{Smoothness Misspecification} \label{ec-sec:misspec}

Despite hyperparameter tuning via cross-validation,
the smoothness parameter may still be misspecified in the KRR-driven method.
We conduct a numerical study on the issue of smoothness misspecification, which serves as a complement to the analysis of the impact of hyperparameter tuning in Section~\ref{ec-sec:hyper-tuning}.

We follow the same experimental setup as in Section~\ref{sec:test-function}. Let $\nu_0$ denote the smoothness of the unknown function $f$, and let $\nu$ represent the smoothness used when implementing the KRR-driven method. (Note that these two values are identical, and we do not differentiate them in Section~\ref{sec:test-function}.) We consider two distinct cases of misspecification: undersmoothing ($\nu < \nu_0$) and oversmoothing ($\nu > \nu_0$). The former case implies that the kernel used in KRR is less smooth than the function to be estimated, while the latter suggests the opposite.
Furthermore, to emphasize the implications of smoothness misspecification, we set the difference between $\nu$ and $\nu_0$ as large as possible, with one being $0.5$ while the other being $\infty$.\footnote{The Mat\'ern kernel with an infinite smoothness is identical to the Gaussian kernel; see \citet[page~85]{RasmussenWilliams06_ec}.}
In practice, however, $\nu$ is obtained as an estimate of $\nu_0$ using cross-validation. It is unlikely that the difference between the two values would be as dramatic as we set in this experiment. Consequently, the impact of smoothness misspecification in the presence of cross-validation may be less significant than what is demonstrated in the following results.

\begin{figure}[ht]
    \FIGURE{
    $
    \begin{array}{c}
    \includegraphics[width=\textwidth]{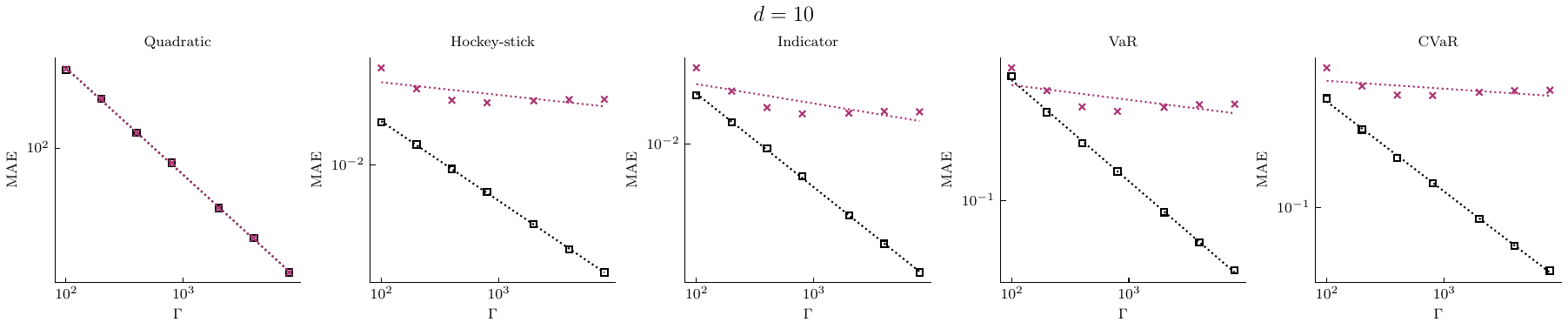} \\
    \includegraphics[width=\textwidth]{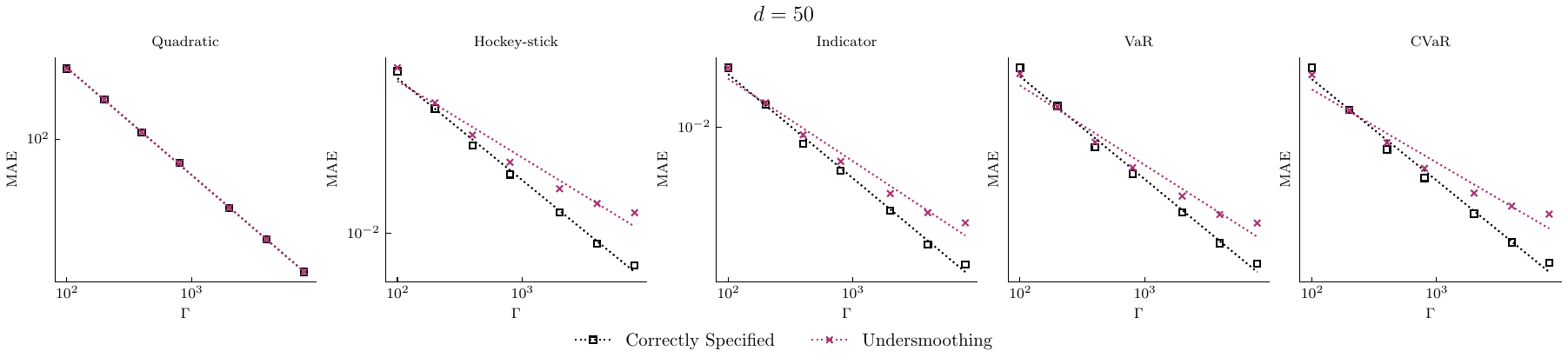}
    \end{array}
    $
    }
    {Impact of Using  Undersmoothed Kernels.    \label{fig:sec6.1-misspec-under}
    }
    {The true smoothness is $\nu_0 = \infty$, but is misspecified as $\nu = 0.5$.}
\end{figure}

\begin{figure}[ht]
    \FIGURE{
    $
    \begin{array}{c}
    \includegraphics[width=\textwidth]{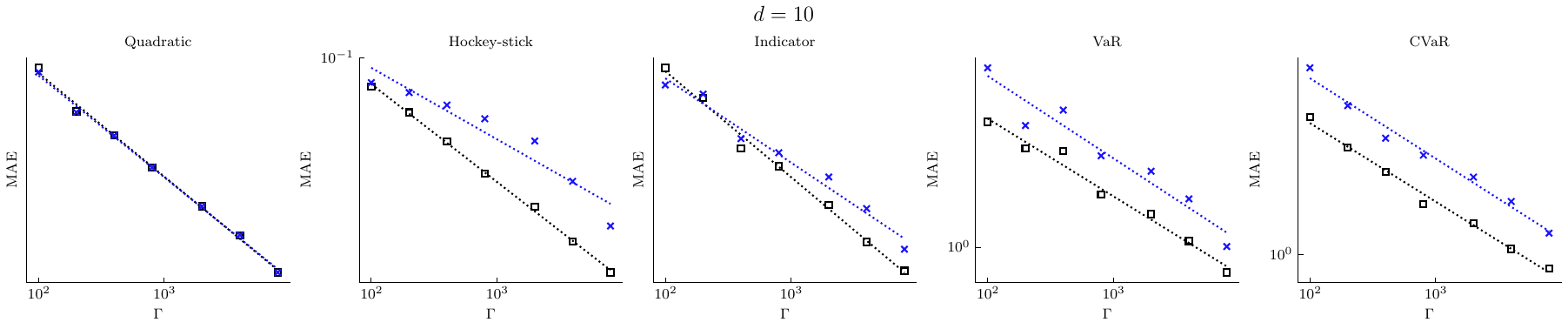} \\
        \includegraphics[width=\textwidth]{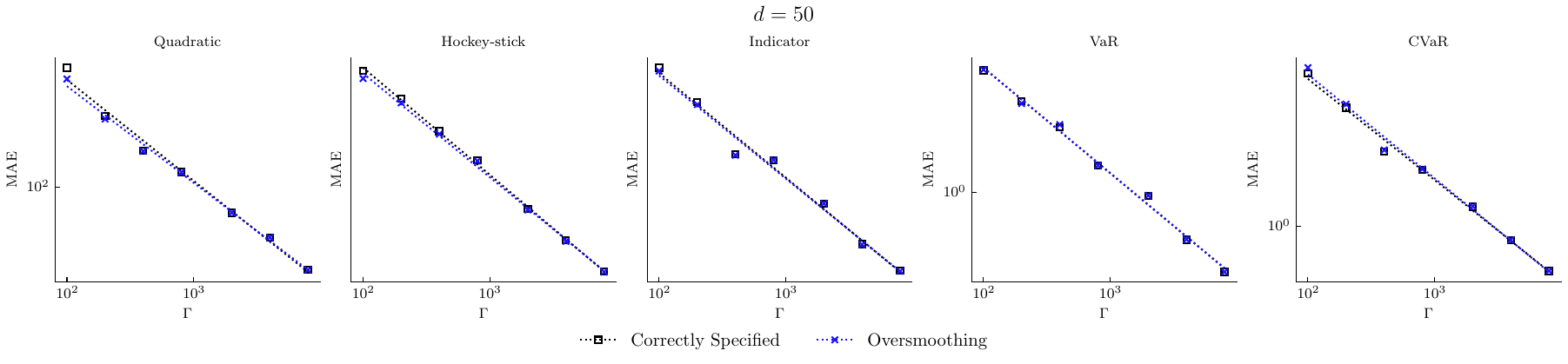}
    \end{array}
    $
    }
    {Impact of Using  Oversmoothed Kernels.     \label{fig:sec6.1-misspec-over}
    }
    {The true smoothness is $\nu_0 = 0.5$, but is misspecified as $\nu = \infty$. }
\end{figure}

Figures~\ref{fig:sec6.1-misspec-under} and \ref{fig:sec6.1-misspec-over} show the MAEs of the KRR-driven method for the undersmoothing and oversmoothing cases, respectively.
In each figure, we compare the performances using a misspecified kernel with those using a correctly specified kernel.
There are several findings.
First, as expected, smoothness misspecification (either undersmoothing or oversmoothing) has a notable impact on the performance, reducing the convergence rate of the KRR-driven method in most situations except when  $\mathcal{T}(\cdot) = \E[\eta(\cdot)]$ with $\eta$ being a quadratic function.

Second, the impact of misspecification on the convergence rate diminishes in higher dimensions.
This finding is consistent with Figure~\ref{fig:sec6.1-tuning-finalize}, which illustrates that the difference in performance between the KRR-driven method with correctly specified hyperparameters and the one with estimated hyperparameters diminishes as the dimensionality increases.
The intuitive reason for both findings is the same: both smoothness and dimensionality have a combined influence on the KRR's convergence rate, represented as a ratio of the former to the latter.
As a result, for the same level of smoothness, increased dimensionality leads to a reduced impact of smoothness on the convergence rate.

Lastly, by comparing the two figures, we observe that the impact of oversmoothing tends to be less significant than that of undersmoothing. This is somewhat consistent with recent studies in the machine learning literature.
Loosely speaking, in the oversmoothing case ($\nu > \nu_0$), one can establish an upper bound on the convergence rate of KRR using smoothness $\nu$ that matches the minimax lower bound on the convergence rate for estimating an unknown function with smoothness $\nu_0$. However, this result does not hold for the undersmoothing case. For further details, we refer to \cite{TuoWang20_ec} and the references therein.

\subsection{Cross-validation: $\mathcal{T}$-dependent versus $\mathcal{T}$-independent}\label{ec-sec:dependence-vs-indep}

We use the portfolio risk management problem in Section~\ref{sec:prm} to demonstrate the differences between the $\mathcal{T}$-dependent and $\mathcal{T}$-independent LOOCV approaches.\footnote{We have not conducted similar experiments for the input uncertainty quantification problem in Section~\ref{sec:uq}, as the two types of LOOCV are equivalent when $\mathcal{T}$ assumes the form of VaR or CVaR. These are precisely the functionals utilized in Section~\ref{sec:uq}, making a comparison between the two LOOCV types unnecessary in this context.}
Our comparison focuses on the estimation accuracy of the KRR-driven method, induced by each approach, as well as the computational overhead associated with them.
In Section~\ref{sec:prm}, we investigate five different forms of $\mathcal{T}$: $\mathcal{T}(\cdot)=\E[\eta(\cdot)]$ with $\eta$ being a quadratic, hockey-stick, or indicator function, and $\mathcal{T}(\cdot) = \VaR_\tau(\cdot)$ or $\mathcal{T}(\cdot) = \CVaR_\tau(\cdot)$ with $\tau=0.99$. According to the analysis in Section~\ref{sec:LOO}, for the latter two of these functionals, the two types of LOOCV are equivalent, while for the first three, such equivalence does not hold.
Accordingly, when assessing the estimation accuracy of the KRR-driven method induced by the two types of LOOCV, our comparison is primarily centered on the first three functionals.

\begin{table}[ht]
            \TABLE{RRMSE ($\%$) of the KRR-driven Method with Different Types of LOOCV for the Portfolio Risk Management Problem. \label{tab:prm-indcv-RRMSE}}
            {\begin{tabular}{@{\extracolsep{10pt}} c @{\extracolsep{10pt}} l @{\extracolsep{10pt}} r r @{\extracolsep{10pt}} r r}
            \toprule
            & \multirow{2}{*}{$\mathcal{T}$} & \multicolumn{2}{c}{$\Gamma = 10^4$} & \multicolumn{2}{c}{$\Gamma = 10^5$} \\
            \cmidrule{3-6}
            & & $\mathcal{T}$-dependent & $\mathcal{T}$-independent & $\mathcal{T}$-dependent & $\mathcal{T}$-independent \\
            \cmidrule{1-2} \cmidrule{3-4} \cmidrule{5-6}
            \multirow{3}{*}{$q = 10$} & Quadratic & 3.67 & 6.17 & 1.21 & 5.78 \\
            & Hockey-stick & 3.23 & 6.74 & 1.44 & 6.51 \\
            & Indicator & 2.44 & 4.48 & 0.77 & 3.99 \\
            \cmidrule{1-2} \cmidrule{3-4} \cmidrule{5-6}
            \multirow{3}{*}{$q = 20$} & Quadratic & 3.46 & 3.61 & 1.13 & 1.53 \\
            & Hockey-stick & 3.92 & 7.86 & 1.23 & 7.03 \\
            & Indicator & 2.93 & 6.42 & 0.93 & 5.92 \\
            \cmidrule{1-2} \cmidrule{3-4} \cmidrule{5-6}
            \multirow{3}{*}{$q = 50$} & Quadratic & 3.30 & 5.42 & 1.08 & 3.66 \\
            & Hockey-stick & 4.23 & 8.22 & 1.35 & 7.89 \\
            & Indicator & 3.10 & 7.49 & 1.00 & 7.13 \\
            \cmidrule{1-2} \cmidrule{3-4} \cmidrule{5-6}
            \multirow{3}{*}{$q = 100$} & Quadratic & 3.37 & 8.84 & 1.10 & 8.38 \\
            & Hockey-stick & 4.82 & 10.86 & 1.57 & 9.55 \\
            & Indicator & 3.71 & 10.55 & 1.19 & 9.96 \\
            \bottomrule
            \end{tabular}
            }
            {The results for $\mathcal{T}$-dependent LOOCV are taken from Table~\ref{tab:prm-results-matern}. On average, the RRMSE ratio of the $\mathcal{T}$-dependent LOOCV to the $\mathcal{T}$-independent LOOCV is 0.50 for $\Gamma=10^4$ and is 0.21 for $\Gamma=10^5$.}
        \end{table}

First, the $\mathcal{T}$-dependent LOOCV allows the KRR-driven method to achieve significantly better estimation accuracy compared to the $\mathcal{T}$-independent LOOCV in all cases studied. (On average, for $\Gamma=10^4$, the decrease is 50\%, while for $\Gamma=10^5$, the average reduction approaches 80\%.)
This observation aligns with the findings of \cite{HongJunejaLiu17}. They developed a nested simulation method based on $k$-nearest neighbors ($k$NN), where $k$ serves as a hyperparameter in their approach. Using the standard (i.e., $\mathcal{T}$-independent) LOOCV, they tuned the value of $k$. Although they did not report specific numerical results, they noted that ``different $k$'s may be more appropriate for different functions $g$." (In their paper, the notation $g$ represents the same concept as $\eta$ in our work.)

Second, the advantage of the $\mathcal{T}$-dependent LOOCV becomes more pronounced as the dimensionality of the problem (i.e., $q$) and the simulation budget $\Gamma$ increase.
For example, consider the case when $\mathcal{T}$ corresponds to an indicator function.
In this case, for $q=10$ and $\Gamma=10^4$, the ratio of the RRMSE associated with the $\mathcal{T}$-dependent LOOCV to the one associated with the $\mathcal{T}$-independent LOOCV is {$2.44/4.48 \approx 0.54$};
in contrast, for $q=100$ and $\Gamma=10^5$, the ratio is further decreased to {$1.19/9.96\approx 0.12$}.

The advantage of the $\mathcal{T}$-dependent LOOCV, as demonstrated in Table~\ref{tab:prm-indcv-RRMSE}, may come at the cost of potentially longer computational time. This additional computational time does not originate from the $\mathcal{T}$-dependent procedure itself. In fact, the execution time for both types of LOOCV is essentially the same when only one functional is of interest.
Instead, the increased computational time may arise if one is interested in multiple functionals simultaneously. In such cases, it may be necessary to re-run the procedure multiple times, once for each functional.

\begin{table}[ht]
    \TABLE{Running Time (in sec.) of the KRR-driven Method for the Portfolio Risk Management Problem. \label{tab:prm-indcv-wct}}
    {
    \begin{tabular}{@{\extracolsep{10pt}} l @{\extracolsep{10pt}} r r @{\extracolsep{10pt}} r r}
    \toprule
    & \multicolumn{2}{c}{$\Gamma = 10^4$} & \multicolumn{2}{c}{$\Gamma = 10^5$} \\
    \cmidrule{2-3} \cmidrule{4-5}
    & Simulation & Estimation & Simulation & Estimation \\
    \cmidrule{1-1} \cmidrule{2-3} \cmidrule{4-5}
    $q = 10$ & 75.10 & 2.03 & 763.16 & 294.09 \\
    $q = 20$ & 76.88 & 2.63 & 794.76 & 341.53 \\
    $q = 50$ & 81.91 & 4.05 & 832.44 & 488.05 \\
    $q = 100$ & 90.31 & 6.44 & 914.77 & 876.58 \\
    \bottomrule
    \end{tabular}
    }
    {This table is part of Table~\ref{tab:prm-wct}. The reported running times represent the average for one functional $\mathcal{T}$.}
\end{table}

Table~\ref{tab:prm-indcv-wct} presents the running times of the KRR-driven method for the portfolio risk management problem. In this table, the numbers under the ``Simulation'' columns  indicate the time duration necessary for generating simulation samples. Meanwhile, the numbers under the ``Estimation'' columns  represent the time required to execute either type of LOOCV \emph{once}, involving 10 searches over the space of hyperparameter $\Xi$.
Consequently, when focusing on the three functionals in Table~\ref{tab:prm-indcv-RRMSE} for which the two types of LOOCV are not equivalent, the total time required for the estimation step using the $\mathcal{T}$-dependent LOOCV would be three times greater than that using the $\mathcal{T}$-independent LOOCV. This corresponds to triple the values listed under the ``Estimation'' columns in Table~\ref{tab:prm-indcv-wct}.

For example, consider the case of $\Gamma=10^4$.
Using the $\mathcal{T}$-independent LOOCV, the total computational overhead for estimating $\theta=\mathbb{E}[\eta(\cdot)]$ with $\eta$ being the three functions in Table~\ref{tab:prm-indcv-RRMSE} is less than 7 seconds for all the values of $q$ studied. In contrast, employing the $\mathcal{T}$-dependent LOOCV results in a total computational overhead of less than {$7\times 3 = 21$} seconds.
The increased overhead is justifiable considering the average $50\%$ reduction in RRMSE achieved through the use of the $\mathcal{T}$-dependent LOOCV.
When $\Gamma=10^5$, the computational overhead for both types of LOOCV substantially increases.
Nevertheless, the use of the $\mathcal{T}$-dependent LOOCV achieves nearly an 80\% reduction in RRMSE on average, compared to the $\mathcal{T}$-independent LOOCV.
The former approach is generally preferable, especially when faced with significant simulation time, unless there are many functionals that must be examined simultaneously.

In a more general context,
the decision to use either the $\mathcal{T}$-dependent or $\mathcal{T}$-independent versions of LOOCV should be based on factors such as the number of functions of interest, the estimation time for each functional, and the simulation time required.
Recall that the two types of LOOCV are equivalent for some functionals (category~1) but not for others (category~2).
For instance, category~1 may include $\mathcal{T}(\cdot) = \VaR_\tau(\cdot)$ or $\mathcal{T}(\cdot) = \CVaR_\tau(\cdot)$ for different risk levels $\tau$, or $\mathcal{T}(\cdot) = \E[\eta(\cdot)]$ with $\eta$ being a linear function; category~2 may include $\mathcal{T}(\cdot) = \E[\eta(\cdot)]$ with $\eta$ being a nonlinear function.
To facilitate our discussion, let us assume that among all the functionals of interest, $N_1$ of them belong to category~1, while $N_2$ of them belong to category~2. Let $D$ denote the time required to execute either the $\mathcal{T}$-dependent or the $\mathcal{T}$-independent LOOCV.
Given simulation samples, the total time required to construct an estimate of $\theta$ for each $\mathcal{T}$ of interest using the KRR-driven method can be expressed as follows.
It is $(\ind\{N_1 \geq 1\} + N_2) D$ if the $\mathcal{T}$-dependent LOOCV is employed and is
simply $D$ if the $\mathcal{T}$-independent LOOCV is employed.
The $\mathcal{T}$-dependent LOOCV is clearly preferable when $N_2$ is small;
however, the $\mathcal{T}$-independent LOOCV may become more advantageous when $N_2$ is considerably large.

\section{Additional Implementation Details for Section~\ref{sec:prm}} \label{ec-sec:prm-details}

\subsection{Accuracy of the Brute-force Estimation of $\theta$}\label{ec-sec:brute-force}

The true value of $\theta$ is unknown beforehand. Thus, to compare different methods for nested simulation, we adopt a brute-force approach to obtain an accurate approximation of $\theta$ by generating $10^8$ i.i.d. distributed copies of $Z=\E[Y|X]$. Since the magnitude of $\theta$ varies based on different forms of  $\mathcal{T}$ and values of $q$, we present relative standard deviations in Table~\ref{tab:prm-std} to quantify the uncertainty in the estimates.

\begin{table}[ht]
        \TABLE{Relative Standard Deviation of the ``Brute-force'' Estimate of $\theta$ for the Portfolio Risk Management Problem. \label{tab:prm-std}}
        {
        \begin{tabular}{@{\extracolsep{10pt}} l @{\extracolsep{10pt}} r @{\extracolsep{10pt}} r @{\extracolsep{10pt}} r @{\extracolsep{10pt}} r}
        \toprule
        $\mathcal{T}$ & $q = 10$ & $q = 20$ & $q = 50$ & $q = 100$ \\
        \midrule
        Quadratic & 1.63$e$-04 & 1.50$e$-04 & 1.50$e$-04 & 1.48$e$-04  \\
        Hockey-stick & 1.49$e$-04 & 1.75$e$-04 & 1.92$e$-04 & 2.24$e$-04 \\
        Indicator & 1.07$e$-04 & 1.24$e$-04 & 1.38$e$-04 & 1.62$e$-04 \\
        VaR & 1.27$e$-04 & 1.48$e$-04 & 1.63$e$-04 & 1.58$e$-04 \\
        CVaR & 1.31$e$-04 & 1.56$e$-04 & 1.65$e$-04 & 1.73$e$-04 \\
        \bottomrule
        \end{tabular}
        }
        {The relative standard deviation is determined by performing bootstrapping over $10^8$ i.i.d. copies of $Z$. To obtain this value, we generate a total of $B=10^4$ bootstrap samples. For the $b$-th bootstrap sample, we apply the standard nested simulation method to estimate $\theta$, and denote this estimate as $\hat{\theta}^{(b)}$. The relative standard deviation is then obtained by computing the ratio of the standard deviation of the set $\{\hat{\theta}^{(b)}: b=1,\ldots,B\}$ to its mean.}
    \end{table}

\subsection{Generation of $\sigma_{ij}$}\label{ec-sec:vol}
We generate the lower-triangular matrix $(\sigma_{ij})_{i,j=1}^{q}$ by applying the Cholesky decomposition to a covariance matrix $\BFSigma$ that is randomly generated as follows.
We first generate a random correlation matrix $\BFV$ using the Python function (in the SciPy library) \texttt{scipy.stats.random\_correlation.rvs}, which implements an algorithm proposed by \cite{DaviesHigham00_ec}.
The input to this Python function is the eigenvalues of $\BFV$.
We generate these eigenvalues $\lambda_i$, $i = 1, \ldots, q$, from $\mathsf{Uniform}[0.1,0.8]$, and normalize them so that $\sum_i \lambda_i = q$.
The normalization is necessary because the sum of the eigenvalues must equal to the trace of $\BFV$ and because the diagonal entries of $\BFV$ are all ones. Then, we set $\BFSigma = \BFD \BFV \BFD$, where $\BFD$ is a diagonal matrix with the diagonal entries generated from $\mathsf{Uniform}[0.1,0.8]$.

\subsection{Generation of the Running Maximum} \label{ec-sec:run-max}
When simulating the running maximum $\max_{0\leq t\leq T} S_i(t)$,
the naive approximation based on discretization of the sample path is generally inaccurate.
The approach of Brownian bridge approximation works as follows. We first generate a discrete sample path of $S_i(t)$ at equally spaced time points on $[0,T]$, where $t=lh$ for $l=0,1,\ldots,L$, $h = T/L$, and $L=200$. According to the Euler discretization scheme for SDEs, when $h$ is small, the sample path between two adjacent time points, $\{S_i(t):lh \leq t\leq (l+1)h\}$, can be approximated by a Brownian motion. Therefore, conditional on $S_i(lh)$ and $S_i((l+1)h)$, the sample path over this interval can be approximated by a Brownian bridge. Let $\{R_i(t):lh\leq t\leq (l+1)h\}$ denote this Brownian bridge.
Then, its maximum $M_i^{(l)} \coloneqq \max_{lh\leq t\leq (l+1)h} R_i(t)$ can be simulated exactly. We use $\max_{1\leq l\leq L}M_i^{(l)}$ to approximate $\max_{0\leq t\leq T}S_i(t)$. For more details, please refer to \citet[Chapter~6.4]{Glasserman03_ec}.

\section{Proof of Proposition~\ref{prop:rate_l1norm}} \label{sec:rate_l1norm}

\begin{lemma}[Representer Theorem (Theorem~4.2 in \citealt{ScholkopfSmola02_ec})]\label{lemma:representation-thm}
Let $k:\Omega\times\Omega\mapsto\RR$ be a positive definite kernel, $\mathscr{N}_k(\Omega)$ be the RKHS of $k$,
$L:\RR \times \RR \mapsto \RR_+$ be an arbitrary loss function,
$Q:\RR_+ \mapsto \RR$ be a strictly increasing function,
and $\{(\BFx_i,y_i)\}_{i=1}^n$ be a set of training data with $\BFx_i\in\Omega$ and $y_i\in\RR$.
Then, each optimal solution to the  optimization problem
\begin{equation}\label{eq:reg-risk-min}
    \min_{g\in\mathscr{N}_k(\Omega)}\frac{1}{n}\sum_{i=1}^n L(y_i, g(\BFx_i)) + Q(\|g\|_{\mathscr{N}_k(\Omega)}),
\end{equation}
admits a representation of the form $g^*=\sum_{i=1}^n \beta_i^* k(\BFx_i,\cdot)$ for some constants $\beta_i^*\in\RR$, $i=1,\ldots,n$.
\end{lemma}

\begin{lemma}[Theorem~14.4-1 in \citealt{BishopFienbergHolland07_ec}]\label{lemmacheby}
Let $Z_1,\ldots,Z_n$ be zero-mean random variables such that $\Var(Z_i)<\infty$ for all $i=1,\ldots,n$. Then,
$Z_n  = O_{\pr}(\sqrt{\Var(Z_n)})$.
\end{lemma}

\proof{Proof of Proposition~\ref{prop:rate_l1norm}.}
Let $\BFf = (f(\BFx_1),\ldots,f(\BFx_n))^\intercal$ and $\bar \BFepsilon = (\bar \epsilon_1,\ldots,\bar \epsilon_n)^\intercal$, where $\bar \epsilon_i=\frac{1}{m}\sum_{j=1}^m \epsilon_{ij}.$
It follows from the expression of $\hat{f}$  that
$$\hat{f}(\BFx_i) = \BFr(\BFx_i)^\intercal(\BFR+n\lambda \BFI )^{-1}\bar{\BFy} = \BFr(\BFx_i)^\intercal(\BFR+n\lambda \BFI )^{-1}\BFf + \BFr(\BFx_i)^\intercal(\BFR+n\lambda \BFI )^{-1}\bar \BFepsilon.$$
Thus,
\begin{equation}\label{eq:thmsmoothI21}
\begin{aligned}
& \left|\frac{1}{n}\sum_{i=1}^n \varphi(f(\BFx_i))(f(\BFx_i)-\hat{f}(\BFx_i))\right|  \\
    \leq{}& \Biggl|\underbrace{\frac{1}{n}\sum_{i=1}^n \varphi(f(\BFx_i))(f(\BFx_i)-\BFr(\BFx_i)^\intercal(\BFR+n\lambda \BFI )^{-1}\BFf)}_{H_1}\Biggr|
    + \Biggl|\underbrace{\frac{1}{n}\sum_{i=1}^n \varphi(f(\BFx_i))\BFr(\BFx_i)^\intercal(\BFR+n\lambda \BFI )^{-1}\bar \BFepsilon}_{H_2}\Biggr|.
\end{aligned}
\end{equation}

To bound $H_1$, we apply
the Cauchy--Schwarz inequality:
\begin{align}\label{eq:thmsmoothI3X}
    |H_1| \leq{}& \left(\frac{1}{n}\sum_{i=1}^n \varphi(f(\BFx_i))^2\right)^{1/2}\left(\frac{1}{n}\sum_{i=1}^n (f(\BFx_i)-\BFr(\BFx_i)^\intercal(\BFR+n\lambda \BFI )^{-1}\BFf)^2\right)^{1/2}\nonumber\\
    \leq{}& C \left(\frac{1}{n}\sum_{i=1}^n (f(\BFx_i)-\BFr(\BFx_i)^\intercal(\BFR+n\lambda \BFI )^{-1}\BFf)^2\right)^{1/2},
\end{align}
where $C = \sup_{z\in \{f(\BFx): \BFx\in\Omega\}} |g(z)|<\infty$.

Let $f_{\dagger}(\BFx) \coloneqq \BFr(\BFx)^\intercal(\BFR+n\lambda \BFI )^{-1}\BFf$.
By Lemma~\ref{lemma:representation-thm}, it can be checked that $f_{\dagger}$ is the solution to the optimization problem:
\begin{align}\label{eq:thm1opt1}
    f_{\dagger}= \argmin_{g\in \mathscr{N}_\Psi(\Omega)} \frac{1}{n}\sum_{i=1}^n (g(\BFx_i) - f(\BFx_i))^2 + \lambda \|g\|_{\mathscr{N}_\Psi(\Omega)}^2.
\end{align}
Clearly, \eqref{eq:thm1opt1} implies
\begin{align*}%
\frac{1}{n}\sum_{i=1}^n (f(\BFx_i)-\BFr(\BFx_i)^\intercal(\BFR+n\lambda \BFI )^{-1}\BFf)^2
={}& \frac{1}{n}\sum_{i=1}^n (f(\BFx_i) - f_{\dagger}(\BFx_i))^2
\nonumber\\
\leq{}& \frac{1}{n}\sum_{i=1}^n (f(\BFx_i) - f_{\dagger}(\BFx_i))^2 + \lambda \|f_{\dagger}\|_{\mathscr{N}_\Psi(\Omega)}^2\nonumber\\
\leq{}& \frac{1}{n}\sum_{i=1}^n (f(\BFx_i) - f(\BFx_i))^2 + \lambda \|f\|_{\mathscr{N}_\Psi(\Omega)}^2 = \lambda \|f\|_{\mathscr{N}_\Psi(\Omega)}^2.
\end{align*}
Thus,
\begin{align}\label{eq:thmsmoothI3}
   |H_1| = O(\lambda^{1/2}).
\end{align}

Now consider $H_2$. Because $\epsilon_{ij}$'s are zero-mean sub-Gaussian random variables,  $H_2$ is also sub-Gaussian and has mean zero, which implies that $H_2$ has a finite variance.
By Lemma~\ref{lemmacheby},
\begin{align}\label{eq:thmsmoothI4}
    |H_2| =O_{\pr}(\sqrt{\Var(H_2)}).
\end{align}

Let $\BFu_i=(u_1(\BFx_i),\ldots,u_n(\BFx_i))^\intercal=(\BFR+n\lambda \BFI )^{-1}\BFr(\BFx_i)$. Since $|\varphi(f(\BFx_i))|\leq C$ for all $i=1,\ldots,n$, by Assumption~\ref{assump:subG}, we have that
\begin{align}\label{eq:thmsmoothI4var}
    \Var(H_2) ={}& \Var\left(\frac{1}{n}\sum_{i=1}^n \varphi(f(\BFx_i))\BFr(\BFx_i)^\intercal(\BFR+n\lambda \BFI )^{-1}\bar \BFepsilon\right)\nonumber\\
    ={}& \Var\left(\frac{1}{n}\sum_{i=1}^n \varphi(f(\BFx_i))\sum_{j=1}^n u_j(\BFx_i)\bar \epsilon_j\right)\nonumber\\
    \leq{}& \frac{\sigma^2}{m}\sum_{j=1}^n\left(\frac{1}{n}\sum_{i=1}^n \varphi(f(\BFx_i))u_j(\BFx_i)\right)^2.
\end{align}

Let $\BFe_j\in\RR^d$ be a vector of zeros except the $j$-th entry being one.
Direct computation shows that
\begin{align}\label{eq:thmsmoothI4var2}
    & \sum_{j=1}^n\left(\frac{1}{n}\sum_{i=1}^n \varphi(f(\BFx_i)) u_j(\BFx_i)\right)^2\nonumber\\
    ={}& \sum_{j=1}^n\left(\frac{1}{n}\sum_{i=1}^n \varphi(f(\BFx_i))\BFe_j^\intercal(\BFR+n\lambda \BFI )^{-1}\BFr(\BFx_i)\right)^2\nonumber\\
    ={}& \sum_{j=1}^n\left( \BFe_j^\intercal(\BFR+n\lambda \BFI )^{-1}\left( \frac{1}{n}\sum_{i=1}^n\varphi(f(\BFx_i))\BFr(\BFx_i)\right)\right)^2\nonumber\\
    ={}& \sum_{j=1}^n\left(\frac{1}{n}\sum_{i=1}^n\varphi(f(\BFx_i))\BFr(\BFx_i)\right)^\intercal (\BFR+n\lambda \BFI )^{-1}\BFe_j\BFe_j^\intercal(\BFR+n\lambda \BFI )^{-1}\left( \frac{1}{n}\sum_{i=1}^n\varphi(f(\BFx_i))\BFr(\BFx_i)\right)\nonumber\\
    ={}& \left(\frac{1}{n}\sum_{i=1}^n\varphi(f(\BFx_i))\BFr(\BFx_i)\right)^\intercal(\BFR+n\lambda \BFI )^{-2}\left(\frac{1}{n}\sum_{i=1}^n\varphi(f(\BFx_i))\BFr(\BFx_i)\right)\nonumber\\
    \leq{}& \left(\frac{1}{n}\sum_{i=1}^n\varphi(f(\BFx_i))\BFr(\BFx_i)\right)^\intercal  \BFR^{-2}\left(\frac{1}{n}\sum_{i=1}^n\varphi(f(\BFx_i))\BFr(\BFx_i)\right)\nonumber\\
    ={}& \frac{1}{n^2}\sum_{i,j=1}^n \varphi(f(\BFx_i))g(f(\BFx_j)) \BFr(\BFx_i)^\intercal \BFR^{-2}r(\BFx_j) \nonumber\\
    = {}&\frac{1}{n^2}\sum_{i,j=1}^n \varphi(f(\BFx_i))g(f(\BFx_j)) \BFe_i^\intercal \BFe_j \leq \frac{C^2}{n},
\end{align}
where the fourth equality holds because $\sum_{j=1}^n \BFe_j \BFe_j^\intercal = \BFI $, and the last step because $\BFR^{-1}\BFr(\BFx_i)=\BFe_i$.
Then,
plugging \eqref{eq:thmsmoothI4var} and \eqref{eq:thmsmoothI4var2} into \eqref{eq:thmsmoothI4} yields
\begin{align}\label{eq:thmsmoothH2}
   |H_2|=O_{\pr}((mn)^{-1/2}).
\end{align}

The proof is completed by combining \eqref{eq:thmsmoothI21}, \eqref{eq:thmsmoothI3},  and \eqref{eq:thmsmoothH2}.
\Halmos\endproof

\section{Proof of Proposition~\ref{prop:rate_seminorm}}\label{sec:rate_seminorm}

\begin{lemma}[Lemma~A.1 in \citealt{TuoWangWu20_ec}]\label{lemmaefsmall}
Let $\Omega$ be a bounded convex subset of $\RR^d$ and
$\{\epsilon_1,\ldots,\epsilon_n\}$ be independent, zero-mean sub-Gaussian random variables.
Then,
there exist positive constants $C_1$ and $C_2$ such that
for all $t$ large enough,
\begin{align*}%
\Pr\left(\sup_{g\in \mathscr{N}_{\Psi}(\Omega)}\frac{\bigl|\frac{1}{n}\sum_{j=1}^n\epsilon_j g(\BFx_j)\bigr|}{\|g\|_n^{1 - \frac{d}{2\nu+d}}\|g\|_{\mathscr{N}_{\Psi}(\Omega)}^{\frac{d}{2\nu+d}}} \geq tn^{-1/2}\right) \leq C_1\exp(-C_2 t^2),
\end{align*}
where
$\|g\|_n=\left(n^{-1}\sum_{i=1}^n g^2(\BFx_i)\right)^{1/2}$ denotes the empirical semi-norm of $g$
\end{lemma}

\proof{Proof of Proposition~\ref{prop:rate_seminorm}.}

Note that $\hat{f}$ is the solution to
\[
 \min_{g\in \mathscr{N}_{\Psi}(\Omega)}\bigg( \frac{1}{n}\sum_{i=1}^n(\bar{y}_i-g(\BFx_i))^2 + \lambda \|g\|^2_{\mathscr{N}_{\Psi}(\Omega)}\bigg).
\]
Hence,
\begin{align}\label{eq:basiceq}
    \frac{1}{n}\sum_{i=1}^n (\hat{f}(\BFx_i)-\bar y_i)^2 + \lambda \|\hat{f}\|_{\mathscr{N}_\Psi(\Omega)}^2 \leq \frac{1}{n}\sum_{i=1}^n (f(\BFx_i)-\bar y_i)^2 + \lambda \| f\|_{\mathscr{N}_\Psi(\Omega)}^2.
\end{align}

Using the notation $\|\hat{f}-f\|_n^2 = \frac{1}{n}\sum_{i=1}^n (f(\BFx_i)-\hat{f}(\BFx_i))^2$,
and plugging $\bar y_i = f(\BFx_i) + \bar \epsilon_i$ into \eqref{eq:basiceq}, we have that
\begin{align*}
    \|\hat{f}-f\|_n^2 + \lambda \|\hat{f}\|_{\mathscr{N}_\Psi(\Omega)}^2 \leq{}&  \frac{2}{n}\sum_{i=1}^n \bar{\epsilon}_i (\hat{f}-f)(\BFx_i) + \lambda \| f\|_{\mathscr{N}_\Psi(\Omega)}^2.
\end{align*}
Moreover, note that $\bar \epsilon_i$ is $\mathsf{subG}(\sigma^2/m)$.
It then follows from Lemma~\ref{lemmaefsmall}  that
\[
\frac{1}{n}\sum_{i=1}^n \bar{\epsilon}_i (\hat{f}-f)(\BFx_i) = O_{\pr}((mn)^{-1/2})\|\hat{f}-f\|_n^{1-\frac{d}{2\nu + d}}\|\hat{f}-f\|_{\mathscr{N}_\Psi(\Omega)}^{\frac{d}{2\nu + d}}.
\]
Hence,
\begin{align}\label{eq:pf1bineq11}
    \|\hat{f}-f\|_n^2 + \lambda \|\hat{f}\|_{\mathscr{N}_\Psi(\Omega)}^2 \leq{}& O_{\pr}((mn)^{-1/2})\|\hat{f}-f\|_n^{1-\frac{d}{2\nu + d}}\|\hat{f}-f\|_{\mathscr{N}_\Psi(\Omega)}^{\frac{d}{2\nu + d}}+ \lambda \| f\|_{\mathscr{N}_\Psi(\Omega)}^2.
\end{align}
It can be seen that
\eqref{eq:pf1bineq11} implies either
\begin{align}\label{eq:pf1bineqc11}
    \|\hat{f}-f\|_n^2 + \lambda \|\hat{f}\|_{\mathscr{N}_\Psi(\Omega)}^2 \leq{}& O_{\pr}((mn)^{-1/2})\|\hat{f}-f\|_n^{1-\frac{d}{2\nu + d}}\|\hat{f}-f\|_{\mathscr{N}_\Psi(\Omega)}^{\frac{d}{2\nu + d}},
\end{align}
or
\begin{align}\label{eq:pf1bineqc12}
    \|\hat{f}-f\|_n^2 + \lambda \|\hat{f}\|_{\mathscr{N}_\Psi(\Omega)}^2 \leq{}&
    2
    \lambda \| f\|_{\mathscr{N}_\Psi(\Omega)}^2.
\end{align}

\textbf{Case 1:} Assume \eqref{eq:pf1bineqc11} holds. Then, by the triangle inequality,
\begin{align}\label{eq:pf1bineqc11'}
    \|\hat{f}-f\|_n^2 + \lambda \|\hat{f}\|_{\mathscr{N}_\Psi(\Omega)}^2 \leq{}& O_{\pr}((mn)^{-1/2})\|\hat{f}-f\|_n^{1-\frac{d}{2\nu + d}}\bigl(\|\hat{f}\|_{\mathscr{N}_\Psi(\Omega)} + \|f\|_{\mathscr{N}_\Psi(\Omega)}\bigr) ^{\frac{d}{2\nu + d}}.
\end{align}
Next, we  analyze \eqref{eq:pf1bineqc11'} separately depending on whether $\| f\|_{\mathscr{N}_\Psi(\Omega)}\leq \|\hat{f}\|_{\mathscr{N}_\Psi(\Omega)}$.

If $\| f\|_{\mathscr{N}_\Psi(\Omega)}\leq \|\hat{f}\|_{\mathscr{N}_\Psi(\Omega)}$, then \eqref{eq:pf1bineqc11'} implies that
\begin{align*}
    \|\hat{f}-f\|_n^2 + \lambda \|\hat{f}\|_{\mathscr{N}_\Psi(\Omega)}^2 \leq{}& O_{\pr}((mn)^{-1/2})\|\hat{f}-f\|_n^{1-\frac{d}{2\nu + d}}\|\hat{f}\|_{\mathscr{N}_\Psi(\Omega)}^{\frac{d}{2\nu + d}}.
\end{align*}
Therefore,
\begin{align}
\|\hat{f}-f\|_n^2 \leq{}& O_{\pr}((mn)^{-1/2})\|\hat{f}-f\|_n^{1-\frac{d}{2\nu + d}}\|\hat{f}\|_{\mathscr{N}_\Psi(\Omega)}^{\frac{d}{2\nu + d}}, \label{eq:pf1bineqc111} \\
\lambda \|\hat{f}\|_{\mathscr{N}_\Psi(\Omega)}^2 \leq{}& O_{\pr}((mn)^{-1/2})\|\hat{f}-f\|_n^{1-\frac{d}{2\nu + d}}\|\hat{f}\|_{\mathscr{N}_\Psi(\Omega)}^{\frac{d}{2\nu + d}}. \label{eq:pf1bineqc111'}
\end{align}
Solving the system of inequalities \eqref{eq:pf1bineqc111}--\eqref{eq:pf1bineqc111'} yields
\begin{align}
    \|\hat{f}-f\|_n^2 ={}& O_{\pr}\left((mn)^{-1}\lambda^{-\frac{d}{2\nu+d}}\right), \label{eq:pf1bineqc11X}\\
    \|\hat{f}\|_{\mathscr{N}_\Psi(\Omega)}^2={}&  O_{\pr}\left((mn)^{-1}\lambda^{-\frac{2(\nu+d)}{2\nu+d}}\right).\label{eq:pf1bineqc11X'}
\end{align}

If $\| f\|_{\mathscr{N}_\Psi(\Omega)}> \|\hat{f}\|_{\mathscr{N}_\Psi(\Omega)}$,  then \eqref{eq:pf1bineqc11'} implies that
\begin{align*}
    \|\hat{f}-f\|_n^2 + \lambda \|\hat{f}\|_{\mathscr{N}_\Psi(\Omega)}^2 \leq{}& O_{\pr}((mn)^{-1/2})\|\hat{f}-f\|_n^{1-\frac{d}{2\nu + d}}\|f\|_{\mathscr{N}_\Psi(\Omega)}^{\frac{d}{2\nu + d}}.
\end{align*}
Hence, noting that $\|f\|_{\mathscr{N}_\Psi(\Omega)}$ is a constant, we have
\begin{equation}\label{eq:pf1bineqc21}
\|\hat{f}-f\|_n^2 \leq     O_{\pr}((mn)^{-1/2})\|\hat{f}-f\|_n^{1-\frac{d}{2\nu + d}}.
\end{equation}
Solving \eqref{eq:pf1bineqc21} yields
\begin{align}\label{eq:pf1bineqc21X}
    \|\hat{f}-f\|_n^2 = O_{\pr}\left((mn)^{-\frac{2\nu+d}{2\nu+2d}}\right).
\end{align}

\textbf{Case 2:} Assume \eqref{eq:pf1bineqc12} holds. Then,
\begin{align}
    \|\hat{f}-f\|_n^2 \leq{}& 2\lambda \| f\|_{\mathscr{N}_\Psi(\Omega)}^2 = O_{\pr}(\lambda). \label{eq:pf1bineqc12'}
\end{align}

Combining the two cases, i.e., combining \eqref{eq:pf1bineqc11X}, \eqref{eq:pf1bineqc21X}, and \eqref{eq:pf1bineqc12'}, we conclude that
\begin{align*} %
   \|\hat{f}-f\|_n^2 = O_{\pr}\left((mn)^{-1}\lambda^{-\frac{d}{2\nu+d}} + (mn)^{-\frac{2\nu+d}{2\nu+2d}} + \lambda\right). \Halmos
\end{align*}
\endproof

\section{Proof of Theorem~\ref{thm:gsmooth}}\label{app:pfthmgs}

\proof{Proof of Theorem~\ref{thm:gsmooth}.}
By the triangle inequality, we have
\begin{align}\label{eq:thmsmootha}
    |\hat \theta_{n,m}-\theta| ={}& \left|\E [\eta(f(X))] -\frac{1}{n}\sum_{i=1}^n \eta(f(\BFx_i))+ \frac{1}{n}\sum_{i=1}^n \eta(f(\BFx_i))- \frac{1}{n}\sum_{i=1}^n \eta(\hat{f}(\BFx_i))\right|\nonumber\\
    \leq{}& \underbrace{\left|\E [\eta(f(X))] -\frac{1}{n}\sum_{i=1}^n \eta(f(\BFx_i))\right|}_{I_1}
    + \underbrace{ \left|\frac{1}{n}\sum_{i=1}^n \bigl[\eta(f(\BFx_i))- \eta(\hat{f}(\BFx_i)) \bigr] \right| }_{I_2}.
\end{align}

The reproducing property of RKHSs implies that for any $\BFx\in \Omega$,
\begin{equation}
\begin{aligned}\label{eq:reproP}
    |f(\BFx)| = |\langle f, \Psi(\BFx - \cdot)\rangle_{\mathscr{N}_{\Psi}(\Omega)}| \leq{}& \|f\|_{\mathscr{N}_{\Psi}(\Omega)} \|\Psi(\BFx - \cdot )\|_{\mathscr{N}_{\Psi}(\Omega)} \\ ={}&  \|f\|_{\mathscr{N}_{\Psi}(\Omega)} \Psi(\BFx - \BFx)
    = \|f\|_{\mathscr{N}_{\Psi}(\Omega)}\Psi(\BFzero),
\end{aligned}
\end{equation}
which implies both $\|f\|_{\mathscr{L}_\infty(\Omega)} $ and $\|\eta(f(\cdot))\|_{\mathscr{L}_\infty(\Omega)}$ are finite.
Therefore, $\eta(f(\BFx_i))$'s are bounded random variables, thereby being sub-Gaussian.
Then, the central limit theorem  implies that
\begin{align}\label{eq:thmsmoothI1}
    I_1 = O_{\pr}(n^{-1/2}).
\end{align}

It remains to bound $I_2$.  It follows from Taylor's expansion and the triangle inequality that
\begin{align}\label{eq:thmsmoothI2}
    I_2 ={}& \left|\frac{1}{n}\sum_{i=1}^n \eta'(f(\BFx_i))(f(\BFx_i)-\hat{f}(\BFx_i)) + \frac{1}{2n}\sum_{i=1}^n \eta''(\tilde{z}_i)(f(\BFx_i)-\hat{f}(\BFx_i))^2 \right|\nonumber\\
    \leq{}& \underbrace{\left|\frac{1}{n}\sum_{i=1}^n \eta'(f(\BFx_i))(f(\BFx_i)-\hat{f}(\BFx_i))\right|}_{I_{21}}
    + \underbrace{\left|\frac{1}{2n}\sum_{i=1}^n \eta''(\tilde{z}_i)(f(\BFx_i)-\hat{f}(\BFx_i))^2 \right|}_{I_{22}},
\end{align}
where $\tilde{z}_i$ is a value between $f(\BFx_i)$ and $\hat{f}(\BFx_i)$.

Because $|\eta'(f(\BFx))|$ is bounded for all $\BFx\in\Omega$, it follows from Proposition~\ref{prop:rate_l1norm} that
\begin{align}\label{eq:thmsmoothI21X}
    I_{21} = O_{\pr}(\lambda^{1/2} + (mn)^{-1/2}).
\end{align}

Let $C = \sup_{z\in \{f(\BFx): \BFx\in\Omega\}} |\eta''(z)|<\infty$.
The term $I_{22}$ can be bounded using Proposition~\ref{prop:rate_seminorm}:
\begin{align}\label{eqpf1I22smoo}
    I_{22} \leq{}& \frac{C}{2n}\sum_{i=1}^n (f(\BFx_i)-\hat{f}(\BFx_i))^2
    = O_{\pr}\left((mn)^{-1}\lambda^{-\frac{d}{2\nu+d}} + (mn)^{-\frac{2\nu+d}{2\nu+2d}} + \lambda\right).
\end{align}

Then, we combine
\eqref{eq:thmsmootha}, \eqref{eq:thmsmoothI1}, \eqref{eq:thmsmoothI2}, \eqref{eq:thmsmoothI21X}, and \eqref{eqpf1I22smoo}.
This yields
\begin{align}\label{eq:thmsmoothar}
    |\hat \theta_{n,m} - \theta|={}& O_{\pr}\left(n^{-1/2} +\lambda^{1/2}+(mn)^{-1/2} + (mn)^{-1}\lambda^{-\frac{d}{2\nu+d}} + (mn)^{-\frac{2\nu+d}{2\nu+2d}}+\lambda\right)\nonumber\\
    ={}& O_{\pr}\left(n^{-1/2} + \lambda^{1/2} + (mn)^{-1}\lambda^{-\frac{d}{2\nu + d}}\right),
\end{align}
where the second inequality holds because $\lambda = O(1)$, $(mn)^{-1/2}\leq n^{-1/2}$, and $(mn)^{-\frac{2\nu+d}{2\nu+2d}}\leq n^{-1/2}$.

If $\nu\geq d/2$, then we set  $n\asymp \Gamma$ and $\lambda \asymp \Gamma^{-1}$.
Because $\frac{2\nu}{2\nu+d}\geq \frac{1}{2}$ in this case,
\eqref{eq:thmsmoothar} becomes
\begin{align*}
    |\hat \theta_{n,m} - \theta|={}  O_{\pr}\left(\Gamma^{-1/2} + \Gamma^{-1/2} +  \Gamma^{-\frac{2\nu}{2\nu + d}}\right)  =  O_{\pr}\left(\Gamma^{-1/2}\right).
\end{align*}

If $0<\nu < d/2$, then we set $n\asymp \Gamma^{\frac{2(2\nu+d)}{2\nu+3d}}$ and $\lambda \asymp \Gamma^{-\frac{2(2\nu+d)}{2\nu+3d}}$. (This choice of $n$ is possible because $\frac{2(2\nu+d)}{2\nu+3d} < 1$ for $\nu < d/2$.)
This makes \eqref{eq:thmsmoothar} become
\[
|\hat \theta_{n,m} - \theta|={}  O_{\pr}\left(\Gamma^{-\frac{2\nu+d}{2\nu+3d}} + \Gamma^{-\frac{2\nu+d}{2\nu+3d}} +  \Gamma^{-\frac{2\nu+d}{2\nu+3d}}\right)  =  O_{\pr}\left(\Gamma^{-\frac{2\nu+d}{2\nu+3d}}\right).  \Halmos
\]
\endproof

\section{Proof of Proposition~\ref{prop:rho_n}}\label{sec:rho_n}

\begin{lemma}\label{lemmaGNinter}
Let $\Omega$ be a bounded convex subset of $\RR^d$ and
$0 < s_1< s_2$. Then, there exists a constant $C>0$  such that for any $g\in \mathscr{H}^{s_2}(\Omega)$,
    \begin{align*}
        \|g\|_{\mathscr{L}_\infty(\Omega)}\leq & C\|g\|_{\mathscr{L}_2(\Omega)}^{1-\frac{s_1}{s_2}}\|g\|_{\mathscr{H}^{s_2}(\Omega)}^{\frac{s_1}{s_2}}.
    \end{align*}
\end{lemma}

\proof{Proof of Lemma~\ref{lemmaGNinter}.}
This is a direct result of applying the Gagliardo--Nirenberg interpolation inequality to the Sobolev spaces (see, e.g., \citealt{BrezisMironescu19_ec}).
\Halmos \endproof

In order to quantify the capacity of a function class, we need the following two definitions of entropy number and bracket entropy number (see \citealt{geer2000empirical_ec} for more discussions).

\begin{definition}[Entropy Number]\label{def:metric-entropy}
Let $\mathscr{G}$ be a function space equipped with a norm $\|\cdot\|$, and $\mathscr{G}_0\subset \mathscr{G}$ be a function class.
For any $\varepsilon>0$, let $\mathscr{B}_\varepsilon(h,\|\cdot\|)\coloneqq\{g\in\mathscr{G}: \|g-h\|\leq\varepsilon\}$ be an $\varepsilon$-ball that is centered at $h\in\mathscr{G}$.
The \emph{covering number}  $\mathcal{N} (\varepsilon,\mathscr{G}_0,\|\cdot\|)$ is defined as
\[\mathcal{N} (\varepsilon,\mathscr{G}_0,\|\cdot\|)\coloneqq\min\biggl\{n: \ \mbox{There exist } g_1,\ldots, g_n\in \mathscr{G}_0 \mbox{ such that } \mathscr{G}_0\subseteq\bigcup_{i=1}^n \mathscr{B}_{\varepsilon}(g_i,\|\cdot\|)\biggr\}.\]
Then, $\mathcal{H} (\varepsilon,\mathscr{G}_0,\|\cdot\|) \coloneqq \log_2\mathcal{N} (\varepsilon,\mathscr{G}_0,\|\cdot\|)$ is called the \emph{entropy number} of $\mathscr{G}_0$.
\end{definition}

\begin{definition}[Bracket Entropy Number]\label{def:metric-bracket-entropy}
Let $\mathscr{G}$ and $\mathscr{G}_0\subset \mathscr{G}$ be in Definition \ref{def:metric-entropy}.
For any $\varepsilon>0$, let $\mathcal{N} _{[\;]}(\varepsilon,\mathscr{G}_0,\|\cdot\|)$ be the smallest value of $n$ for which there exist pairs of functions $\{g_j^L,g_j^U\}\subset \mathscr{G}_0$ such that $\|g_j^U-g_j^L\|\leq \varepsilon$ for all $j=1,...,n$, and such that for each $g\in \mathscr{G}_0$, $g_j^L\leq g\leq g_j^U$. Then, $\mathcal{H} _{[\;]}(\varepsilon,\mathscr{G}_0,\|\cdot\|) \coloneqq \log_2\mathcal{N} _{[\;]}(\varepsilon,\mathscr{G}_0,\|\cdot\|)$ is called the \emph{bracket entropy number} of $\mathscr{G}_0$.
\end{definition}

\begin{lemma}[Lemma~2.1 in \citealt{geer2000empirical_ec}]\label{lemenandben}
Let $\Omega$ be a bounded convex subset of $\RR^d$,
$\mathscr{G}=\{g:\Omega\mapsto \RR: \|g\|_{\mathscr{L}_\infty}<\infty\}$,
and $\mathscr{G}_0\subset \mathscr{G}$.
Then,
for $p =1,2$ and any $\varepsilon>0$,
there exists a constant $C>0$ such that
\[
\mathcal{H} (\varepsilon,\mathscr{G}_0,\|\cdot\|_{\mathscr{L}_p})\leq \mathcal{H} _{[\;]}(\varepsilon,\mathscr{G}_0,\|\cdot\|_{\mathscr{L}_p}) \qq{and} \mathcal{H} _{[\;]}(\varepsilon,\mathscr{G}_0,\|\cdot\|_{\mathscr{L}_p})\leq C\mathcal{H} (\varepsilon/2,\mathscr{G}_0,\|\cdot\|_{\mathscr{L}_\infty}). \]
\end{lemma}

\begin{lemma}[Lemma~5.16 in \citealt{geer2000empirical_ec}]\label{lemmageer}
Let $\Omega$ be a bounded convex subset of $\RR^d$,
$\mathscr{G}=\{g:\Omega\mapsto \RR: \|g\|_{\mathscr{L}_2}<\infty\}$,
and
$\mathscr{G}_0\subset \mathscr{G}$.
Suppose that the sequence $\{\delta_n>0:n\geq 1\}$ satisfies $n\delta_n^2\geq \mathcal{H} _{[\;]}(\delta_n,\mathscr{G}_0,\|\cdot\|_{\mathscr{L}_2})$ for all $n\geq 1$, and $n\delta_n^2\rightarrow \infty$ as $n\to\infty$. Then,  for any $C>0$  and $t\in(0,1)$,
    \begin{align*}
        \limsup_{n\rightarrow\infty }\pr\left(\sup_{g\in \mathscr{G}_0,\|g\|_{\mathscr{L}_2}>C t^{-1}\delta_n}\left|\frac{\|g\|_n}{\|g\|_{\mathscr{L}_2}}-1\right|> t\right)=0.
    \end{align*}
\end{lemma}

\begin{lemma}\label{corogeer}
Let $\Omega$ be a bounded convex subset of $\RR^d$ and
$\mathscr{B} = \{g\in \mathscr{H}^{s}(\Omega): \|g\|_{\mathscr{H}^{s}(\Omega)}\leq 1\}$. Then,  $\|g\|_{\mathscr{L}_2(\Omega)}=O_{\pr}(n^{-\frac{s}{2s+d}} + \|g\|_n)$ for all $g\in \mathscr{B}$.
\end{lemma}
\proof{Proof of Lemma~\ref{corogeer}.}
Note that $\mathscr{B}$ is the unit ball in the Sobolev space $\mathscr{H}^{s}(\Omega)$.
It can be bounded by Lemma~\ref{lemenandben}
\begin{align*}
\mathcal{H} _{[\;]}(\delta_n,\mathscr{B},\|\cdot\|_{\mathscr{L}_2})\leq C\mathcal{H} (\delta_n/2,\mathscr{B},\|\cdot\|_{\mathscr{L}_\infty(\Omega)})\leq  C_1 \delta_n^{-s/d},
\end{align*}
for some constants $C,C_1>0$,
where the second inequality follows from the theorem on page 105 of \citet{edmunds2008function_ec}.
Now, we take $\delta_n = C_1 n^{-\frac{s}{2s+d}}$ such that
\[
n\delta_n^2\geq \mathcal{H} _{[\;]}(\delta_n,\mathscr{B},\|\cdot\|_{\mathscr{L}_2}), \]
and then apply Lemma~\ref{lemmageer}.
This leads to
\begin{align*}
        \limsup_{n\rightarrow\infty }\pr\left(\sup_{g\in \mathscr{B},\|g\|_{\mathscr{L}_2}>C_1t^{-1} n^{-\frac{s}{2s+d}}}\biggl|\frac{\|g\|_n}{\|g\|_{\mathscr{L}_2}}-1\biggr|> t\right)=0.
\end{align*}
Hence, $\|g\|_{\mathscr{L}_2}=O_{\pr}(\max\{n^{-\frac{s}{2s+d}},\|g\|_n\}) =  O_{\pr}(n^{-\frac{s}{2s+d}} + \|g\|_n)$.
\Halmos \endproof

To prove Proposition~\ref{prop:rho_n}, note that,
by the expression of $\hat{f}$,
\begin{align}\label{eq:thmnshsC3dbway2}
    |f(\BFx_i) - \hat{f}(\BFx_i)| \leq{}& \Bigl|\underbrace{f(\BFx_i)-\BFr(\BFx_i)^\intercal(\BFR+n\lambda \BFI )^{-1}\BFf}_{M_1(\BFx_i)}\Bigr|
    + \Bigl|\underbrace{\BFr(\BFx_i)^\intercal(\BFR+n\lambda \BFI )^{-1}\bar \BFepsilon}_{M_2(\BFx_i)}\Bigr|.
\end{align}
Hence, it suffices to bound $\max_{1\leq i\leq n} |M_1(\BFx_i)|$ and $\max_{1\leq i\leq n} |M_2(\BFx_i)|$, respectively.

\begin{lemma}\label{lemma:M_1}
Suppose $f\in\mathscr{N}_\Psi(\Omega)$ and
Assumption~\ref{assump:support} holds.  Then,
\[
\max_{1\leq i\leq n}|M_1(\BFx_i)| =  O_{\pr}\left(\bigl(n^{-\frac{\nu}{2\nu + 2d}} + \lambda^{\frac{\nu}{2\nu+d}}\bigr) \wedge (n\lambda)^{1/2}\right).
\]
\end{lemma}
\proof{Proof of Lemma~\ref{lemma:M_1}.}

Let $f_{\dagger}(\BFx) \coloneqq \BFr(\BFx)^\intercal(\BFR+n\lambda \BFI)^{-1}\BFf$ be the solution to \eqref{eq:thm1opt1}.
The term $M_1(\BFx_i)$ can be bounded in two different ways, and we will take the smaller one as the upper bound.

\textbf{First way to bound $M_1(\BFx_i)$.}
Note that $|f(\BFx) - f_{\dagger}(\BFx)|\leq \|f- f_{\dagger}\|_{\mathscr{L}_\infty(\Omega)}$ for all $\BFx\in\Omega$, so $\max_{1\leq i\leq n}|M_1(\BFx_i)| \leq \|f- f_{\dagger}\|_{\mathscr{L}_\infty(\Omega)}$. Because $f_{\dagger}(\BFx)$ is the solution to \eqref{eq:thm1opt1}, we have that
\begin{align*}
\frac{1}{n}\sum_{i=1}^n (f(\BFx_i) - f_{\dagger}(\BFx_i))^2 + \lambda \|f_{\dagger}\|_{\mathscr{N}_\Psi(\Omega)}^2
\leq{} \frac{1}{n}\sum_{i=1}^n (f(\BFx_i) - f(\BFx_i))^2 + \lambda \|f\|_{\mathscr{N}_\Psi(\Omega)}^2 = \lambda \|f\|_{\mathscr{N}_\Psi(\Omega)}^2,
\end{align*}
which implies
\begin{align}
    \|f-f_{\dagger}\|_n ={}& O_{\pr}(\lambda^{1/2}),\label{eq:opt1error1}\\
    \|f_{\dagger}\|_{\mathscr{N}_\Psi(\Omega)} \leq{} & \|f\|_{\mathscr{N}_\Psi(\Omega)}.\label{eq:opt1error2}
\end{align}
Because of the norm equivalence between $\mathscr{N}_\Psi(\Omega)$ and  $\mathscr{H}^{\nu+d/2}(\Omega)$,
there exists a constant $C_1$ such that
\begin{align}\label{eq:opt1error2X}
    \|f-f_{\dagger}\|_{\mathscr{H}^{\nu+d/2}(\Omega)}\leq C_1\|f-f_{\dagger}\|_{\mathscr{N}_\Psi(\Omega)}\leq C_1(\|f\|_{\mathscr{N}_\Psi(\Omega)}+\|f_{\dagger}\|_{\mathscr{N}_\Psi(\Omega)}) \leq 2C_1 \|f\|_{\mathscr{N}_\Psi(\Omega)},
\end{align}
where the last equality is by \eqref{eq:opt1error2}.

Let $g=\frac{f-f_{\dagger}}{2C_1\|f\|_{\mathscr{N}_\Psi(\Omega)} }$.
Then, $\|g\|_{\mathscr{H}^{\nu+d/2}(\Omega)}\leq 1$.
It follows immediately from Lemma~\ref{corogeer} that
\[
\|g\|_{\mathscr{L}_2(\Omega)}=O_{\pr}\left(n^{-\frac{\nu + d/2}{2\nu + 2d}} + \|g\|_n\right),
\]
which, together with \eqref{eq:opt1error1}, implies
\begin{align}\label{eq:opt1errorL21}
    \|f-f_{\dagger}\|_{\mathscr{L}_2(\Omega)} = O_{\pr}\left(n^{-\frac{\nu + d/2}{2\nu + 2d}} + \|f-f_{\dagger}\|_n\right) = O_{\pr}\left(n^{-\frac{\nu + d/2}{2\nu + 2d}} + \lambda^{1/2}\right).
\end{align}
By Lemma~\ref{lemmaGNinter}, it follows from \eqref{eq:opt1error2X} and \eqref{eq:opt1errorL21} that for there exists a constant $C_2>0$ such that
\begin{align}
    \max_{1\leq i\leq n}|M_1(\BFx_i)| \leq\|f-f_{\dagger}\|_{\mathscr{L}_\infty(\Omega)}\leq{} & C_2 \|f-f_{\dagger}\|_{\mathscr{L}_2(\Omega)}^{1-\frac{d}{2\nu+d}}\|f-f_{\dagger}\|_{\mathscr{N}_\Psi(\Omega)}^{\frac{d}{2\nu+d}} \nonumber\\
    ={}& O_{\pr}\left(n^{-\frac{\nu}{2\nu + 2d}} + \lambda^{\frac{\nu}{2\nu+d}}\right).\label{eq:thmnshsC3fhatinf}
\end{align}

\textbf{Second way to bound $M_1(\BFx_i)$.}
A second way to bound $M_1(\BFx_i)$ is to work on $M_1(\BFx_i)$ directly without using the $\mathscr{L}_\infty$ bound. For $M_1(\BFx_i)$, Lemma~F.8 in \cite{wang2020inference_ec} asserts that
\[
(f(\BFx) - \BFr(\BFx)^{\intercal}(\BFR+ n\lambda \BFI )^{-1}\BFf)^2 \leq (\Psi(\BFzero) -  \BFr(\BFx)^{\intercal}(\BFR+ n\lambda \BFI )^{-1}\BFr(\BFx))\|f\|_{\mathscr{N}_{\Psi}(\Omega)}^2.
\]
Moreover, note that
\begin{align*}
    \Psi(\BFzero)-\BFr(\BFx_i)^\intercal(\BFR+n\lambda  \BFI )^{-1}\BFr(\BFx_i)
    ={}& \BFr(\BFx_i)^{\intercal}\BFR^{-1}\BFr(\BFx_i) - \BFr(\BFx_i)^{\intercal}(\BFR+n\lambda \BFI )^{-1}\BFr(\BFx_i)\nonumber\\
    ={}& \BFr(\BFx_i)^{\intercal}\BFR^{-1}[(\BFR+n\lambda) \BFI  - \BFR](\BFR+n\lambda \BFI )^{-1}\BFr(\BFx_i)\nonumber\\
    ={}& n\lambda \BFe_i ^{\intercal}(\BFR+n\lambda \BFI )^{-1}\BFr(\BFx_i)\nonumber\\
    \leq{}& n\lambda\sqrt{\BFe_i^{\intercal}\BFe_i}\sqrt{\BFr(\BFx_i)^{\intercal}(\BFR+n\lambda \BFI )^{-2}\BFr(\BFx_i)}\nonumber\\
    \leq{}& n\lambda\sqrt{\BFr(\BFx_i)^{\intercal}\BFR^{-2}\BFr(\BFx_i)} = n\lambda,
\end{align*}
where the first inequality follows from the Cauchy--Schwarz inequality.
Hence,
\begin{align*}%
    |M_1(\BFx_i)|^2 \leq (\Psi(\BFzero)-\BFr(\BFx_i)^\intercal(\BFR+n\lambda \BFI )^{-1}\BFr(\BFx_i))\|f\|_{\mathscr{N}_\Psi(\Omega)}^2
    \leq n\lambda \|f\|_{\mathscr{N}_{\Psi}(\Omega)}^2.
\end{align*}
Because the above bound is uniform for all $\BFx_i$, we have
\begin{equation}\label{eq:thmnshsC3I51}
    \max_{1\leq i\leq n}|M_1(\BFx_i)| = O_{\pr}\left((n\lambda)^{1/2}\right).
\end{equation}
Therefore, combining \eqref{eq:thmnshsC3fhatinf} and \eqref{eq:thmnshsC3I51} completes the proof.
\Halmos \endproof

\begin{lemma}\label{lemma:M_2}
Suppose $f\in\mathscr{N}_\Psi(\Omega)$ and
Assumptions~\ref{assump:subG} and \ref{assump:support} hold.
Then,
\[
\max_{1\leq i\leq n}|M_2(\BFx_i)|  = O_{\pr}\left(  \bigl(r_n^{1/2}\wedge 1\bigr) m^{-1/2}  (\log n)^{1/2} \right),
\]
where $r_n = \lambda^{-1}n^{-\frac{2\nu+d}{\nu+d}}+n^{-1}\lambda^{-\frac{d}{2\nu+d}}$.
\end{lemma}
\proof{Proof of Lemma~\ref{lemma:M_2}.}

We begin with bounding $\Var(M_2(\BFx_i))$.
We do it in two different ways and take the smaller one as the upper bound.

\textbf{First way to bound $\Var(M_2(\BFx_i))$.}
Note that
\begin{align}\label{eq:rhonvarb12}
    \Var(M_2(\BFx_i)) \leq \frac{\sigma^2}{m} \BFr(\BFx_i)^\intercal(\BFR+n\lambda \BFI )^{-2}\BFr(\BFx_i).
\end{align}
Fix $\BFx$. Consider the quadratic function
\begin{align*}
    \mathcal{P}(\BFu) = \Psi(\BFzero) - 2\sum_{i=1}^n\Psi(\BFx-\BFx_i)u_i + \sum_{i,j=1}^n u_iu_j \Psi(\BFx_i-\BFx_j) + n\lambda\|\BFu\|_2^2,
\end{align*}
for $\BFu=(u_1,...,u_n)\in \RR^n$. Clearly, $\BFu_*\coloneqq(\BFR+n\lambda \BFI )^{-1}\BFr(\BFx)$ minimizes $\mathcal{P}(\BFu)$. Since $\Psi$ is positive definite,
\begin{align*}
    \Psi(\BFzero) - 2\sum_{i=1}^n\Psi(\BFx-\BFx_i)u_i + \sum_{i,j=1}^n u_iu_j\Psi(\BFx_i-\BFx_j) \geq 0
\end{align*}
for all $\BFu=(u_1,...,u_n)\in \RR^n$, which implies
\begin{align}\label{eq:upofmalP}
    \mathcal{P}(\BFu_*)\geq{}n\lambda\|\BFu_*\|_2^2=n\lambda\BFr(\BFx)^\intercal(\BFR+n\lambda \BFI )^{-2}\BFr(\BFx).
\end{align}
Direct calculation shows
\[
\mathcal{P}(\BFu_*) = \Psi(\BFzero)-\BFr(\BFx)^\intercal(\BFR+n\lambda \BFI )^{-1}\BFr(\BFx).
\]
In order to obtain an upper bound of $\mathcal{P}(\BFu_*)$, we follow the idea from the proof of Lemma~F.8 in \cite{wang2020inference_ec}.

For a fixed $\BFx$, define $h(\BFt)= \Psi(\BFx-\BFt)$. Let $h_\dagger(\BFt)=\BFr(\BFt)^\intercal(\BFR+n\lambda \BFI )^{-1}\BFr(\BFx)$. Clearly,
\begin{align}\label{eq:rhonupofh1}
    \Psi(\BFzero)-\BFr(\BFx)^\intercal(\BFR+n\lambda \BFI )^{-1}\BFr(\BFx) \leq \|h-h_\dagger\|_{\mathscr{L}_\infty(\Omega)}.
\end{align}
By Lemma~\ref{lemma:representation-thm}, it can be checked that $h_{\dagger}$ is the solution to the optimization problem:
\begin{align}\label{eq:thm1opt1hh}
    h_{\dagger}= \argmin_{g\in \mathscr{N}_\Psi(\Omega)} \frac{1}{n}\sum_{i=1}^n (g(\BFx_i) - h(\BFx_i))^2 + \lambda \|g\|_{\mathscr{N}_\Psi(\Omega)}^2.
\end{align}
Note that $\|h-h_{\dagger}\|_{\mathscr{N}_\Psi(\Omega)}^2$ can be bounded by
\begin{align}\label{eq:opt1errorH21h}
    \|h-h_{\dagger}\|_{\mathscr{N}_\Psi(\Omega)}^2 ={}& \Psi(\BFx-\BFx) - 2\BFr(\BFx)^\intercal(\BFR+n\lambda \BFI )^{-1}\BFr(\BFx) + \BFr(\BFx)^\intercal(\BFR+n\lambda \BFI )^{-1}\BFR(\BFR+n\lambda \BFI )^{-1}\BFr(\BFx)\nonumber\\
    \leq{}& \Psi(\BFzero) - \BFr(\BFx)^\intercal(\BFR+n\lambda \BFI )^{-1}\BFr(\BFx)\nonumber\\
    \leq{}&\|h-h_{\dagger}\|_{\mathscr{L}_\infty(\Omega)}.
\end{align}
An upper bound on $\|h-h_{\dagger}\|_n$ can be obtained by
\begin{align*}%
    \|h-h_{\dagger}\|_n^2 ={} & \|h-h_{\dagger}\|_n^2 + \lambda \|h_{\dagger}\|_{\mathscr{N}_\Psi(\Omega)}^2 - \lambda \|h_{\dagger}\|_{\mathscr{N}_\Psi(\Omega)}^2\nonumber\\
    \leq{} & \|h-h\|_n^2 + \lambda \|h\|_{\mathscr{N}_\Psi(\Omega)}^2 - \lambda \BFr(\BFx)^\intercal(\BFR+n\lambda \BFI )^{-1}\BFR(\BFR+n\lambda \BFI )^{-1}\BFr(\BFx)\nonumber\\
    ={} & \lambda(\Psi(\BFx-\BFx)-\BFr(\BFx)^\intercal(\BFR+n\lambda \BFI )^{-1}\BFR(\BFR+n\lambda \BFI )^{-1}\BFr(\BFx))\nonumber\\
    ={} & \lambda(\Psi(\BFx-\BFx)-\BFr(\BFx)^\intercal(\BFR+n\lambda \BFI )^{-1}\BFr(\BFx)\nonumber\\
    & + \BFr(\BFx)^\intercal(\BFR+n\lambda \BFI )^{-1}\BFr(\BFx)-\BFr(\BFx)^\intercal(\BFR+n\lambda \BFI )^{-1}\BFR(\BFR+n\lambda \BFI )^{-1}\BFr(\BFx))\nonumber\\
    ={} & \lambda(\Psi(\BFx-\BFx)-\BFr(\BFx)^\intercal(\BFR+n\lambda \BFI )^{-1}\BFr(\BFx) + n\lambda\BFr(\BFx)^\intercal(\BFR+n\lambda \BFI )^{-2}\BFr(\BFx))\nonumber\\
    \leq{} & 2\lambda(\Psi(\BFx-\BFx)-\BFr(\BFx)^\intercal(\BFR+n\lambda \BFI )^{-1}\BFr(\BFx))\nonumber\\
    \leq{}&2\lambda\|h-h_\dagger\|_{\mathscr{L}_\infty(\Omega)},
\end{align*}
where the first inequality is because $h_{\dagger}$ is the solution to the optimization problem \eqref{eq:thm1opt1hh}; the second inequality is by \eqref{eq:upofmalP}; the last inequality is by \eqref{eq:rhonupofh1}. Thus,
\begin{align}\label{eq:opt1errorh1}
     \|h-h_{\dagger}\|_n=O_{\pr}(\lambda^{1/2}\|h-h_\dagger\|_{\mathscr{L}_\infty(\Omega)}^{1/2}).
\end{align}
Because of the norm equivalence between $\mathscr{N}_\Psi(\Omega)$ and  $\mathscr{H}^{\nu+d/2}(\Omega)$,
there exists a constant $C_3$ such that
\begin{align}\label{eq:rhonopt1errhX}
    \|h-h_{\dagger}\|_{\mathscr{H}^{\nu+d/2}(\Omega)}\leq C_3\|h-h_{\dagger}\|_{\mathscr{N}_\Psi(\Omega)}\leq C_3\|h-h_{\dagger}\|_{\mathscr{L}_\infty(\Omega)}^{1/2},
\end{align}
where the last inequality is because of \eqref{eq:opt1errorH21h}.
Let $g_1 = \frac{h-h_{\dagger}}{C_3\|h-h_{\dagger}\|_{\mathscr{L}_\infty(\Omega)}^{1/2}}$. Thus, $\|g_1\|_{\mathscr{H}^{\nu+d/2}(\Omega)}\leq 1$.
It follows from Lemma~\ref{corogeer} that
\[
\|g_1\|_{\mathscr{L}_2(\Omega)}=O_{\pr}\left(n^{-\frac{\nu + d/2}{2\nu + 2d}} + \|g_1\|_n\right),
\]
which, by \eqref{eq:opt1errorh1}, implies  that
\begin{align}
    \|h-h_{\dagger}\|_{\mathscr{L}_2(\Omega)} ={}& O_{\pr}\left(n^{-\frac{\nu + d/2}{2\nu + 2d}}\|h-h_{\dagger}\|_{\mathscr{L}_\infty(\Omega)}^{1/2} + \|h-h_{\dagger}\|_n\right) \nonumber \\
    ={}&  O_{\pr}\left(n^{-\frac{\nu + d/2}{2\nu + 2d}}\|h-h_{\dagger}\|_{\mathscr{L}_\infty(\Omega)}^{1/2} + \lambda^{1/2}\|h-h_{\dagger}\|_{\mathscr{L}_\infty(\Omega)}^{1/2}\right).\label{eq:opt1errorL21h}
\end{align}
By Lemma~\ref{lemmaGNinter}, it follows from \eqref{eq:opt1errorH21h} that
\begin{align*}%
    \|h-h_{\dagger}\|_{\mathscr{L}_\infty(\Omega)}\leq{} & C_4 \|h-h_{\dagger}\|_{\mathscr{L}_2(\Omega)}^{1-\frac{d}{2\nu+d}}\|h-h_{\dagger}\|_{\mathscr{N}_\Psi(\Omega)}^{\frac{d}{2\nu+d}}\leq C_4 \|h-h_{\dagger}\|_{\mathscr{L}_2(\Omega)}^{1-\frac{d}{2\nu+d}}\|h-h_{\dagger}\|_{\mathscr{L}_\infty(\Omega)}^{\frac{d}{2(2\nu+d)}}\nonumber\\
    ={}&\|h-h_{\dagger}\|_{\mathscr{L}_\infty(\Omega)}^{1/2}O_{\pr}\left(\left(n^{-\frac{\nu + d/2}{2\nu + 2d}} + \lambda^{1/2}\right)^{\frac{2\nu}{2\nu+d}}\right)
\end{align*}
which implies
\begin{align}\label{eq:rhonhdinf2}
    \|h-h_{\dagger}\|_{\mathscr{L}_\infty(\Omega)}=O_{\pr}\left(n^{-\frac{\nu}{\nu + d}} + \lambda^{\frac{2\nu}{2\nu+d}}\right).
\end{align}
Recall that $\mathcal{P}(u_*)\leq \|h-h_{\dagger}\|_{\mathscr{L}_\infty(\Omega)}$, which, together with \eqref{eq:upofmalP} and \eqref{eq:rhonhdinf2}, leads to
\begin{align*}
    \BFr(\BFx)^\intercal(\BFR+n\lambda \BFI )^{-2}\BFr(\BFx) ={}& O_{\pr}\left((n\lambda)^{-1}n^{-\frac{\nu}{\nu + d}}+(n\lambda)^{-1}\lambda^{\frac{2\nu}{2\nu+d}}\right)\nonumber\\
    ={}&O_{\pr}\left(\lambda^{-1}n^{-\frac{2\nu+d}{\nu+d}}+n^{-1}\lambda^{-\frac{d}{2\nu+d}}\right),
\end{align*}
which, together with \eqref{eq:rhonvarb12}, implies
\begin{equation}\label{eq:rhonvarb2}
    \Var(M_2(\BFx_i)) = O_{\pr}\left(\lambda^{-1}n^{-\frac{2\nu+d}{\nu+d}}+n^{-1}\lambda^{-\frac{d}{2\nu+d}}\right).
\end{equation}

\textbf{Second way to bound $\Var(M_2(\BFx_i))$.}
With an argument similar to \eqref{eq:thmsmoothI4var}, we have
\begin{equation}\label{eq:rhonvarb1}
    \Var(M_2(\BFx_i))  \leq \frac{\sigma^2}{m} \BFr(\BFx_i)^\intercal(\BFR+n\lambda \BFI )^{-2}\BFr(\BFx_i)
   \leq \frac{\sigma^2}{m} \BFr(\BFx_i)^{\intercal}\BFR^{-2}\BFr(\BFx_i)
   = \frac{\sigma^2}{m}.
\end{equation}

Let $r_n = \lambda^{-1}n^{-\frac{2\nu+d}{\nu+d}}+n^{-1}\lambda^{-\frac{d}{2\nu+d}}$ and $s_n=r_n\wedge 1$.
Combining \eqref{eq:rhonvarb2} and \eqref{eq:rhonvarb1} yields
\begin{equation*}%
    \Var(M_2(\BFx_i)) = m^{-1} O_{\pr}(s_n).
\end{equation*}

Therefore, for any $\epsilon>0$,
there exist $M_\epsilon$ and $N_\epsilon$ such that $\pr(\Var(M_2(\BFx_i))>M_\epsilon m^{-1}s_n)<\epsilon$, when $n>N_\epsilon$. Take $N_0=\max\{N_\epsilon, \epsilon^{-1}\}$. Note that $M_2(\BFx_i)$ is sub-Gaussian by Assumption~\ref{assump:subG}.
It follows that for all $n\geq N_0$,
\begin{align*}
    & \pr\left(\max_{1\leq i\leq n} |M_2(\BFx_i)| > 2 \sqrt{  \sigma^2 M_\epsilon m^{-1}s_n \log n} \right) \\
    ={} & \pr\left(\max_{1\leq i\leq n} |M_2(\BFx_i)| > 2 \sqrt{  \sigma^2 M_\epsilon m^{-1}s_n \log n},\Var(M_2(\BFx_i))\leq M_\epsilon m^{-1}s_n \right)\\
    & + \pr\left(\max_{1\leq i\leq n} |M_2(\BFx_i)| > 2 \sqrt{  \sigma^2 M_\epsilon m^{-1}s_n \log n},\Var(M_2(\BFx_i))> M_\epsilon m^{-1}s_n \right)\\
    \leq {}& \pr\left(\mbox{There exists $i=1,\ldots,n$ such that } |M_2(\BFx_i)| > 2\sqrt{  \sigma^2 \Var(M_2(\BFx_i)) \log n}\right) + \epsilon \\
    \leq{}& \sum_{i=1}^n \pr\left(|M_2(\BFx_i)| > 2\sqrt{  \sigma^2 \Var(M_2(\BFx_i)) \log n}\right) +\epsilon \\
    \leq{}& \sum_{i=1}^n 2 \exp\left(- \frac{ 4\sigma^2 \log n}{2  \sigma^2 } \right) +\epsilon  =  \frac{2}{n} +\epsilon \leq 3\epsilon, %
\end{align*}
where the first inequality follows from the  union bound. Hence, \begin{equation*} %
  \max_{1\leq i\leq n}|M_2(\BFx_i)|  = O_{\pr}\left(  \bigl(r_n^{1/2}\wedge 1\bigr) m^{-1/2}  (\log n)^{1/2} \right). \Halmos
\end{equation*}
\endproof

\proof{Proof of Proposition~\ref{prop:rho_n}.}
Applying Lemmas~\ref{lemma:M_1} and \ref{lemma:M_2} to  \eqref{eq:thmnshsC3dbway2} leads to
\begin{equation} \label{eq:rhonM1b1}
    \rho_n=O_{\pr}\left(\bigl(n^{-\frac{\nu}{2\nu + 2d}} + \lambda^{\frac{\nu}{2\nu+d}}\bigr) \wedge (n\lambda)^{1/2} +  \bigl(r_n^{1/2}\wedge 1\bigr) m^{-1/2}  (\log n)^{1/2} \right).
\end{equation}

If we set $\lambda \asymp 1/(mn) $, then
\begin{align*}
    r_n =& mn^{-\frac{\nu}{\nu+d}}+n^{-\frac{2\nu}{2\nu+d}}m^{\frac{2\nu+d}{4\nu+d}} \leq 2mn^{-\frac{\nu}{\nu+d}}.
\end{align*}
Noting $r_n^{1/2}\wedge 1\leq r_n^{1/2}\wedge \sqrt{2}$, we have
\begin{align*} %
    \rho_n={}&O_{\pr}\left(\bigl(n^{-\frac{\nu}{2\nu + 2d}} + (mn)^{-\frac{\nu}{2\nu+d}} \bigr)  \wedge m^{-1/2}+ \bigl(r_n^{1/2}\wedge \sqrt{2} \bigr) m^{-1/2}  (\log n)^{1/2}  \right)\nonumber\\
    ={}&O_{\pr}\left(\bigl( n^{-\frac{\nu}{2\nu + 2d}} \wedge  m^{-1/2}\bigr)  +   \bigl(m^{1/2}n^{-\frac{\nu}{2\nu+2d}}\wedge 1\bigr) m^{-1/2}  (\log n)^{1/2}  \right) \\
    = {}&O_{\pr}\left(\bigl( n^{-\frac{\nu}{2\nu + 2d}} \wedge  m^{-1/2}\bigr) +  \bigl( n^{-\frac{\nu}{2\nu + 2d}} \wedge  m^{-1/2}\bigr) (\log n)^{1/2} \right) \\
    = {}& O_{\pr}\left(\bigl( n^{-\frac{\nu}{2\nu + 2d}} \wedge  m^{-1/2}\bigr)(\log n)^{1/2}  \right). \Halmos
\end{align*}
\endproof

\section{Proof of Theorem~\ref{thm:gnonsmooth}}\label{app:hsplusc4}

\proof{Proof of Theorem~\ref{thm:gnonsmooth}.}
Without loss of generality, we assume $z_0=0$ so that $\eta(z) = z^+$.
To handle the term $I_2$ in the decomposition \eqref{eq:thmsmootha},
we use a smooth approximation of $\eta$ as in \cite{HongJunejaLiu17}.
Let
\begin{align}\label{eq:rhodelhs}
    \eta_\delta(z) = \left(\frac{1}{2}(z+\delta)-\frac{\delta}{\pi}\cos\left(\frac{\pi}{2\delta}z\right)\right)\ind{\{-\delta\leq z\leq \delta\}}+z\ind{\{z\geq \delta\}},
\end{align}
where $\delta>0$ is a parameter to be determined later. It can be verified that
\begin{align}\label{eq:rhodelhsD}
    \eta_\delta'(z) ={}&  \left(\frac{1}{2}+\frac{1}{2}\sin\left(\frac{\pi}{2\delta}z\right)\right)\ind{\{-\delta\leq z\leq \delta\}}+\ind{\{z\geq \delta\}},\nonumber\\
    \eta_\delta''(z) ={}&  \frac{\pi}{4\delta}\cos\left(\frac{\pi}{2\delta}z\right)\ind{\{-\delta\leq z\leq \delta\}}.
\end{align}
Moreover, $|\eta_\delta'(z)|\leq 1$, $|\eta_\delta''(z)|\leq \frac{\pi}{4\delta}$, and there exists a constant $C>0$ such that
\begin{align}\label{eq:apprhohsb}
\left\{
\begin{array}{ll}
     |\eta(z)-\eta_\delta(z)|  \leq C\delta,   &\quad \mbox{ if }  z \in[-\delta,   \delta], \\[0.5ex]
    |\eta(z)-\eta_\delta(z)|    = 0, &\quad  \mbox{ otherwise}.
\end{array}
\right.
\end{align}

By the triangle inequality,  we have
\begin{equation}
\label{eq:thmnshsXa}
   I_2
    \leq
    \underbrace{\left| \frac{1}{n}\sum_{i=1}^n \bigl[ \eta(f(\BFx_i))-  \eta_\delta(f(\BFx_i)) \bigr]\right|}_{J_1}     + \underbrace{\left|\frac{1}{n}\sum_{i=1}^n \bigl[\eta_\delta(f(\BFx_i)) -   \eta_\delta(\hat{f}(\BFx_i)) \bigr] \right|}_{J_2}   + \underbrace{ \left|\frac{1}{n}\sum_{i=1}^n \bigl[ \eta_\delta(\hat{f}(\BFx_i))-  \eta(\hat{f}(\BFx_i))\bigr]\right|}_{J_3}.
\end{equation}

\textbf{Bound for $J_1$.}
It follows from \eqref{eq:apprhohsb} that
\begin{align}\label{eq:thmnshsC3I2}
    J_1 \leq \left| \frac{C}{n}\sum_{i=1}^n \delta \ind{\{f(\BFx_i)\in [-\delta,\delta]\}}\right|
    \leq{}& C \delta \left| \frac{1}{n}\sum_{i=1}^n \ind{\{f(\BFx_i)\in [-\delta,\delta]\}} - \E \bigl[\ind{\{f(\BFx_i)\in [-\delta,\delta]\}}\bigr]\right| \nonumber \\
    & + C \delta \E \ind{\{f(\BFx_i)\in [-\delta,\delta]\}}
    =  O_{\pr}(\delta n^{-1/2}) + O_{\pr}(\delta^{\alpha+1}),
\end{align}
for some constant $C>0$,
where the last step follows from the central limit theorem and the fact that
$\E \ind{\{f(\BFx_i)\in [-\delta,\delta]\}} = \pr(|f(\BFx_i)|\leq \delta))=O(\delta^\alpha)$ by Assumption~\ref{assump:margin}.

\textbf{Bound for $J_2$.}
Applying Taylor's expansion to $\eta_\delta$, we have
\begin{align}\label{eq:eq:thmnshsC3I3}
    J_2 ={}& \underbrace{\left|\frac{1}{n}\sum_{i=1}^n \eta_\delta'(f(\BFx_i))(f(\BFx_i)-\hat{f}(\BFx_i))\right|}_{J_{21}}
    + \underbrace{\left| \frac{1}{2n}\sum_{i=1}^n \eta_\delta''(\tilde{z}_i)(f(\BFx_i)-\hat{f}(\BFx_i))^2 \right|}_{J_{22}},
\end{align}
where $\tilde{z}_i$ is a value between $f(\BFx_i)$ and $\hat{f}(\BFx_i)$.

The first term $J_{21}$ can be bounded using Proposition~\ref{prop:rate_l1norm}, which gives us
\begin{align}\label{eq:thmnshsC3I31}
    J_{21} = O_{\pr}(\lambda^{1/2} + (mn)^{-1/2}).
\end{align}

For $J_{22}$,
by \eqref{eq:rhodelhsD}, we find that
\begin{align}\label{eq:thmnshsC3I32}
    J_{22} ={}& \left| \frac{1}{2n}\sum_{i=1}^n \frac{\pi}{4\delta}\cos\left(\frac{\tilde{z}_i}{2\delta}\pi\right)\ind{\{-\delta\leq \tilde{z}_i\leq \delta\}}(f(\BFx_i)-\hat{f}(\BFx_i))^2 \right| \nonumber\\
    \leq{}& \frac{\pi}{8\delta}\left| \frac{1}{n}\sum_{i=1}^n \ind{\{-\delta\leq \tilde{z}_i\leq \delta\}}(f(\BFx_i)-\hat{f}(\BFx_i))^2 \right|.
\end{align}
Since $\tilde{z}_i$ is a value between $f(\BFx_i)$ and $\hat{f}(\BFx_i)$, $\tilde{z}_i\in [-\delta,\delta]$ implies that $f(\BFx_i)\in [-\delta-\rho_n,\delta+\rho_n]$, where $\rho_n = \max_{1\leq i\leq n}|f(\BFx_i) - \hat{f}(\BFx_i)|$.
By \eqref{eq:thmnshsC3I32}, $J_{22}$ can be further bounded by
\begin{align}\label{eq:thmnshsC3I321}
    J_{22} ={}& O_{\pr}\left(\frac{\pi}{8\delta}\left| \frac{1}{n}\sum_{i=1}^n \ind{\{-\delta-\rho_n \leq f(\BFx_i)\leq \delta+\rho_n\}}(f(\BFx_i)-\hat{f}(\BFx_i))^2 \right|\right) \nonumber\\
    ={}& O_{\pr}\left(\frac{1}{\delta}\left| \frac{\rho_n^2}{n}\sum_{i=1}^n \ind{\{-\delta-\rho_n \leq f(\BFx_i)\leq \delta+\rho_n\}}\right|\right)\nonumber\\
    ={}& O_{\pr}\left(\frac{\rho_n^2}{\delta}(n^{-1/2} + (\delta+\rho_n)^\alpha)\right),
\end{align}
where the last step can be shown similarly for \eqref{eq:thmnshsC3I2}.
Plugging \eqref{eq:thmnshsC3I31} and \eqref{eq:thmnshsC3I321} in \eqref{eq:eq:thmnshsC3I3}, we have
\begin{align}\label{eq:thmnshsC3I3}
    J_2 = O_{\pr}\left(\lambda^{1/2} + (mn)^{-1/2}+\frac{\rho_n^2}{\delta}n^{-1/2} + \frac{\rho_n^2}{\delta}(\delta+\rho_n)^\alpha\right).
\end{align}

\textbf{Bound for $J_3$.}
Following the same argument for \eqref{eq:thmnshsC3I2}, it can be seen that
\begin{align}\label{eq:thmnshsC3I4}
    J_3 \leq \left|\frac{1}{n}\sum_{i=1}^n \delta\ind{\{-\delta\leq \hat{f}(\BFx_i)\leq \delta\}}\right|\leq \left|\frac{1}{n}\sum_{i=1}^n \delta\ind{\{-\delta-\rho_n\leq f(\BFx_i)\leq \delta+\rho_n\}}\right| = O_{\pr}(\delta n^{-1/2}+\delta(\delta+\rho_n)^\alpha).
\end{align}
In \eqref{eq:thmnshsC3I4}, the first inequality holds because \eqref{eq:apprhohsb},
and the second inequality holds because $\hat{f}(\BFx_i)\in [-\delta,\delta]$ implies $f(\BFx_i)\in [-\delta-\rho_n,\delta+\rho_n]$.

\textbf{Putting the bounds together.}
Combining \eqref{eq:thmsmootha}, \eqref{eq:thmsmoothI1}, \eqref{eq:thmnshsXa},  \eqref{eq:thmnshsC3I2}, \eqref{eq:thmnshsC3I3}, and \eqref{eq:thmnshsC3I4},
we have
\begin{align}\label{eq:thmnshsXaX}
    |\hat \theta_{n,m}-\theta| ={}& O_{\pr}\left(n^{-1/2} +\delta n^{-1/2}+\delta^{\alpha+1} + \frac{\rho_n^2}{\delta}n^{-1/2} + \frac{\rho_n^2}{\delta}(\delta+\rho_n)^\alpha + \lambda^{1/2} + (mn)^{-1/2} +\delta(\delta+\rho_n)^\alpha\right)\nonumber\\
    ={}& O_{\pr}\left(n^{-1/2} +\delta^{\alpha+1} + \frac{\rho_n^2}{\delta}n^{-1/2} +\frac{\rho_n^2}{\delta}(\delta+\rho_n)^\alpha + \lambda^{1/2}  +\delta\rho_n^\alpha\right).
\end{align}
Then, by setting $\delta=\rho_n$ and $\lambda \asymp 1/(mn) = \Gamma^{-1}$,
\eqref{eq:thmnshsXaX} becomes
$
     |\hat \theta_{n,m} - \theta | =  O_{\pr}\left(n^{-1/2} + \rho_n^{\alpha+1}\right).
$

Lastly, we apply Proposition~\ref{prop:rho_n} to conclude that: if $\nu \geq \frac{d}{\alpha+1}$, we set $n=\Gamma$ to obtain the rate $\max\{\Gamma^{-1/2}, \Gamma^{-\frac{\nu(\alpha+1)}{2\nu+2d}}(\log \Gamma)^{\frac{\alpha+1}{2}}\}$;
if $\nu < \frac{d}{\alpha+1}$, we set $n\asymp \Gamma^{\frac{\alpha+1}{\alpha+2}}$ to obtain the rate $\Gamma^{-\frac{\alpha+1}{2(\alpha+2)}}(\log \Gamma)^{\frac{\alpha+1}{2}}$.
\Halmos \endproof

\section{Proof of Theorem~\ref{thm:gnonsmoothin}}\label{app:ind}

\proof{Proof of Theorem~\ref{thm:gnonsmoothin}.}
Without loss of generality, we assume $z_0=0$ so that $\eta(z) = \ind{\{z\geq 0\}}$.
Again, we work on the decomposition \eqref{eq:thmsmootha}.
For the term $I_2$,
note that if $\eta(f(\BFx_i))\neq \eta(\hat{f}(\BFx_i))$, then we must have $f(\BFx_i)\in [-\rho_n,\rho_n]$, where $\rho_n = \max_{1\leq i\leq n}|f(\BFx_i) - \hat{f}(\BFx_i)|$. Therefore,
\begin{align}\label{eq:pfgdindapI2}
    I_2 ={} & \left|\frac{1}{n}\sum_{i=1}^n \left(\eta(f(\BFx_i))-\eta(\hat{f}(\BFx_i))\right)\ind{\{f(\BFx_i)\in [-\rho_n,\rho_n]\}}\right|\leq \frac{1}{n}\sum_{i=1}^n \ind{\{f(\BFx_i)\in [-\rho_n,\rho_n]\}}\nonumber\\
    = {}& O_{\pr}(n^{-1/2}+\rho_n^\alpha),
\end{align}
where the inequality holds because $|\eta(a)-\eta(b)|\leq 1$ for all $a,b\in\RR$, and the last step from \eqref{eq:thmnshsC3I2}.

Combining \eqref{eq:thmsmootha}, \eqref{eq:thmsmoothI1}, and \eqref{eq:pfgdindapI2}, we conclude that
$
    |\hat \theta_{n,m}-\theta| = O_{\pr}(n^{-1/2}+\rho_n^\alpha).
$
Lastly, we apply Proposition~\ref{prop:rho_n} to conclude that: if $\nu \geq \frac{d}{\alpha}$, we set $n=\Gamma$ to obtain the rate $\max\{\Gamma^{-1/2}, \Gamma^{-\frac{\nu\alpha}{2\nu+2d}}(\log \Gamma)^{\frac{\alpha}{2}}\} = \Gamma^{-\frac{\nu\alpha}{2\nu+2d}}(\log \Gamma)^{\frac{\alpha}{2}}$ because $\alpha\leq 1$;
if $\nu < \frac{d}{\alpha}$, we set $n\asymp \Gamma^{\frac{\alpha}{\alpha+1}}$ to obtain the rate $\Gamma^{-\frac{\alpha}{2(\alpha+1)}}(\log \Gamma)^{\frac{\alpha}{2}}$.
\Halmos \endproof

\section{Proof of Theorem~\ref{thm:gvarcvar}}\label{app:pfgvar}

\begin{lemma}[Refined Hoeffding's Inequality for Bernoulli Random Variables]\label{lemma:Hoeffding-Bernoulli}
Let $Z_1,\ldots,Z_n$ be independent Bernoulli random variables with parameter $p\in(0,1)$ such that $\pr(Z_i = 1) = 1- \pr(Z_i= 0) = p$, $i=1,\ldots,n$. Then,
\[
\pr\biggl(\biggl|\frac{1}{n}\sum_{i=1}^n (Z_i-p)\biggr| \geq t\biggr)\leq 2\exp\biggl(-\frac{n t^2}{2 p  }\biggr),\quad \forall t > 0.
\]
\end{lemma}

\proof{Proof of Lemma~\ref{lemma:Hoeffding-Bernoulli}.}
Note that $\E[Z_i] = p$ and  for any $s\in\RR$,
\begin{align*}
    \log\E[\exp(s (Z_i - p))] ={}& \log \left( p e^{s(1-p)}  + (1-p) e^{-s p} \right)
    = -sp + \log \left(1 + (e^s -1 )p\right) \\
    \leq{}& -sp + (e^s -1)p
    \leq  -sp + \left(s + \frac{s^2}{2}\right) p
    =\frac{s^2 p}{2},
\end{align*}
where the first inequality holds because
$\log(1+x) \leq x$ for all $x>-1$ and $(e^s -1 )p > -1$ for all $s\in\RR$,
and the second inequality holds because $e^x \geq 1 + x + \frac{x^2}{2}$ for all $x\in\RR$. Hence,
$Z_i \sim \mathsf{subG}(p)$ by definition.
The proof is completed by
applying Hoeffding’s inequality for sub-Gaussian random variables \citep[Proposition 2.5]{Wainwright19_ec}.
\Halmos\endproof

\proof{Proof of Theorem~\ref{thm:gvarcvar}: The Case of VaR.}
Let $\zeta_{\VaR} \coloneqq \VaR_\tau(f(X))$
and $\hat{\zeta} \coloneqq  \hat{f}_{(\lceil \tau n\rceil)}$ be its KRR-driven estimator.
Moreover, let $\mathtt{G}(z)$, $\mathtt{G}_n(z)$, and $\hat{\mathtt{G}}_n(z)$ denote the cumulative distribution function (CDF) of $f(X)$, the empirical CDF of $f(X)$, and the empirical CDF of $\hat{f}(X)$, respectively:
\begin{align*} %
    \mathtt{G}(z) \coloneqq \pr(f(X)\leq z),\quad
    \mathtt{G}_n(z) \coloneqq \frac{1}{n}\sum_{i=1}^n \ind{\{f(\BFx_i)\leq z\}},\qq{and}
    \hat{\mathtt{G}}_n(z)  \coloneqq  \frac{1}{n}\sum_{i=1}^n \ind{\{\hat{f}(\BFx_i)\leq z\}}.
\end{align*}
Then, $\mathtt{G}(\zeta_{\VaR}) = \tau$ and $\hat{\mathtt{G}}_n(\hat{\zeta}) = \frac{\lceil \tau n\rceil}{n}$.

Note that
\begin{align*}
    |\mathtt{G}(\hat{\zeta}) - \mathtt{G}(\zeta_{\VaR})|
    ={}& \left|\E\bigl[ \ind{\{f(X)\leq \hat{\zeta}\}}\bigr] - \E\bigl[ \ind{\{f(X)\leq \zeta_{\VaR}\}}\bigr]\right| \\
    ={}&  \E\bigl[ \ind{\{f(X)\in [\min(\hat{\zeta},\zeta_{\VaR}),\max(\hat{\zeta},\zeta_{\VaR})]\}}\bigr] \\
    ={}& \pr\left(\left|f(X) - \frac{(\hat{\zeta}+\zeta_{\VaR})}{2}\right|\leq \frac{|\hat{\zeta}-\zeta_{\VaR}|}{2}\right)  \\
    \geq{}&  C_2|\hat{\zeta}-\zeta_{\VaR}|^{\gamma},
\end{align*}
for some constant $C_2>0$, where the last step follows from  Assumption~\ref{assump:margin-strong}.
Thus,
\begin{equation}\label{eq:pfgvar3}
    |\hat{\zeta}-\zeta_{\VaR}| = O\left(|\mathtt{G}(\hat{\zeta}) - \mathtt{G}(\zeta_{\VaR})|^{1/\gamma}\right).
\end{equation}

Next, we bound $|\mathtt{G}(\hat{\zeta}) - \mathtt{G}(\zeta_{\VaR})|$.
Note that
\begin{align} \label{eq:pfgvar1}
    |\mathtt{G}(\hat{\zeta}) - \mathtt{G}(\zeta_{\VaR})|
    \leq{}& \underbrace{|\mathtt{G}(\hat{\zeta}) - \mathtt{G}_n(\hat{\zeta})| }_{V_1}
    + \underbrace{| \mathtt{G}_n(\hat{\zeta}) - \hat{\mathtt{G}}_n(\hat{\zeta})| }_{V_2}
    + \underbrace{| \hat{\mathtt{G}}_n(\hat{\zeta}) - \mathtt{G}(\zeta_{\VaR})|}_{V_3}.
\end{align}

Let $\rho_n = \max_{1\leq i\leq n}|f(\BFx_i) - \hat{f}(\BFx_i)|$.
Because $\mathtt{G}(-\|f\|_{\mathscr{L}_\infty}) = 0$ and $\mathtt{G}(\|f\|_{\mathscr{L}_\infty}) = 1$, and $|\hat{\zeta}| = |\hat{f}_{(\lceil \tau n\rceil)}|\leq \|f\|_{\mathscr{L}_\infty} + \rho_n$ with $\rho_n\rightarrow 0$ in probability by Proposition~\ref{prop:rho_n}, we can focus on the case $z\in [-2\|f\|_{\mathscr{L}_\infty},2\|f\|_{\mathscr{L}_\infty}]$ when analyzing $V_i$, $i=1,2,3$.
For simplicity, define $\mathcal{I}\coloneqq [-2\|f\|_{\mathscr{L}_\infty},2\|f\|_{\mathscr{L}_\infty}]$.

\textbf{Bound for $V_1$.}
By the Dvoretzky--Kiefer--Wolfowitz inequality \citep{Massart90_ec},
\begin{align*}
    \pr\biggl(\sup_{z\in
    \mathcal{I}
    }|\mathtt{G}_n(z)-\mathtt{G}(z)|> t\biggr)
    \leq \pr\biggl(\sup_{z\in \RR
    }|\mathtt{G}_n(z)-\mathtt{G}(z)|> t\biggr)
    \leq 2e^{-2nt^2}, \quad \forall t>0.
\end{align*}
This yields $\|\mathtt{G}-\mathtt{G}_n\|_{\mathscr{L}_\infty(
\mathcal{I}
)} = O_{\pr}(n^{-1/2})$.
Hence,
\begin{align}\label{eq:V1}
    V_1 = O_{\pr} (\|\mathtt{G}-\mathtt{G}_n\|_{\mathscr{L}_\infty(
    \mathcal{I}
    )}) = O_{\pr}(n^{-1/2}).
\end{align}

\textbf{Bound for $V_2$.}
Let $\rho_n = \max_{1\leq i\leq n}|f(\BFx_i) - \hat{f}(\BFx_i)|$.
Note that
\begin{align}\label{eq:pfgvardifFFn1}
    |\mathtt{G}_n(z)-\hat{\mathtt{G}}_n(z)| ={}& \left|\frac{1}{n}\sum_{i=1}^n (\ind{\{f(\BFx_i)\leq z\}} - \ind{\{\hat{f}(\BFx_i)\leq z\}})\right| \nonumber \\
    \leq{}& \frac{1}{n}\sum_{i=1}^n \ind{\{f(\BFx_i)\in [z-\rho_n,z+\rho_n]\}}  \nonumber \\
    ={}&  O_{\pr} \left( \frac{1}{n}\sum_{i=1}^n \ind{\{f(\BFx_i)\in [z-l_n,z+l_n]\}} \right),
\end{align}
where the last step follows from \eqref{eq:rhonM1b1} and
\begin{equation}\label{eq:l_n_def}
l_n = \bigl( n^{-\frac{\nu}{2\nu + 2d}} + \lambda^{\frac{\nu}{2\nu+d}} \bigr) \wedge (n\lambda)^{1/2} + \bigl(r_n^{1/2}\wedge 1\bigr) m^{-1/2}  (\log n)^{1/2},
\end{equation}
with $r_n = \lambda^{-1}n^{-\frac{2\nu+d}{\nu+d}}+n^{-1}\lambda^{-\frac{2\nu+d}{4\nu+d}}$.

Let $\delta_n = \max\{n^{-1/(2\beta)}, l_n\}$ and
$M_n=\lceil 4\|f\|_{\mathscr{L}_\infty}/\delta_n\rceil$.
Consider a partition of
$\mathcal{I}$
as follows:
let $-2\|f\|_{\mathscr{L}_\infty}=\zeta_0< \zeta_1 < \cdots < \zeta_{M_n}=2\|f\|_{\mathscr{L}_\infty}$ with $\zeta_{j+1}-\zeta_j = 4\|f\|_{\mathscr{L}_\infty}/M_n$, $j=0,\ldots,M_n-1$.
Clearly, $\zeta_{j+1}-\zeta_j\leq \delta_n$.

For any
$z\in\mathcal{I}$,
we take $j_* = \argmin_{0\leq j\leq M_n} |z-\zeta_j|$ and note that  $|z-\zeta_{j_*(z)}|\leq \delta_n$. Thus,
\begin{align}
   &\frac{1}{n}\sum_{i=1}^n \ind{\{f(\BFx_i)\in [z-l_n,z+l_n]\}}  \nonumber \\
   \leq{}& \left|\frac{1}{n}\sum_{i=1}^n (\ind{\{f(\BFx_i)\in [z-l_n,z+l_n]\}}-\ind{\{f(\BFx_i)\in [\zeta_{j_*(z)}-l_n,\zeta_{j_*(z)}+l_n]\}})\right| \nonumber \\
   & + \frac{1}{n}\sum_{i=1}^n \ind{\{f(\BFx_i)\in [\zeta_{j_*(z)}-l_n,\zeta_{j_*(z)}+l_n]\}}\nonumber\\
    \leq{}& \frac{1}{n}\sum_{i=1}^n \ind{\{f(\BFx_i)\in [\zeta_{j_*(z)}-l_n-\delta_n,\zeta_{j_*(z)}+l_n +\delta_n]\}} + \frac{1}{n}\sum_{i=1}^n \ind{\{f(\BFx_i)\in [\zeta_{j_*(z)}-l_n,\zeta_{j_*(z)}+l_n]\}} \nonumber \\
     \leq{}& \frac{2}{n}\sum_{i=1}^n \ind{\{f(\BFx_i)\in [\zeta_{j_*(z)}-l_n-\delta_n,\zeta_{j_*(z)}+l_n +\delta_n]\}},
    \label{eq:pfgvardifFFn1'}
\end{align}
where the first inequality holds because of the triangle inequality, and the second because $|z-\zeta_{j_*(z)}|\leq \delta_n$.
It follows from \eqref{eq:pfgvardifFFn1} and \eqref{eq:pfgvardifFFn1'} that
\begin{align}
    \sup_{z\in
    \mathcal{I}
    } |\mathtt{G}_n(z)-\hat{\mathtt{G}}_n(z)|
    = O_{\pr}\biggl( \max_{0\leq j \leq M_n}  \underbrace{\frac{2}{n}\sum_{i=1}^n \ind{\{f(\BFx_i)\in [\zeta_j-l_n-\delta_n,\zeta_j+l_n +\delta_n]\}} }_{Q_{n,j}} \biggr). \label{eq:max_Q_nj}
\end{align}

Let $p_n = \pr\left( |f(X) - \zeta_j| \leq l_n +\delta_n \right)$.
Then, Assumption~\ref{assump:margin-strong} implies that
\begin{equation}\label{eq:p_n_bound}
    p_n \leq C_1 (l_n + \delta_n)^\beta \leq C_1 \left(l_n + (n^{-1/(2\beta)} + l_n)\right)^\beta
    = O\left(l_n^\beta + n^{-1/2}\right),
\end{equation}
for some constant $C_1>0$,
where second inequality follows from the definition of $\delta_n$.

By the definition of $l_n$ in \eqref{eq:l_n_def}, if $\nu \geq \frac{d}{\beta}$, by Proposition~\ref{prop:rho_n}, setting $n=\Gamma$ (so that $m=1$) yields
\[
l_n = O\left(n^{-\frac{\nu}{(2\nu+2d)}}(\log n)^{1/2} \right),
\]
which, together with \eqref{eq:p_n_bound}, implies that
\begin{equation}\label{eq:p_n_bound_1}
    p_n = O\left(n^{-\frac{\nu}{(2\nu+2d)}}(\log n)^{1/2} + n^{-1/2}\right)
    = O\left(n^{-\frac{\nu}{(2\nu+2d)}}(\log n)^{1/2}\right).
\end{equation}
Otherwise, if $\nu < \frac{d}{\beta}$, then setting $n\asymp \Gamma^{\frac{\beta}{\beta+1}}$ (so that $m\asymp \Gamma^{\frac{1}{\beta+1}} \asymp n^{1/\beta}$) yields
\begin{align*}
l_n = O\left(n^{-1/(2\beta)}(\log n)^{1/2}\right),
\end{align*}
which, together with \eqref{eq:p_n_bound}, implies that
\begin{equation}\label{eq:p_n_bound_2}
    p_n = O\left(n^{-1/2}(\log n)^{\beta/2} + n^{-1/2}\right)
    = O\left(n^{-1/2}(\log n)^{\beta/2}\right).
\end{equation}

Because $\BFx_1,\ldots,\BFx_n$ are i.i.d.,  we have
\begin{align}\label{eq:EQ_nj}
    \E Q_{n,j} = \frac{2}{n} \sum_{i=1}^n \E \ind{\{f(\BFx_i)\in [\zeta_j-l_n-\delta_n,\zeta_j+l_n +\delta_n]\}}  =  2 p_n.
\end{align}
Applying the union bound yields
\begin{align}
     & \pr\left(\max_{0\leq j\leq M_n} (Q_{n,j} - 2 p_n) > 2 n^{-1/2}\right)   \nonumber \\
    ={}& \pr\left(\mbox{There exists $j=0,\ldots,M_n$ such that } Q_{n,j} - p_n > 2 n^{-1/2} \right) \nonumber \\
    \leq{}& \sum_{0\leq j\leq M_n} \pr\left(Q_{n,j} - 2 p_n > 2 n^{-1/2}\right)  \nonumber \\
    = {}& \sum_{0\leq j\leq M_n} \pr\left(\frac{1}{n}\sum_{i=1}^n \left(\ind{\{f(\BFx_i)\in [\zeta_j-l_n-\delta_n,\zeta_j+l_n +\delta_n]\}} -  p_n \right)>  n^{-1/2}\right)  \nonumber \\
    \leq{}& 2(M_n+1) \exp\left(-\frac{1}{2 p_n}\right) \nonumber \\
    \leq{}& 2\left(
    4\|f\|_{\mathscr{L}_\infty}
    n^{1/(2\beta)} + 1\right) \exp\left(-\frac{1}{2 p_n}\right) \to 0, \label{eq:max_Q_nj'}
\end{align}
as $n\to\infty$,
where the second inequality follows from Lemma~\ref{lemma:Hoeffding-Bernoulli}, the third from the definitions of $M_n$ and $\delta_n$,
and the convergence to zero from \eqref{eq:p_n_bound_1} and \eqref{eq:p_n_bound_2}.

It follows from \eqref{eq:max_Q_nj} and \eqref{eq:max_Q_nj'} that
\begin{align*}
    \pr\left( \|\mathtt{G}_n-\hat{\mathtt{G}}_n\|_{\mathscr{L}_\infty(
    \mathcal{I}
    )}  > 2 p_n + 2 n^{-1/2}\right)
\leq{}& \pr\left( \max_{0\leq j\leq M_n} Q_j > 2 p_n + 2 n^{-1/2}\right)   \\
={}& \pr\left(\max_{0\leq j\leq M_n} (Q_{n,j} - 2 p_n) > 2 n^{-1/2}\right)   \to 0,
\end{align*}
as $n\to\infty$.
Hence,
$\|\mathtt{G}_n-\hat{\mathtt{G}}_n\|_{\mathscr{L}_\infty(
\mathcal{I}
)}  = O_{\pr}(p_n + n^{-1/2}) = O_{\pr}(l_n^{\beta} + n^{-1/2})$, implies that
\begin{align}\label{eq:V2}
    V_2 = O_{\pr} \left(\|\mathtt{G}_n-\hat{\mathtt{G}}_n\|_{\mathscr{L}_\infty(
    \mathcal{I}
    )} \right) = O_{\pr}(l_n^{\beta} + n^{-1/2}).
\end{align}

\textbf{Bound for $V_3$.}
\begin{align}\label{eq:V3}
    V_3 = |\hat{\mathtt{G}}_n(\hat{\zeta}) - \mathtt{G}(\zeta_{\VaR})| = \left|\frac{\lceil \tau n\rceil}{n} - \tau\right| \leq n^{-1}.
\end{align}

\textbf{Putting the bounds together.}
Plugging \eqref{eq:V1}, \eqref{eq:V2}, and \eqref{eq:V3} into \eqref{eq:pfgvar1} leads to
\begin{align*}%
    |\mathtt{G}(\hat{\zeta}) - \mathtt{G}(\zeta_{\VaR})| ={}&
    O_{\pr}(n^{-1/2}) +   O_{\pr}(l_n^{\beta} + n^{-1/2}) + O(n^{-1})  \\
    ={}& O_{\pr}(l_n^{\beta} + n^{-1/2}),
\end{align*}
which, together with \eqref{eq:pfgvar3}, implies that
\begin{align}\label{eq:VaR-rate}
    |\hat{\zeta}-\zeta_{\VaR}| = O_{\pr}(l_n^{\beta/\gamma} + n^{-1/(2\gamma)}).
\end{align}
Lastly, we apply Proposition~\ref{prop:rho_n} to conclude that: if $\nu \geq \frac{d}{\beta}$, we set $n=\Gamma$ to obtain the rate $\Gamma^{-\frac{\nu\beta}{(2\nu+2d)\gamma}}(\log \Gamma)^{\frac{\beta}{2\gamma}}$;
if $\nu < \frac{d}{\beta}$, we set $n\asymp \Gamma^{\frac{\beta}{\beta+1}}$ to obtain the rate $\Gamma^{-\frac{\beta}{2\gamma(\beta+1)}}(\log \Gamma)^{\frac{\beta}{2\gamma}}$.  \Halmos\endproof

\proof{Proof of Theorem~\ref{thm:gvarcvar}: The Case of CVaR.}
Let $\hat z = n^{-1}\sum_{i=1}^n (\hat{f}(\BFx_i)-\hat{f}_{(\lceil \tau n\rceil)})^+$.
By the triangle inequality, we have
\begin{align}\label{eq:pfgcvar1}
    |\hat\theta_{n,m}-\CVaR_\tau(f(X))| \leq{}& |\hat{\zeta} - \zeta_{\VaR}| + (1-\tau)^{-1}|\hat z - \E[(f(X)-\VaR_\tau(f(X)))^+]|.
\end{align}

Note that
\begin{equation}\label{eq:pfgcvar2}
\begin{aligned}
    |\hat z - \E[f(X)-\VaR_\tau(f(X)))^+]|
    \leq{}& \underbrace{\left|\frac{1}{n}\sum_{i=1}^n (\hat{f}(\BFx_i)-\hat{f}_{(\lceil \tau n\rceil)})^+ - \frac{1}{n}\sum_{i=1}^n (f(\BFx_i)-\zeta_{\VaR})^+ \right|}_{W_1} \\
    & + \underbrace{\left|\frac{1}{n}\sum_{i=1}^n (f(\BFx_i)-\zeta_{\VaR})^+ - \E[(f(X)-\VaR_\tau(f(X)))^+]\right|}_{W_2}.
\end{aligned}
\end{equation}

For $W_1$,
applying the basic inequality $|\max(a,0)-\max(b,0)|\leq |a-b|$ yields
\begin{align}
    W_1 \leq{}& \frac{1}{n}\sum_{i=1}^n \left| (\hat{f}(\BFx_i)-\hat{f}_{(\lceil \tau n\rceil)})  - (f(\BFx_i) - \zeta_{\VaR} )\right|
    \leq  \frac{1}{n}\sum_{i=1}^n \left(|\hat{f}(\BFx_i)- f(\BFx_i)| + |\hat{\zeta}  - \zeta_{\VaR}|\right) \nonumber \\
    \leq{}&   \rho_n + |\hat{\zeta}-\zeta_{\VaR}| = O_{\pr}(l_n) + |\hat{\zeta}-\zeta_{\VaR}| ,\label{eq:pfgcvarI1}
\end{align}
where  $\rho_n = \max_{1\leq i\leq n}|f(\BFx_i) - \hat{f}(\BFx_i)|$, $l_n$ is given by \eqref{eq:l_n_def}, and the last step follows from Proposition~\ref{prop:rho_n}.

For $W_2$, the central limit theorem implies that
\begin{align}\label{eq:pfgcvarI2}
    W_2 = O_{\pr}(n^{-1/2}).
\end{align}

Combining \eqref{eq:pfgcvar1}--\eqref{eq:pfgcvarI2} yields
\begin{align}\label{eq:pfgcvarbd}
    |\hat\theta_{n,m}-\CVaR_\tau(f(X))| = O_{\pr}(l_n+|\hat{\zeta}-\zeta_{\VaR}| + n^{-1/2}).
\end{align}
Plugging \eqref{eq:VaR-rate} into \eqref{eq:pfgcvarbd}, we have
\begin{align}
    |\hat\theta_{n,m}-\CVaR_\tau(f(X))| ={}& O_{\pr}(l_n+l_n^{\beta/\gamma} + n^{-1/(2\gamma)}+ n^{-1/2}) \nonumber \\
    ={}& O_{\pr}(l_n^{\beta/\gamma} + n^{-1/(2\gamma) }+ n^{-1/2}),\label{eq:pfgcvarbd2}
\end{align}
where the second equality holds because $\beta \leq \gamma$ by Assumption~\ref{assump:margin-strong}.
Therefore, since $\gamma\geq 1$, the convergence rate is $O_{\pr}(l_n^{\beta/\gamma} + n^{-1/(2\gamma) })$, which is the same as that of $|\hat{\zeta}-\zeta_{\VaR}|$, given by the case of VaR in Theorem~\ref{thm:gvarcvar}. \Halmos\endproof

\end{document}